\documentclass[]{jfm}

\usepackage{graphicx}
\usepackage{newtxtext}
\usepackage{newtxmath}
\usepackage{natbib}
\usepackage{hyperref}
\hypersetup{
	colorlinks = true,
	urlcolor   = blue,
	citecolor  = black,
}

\newcommand{\RomanNumeralCaps}[1]
\linenumbers

\usepackage{amsmath,accents}
\usepackage[labelfont={small}, font={normal}]{subfig}
\usepackage{graphicx}
\usepackage{epstopdf}
\usepackage{amssymb}
\usepackage{xcolor}

\title{Rotation of a fiber in simple shear flow of a dilute polymer solution}
\author{Arjun Sharma\aff{1}
	\corresp{\email{as3833@cornell.edu}},
	\and Donald L. Koch\aff{2}}

\affiliation{\aff{1}Sibley School of Mechanical and Aerospace Engineering, Cornell University, Ithaca, NY, 14853, USA
	\aff{2}Robert Frederick Smith School of Chemical and Biomolecular Engineering, Cornell University, Ithaca, NY, 14853, USA}
\begin{document}
	\maketitle
\begin{abstract}
The motion of a freely rotating prolate spheroid in a simple shear flow of a dilute polymeric solution is examined in the limit of large particle aspect ratio, $\kappa$. A regular perturbation expansion in the polymer concentration, $c$, a generalized reciprocal theorem, and slender body theory to represent the velocity field of a Newtonian fluid around the spheroid are used to obtain the $\mathcal{O}(c)$ correction to the particle's orientational dynamics. The resulting dynamical system predicts a range of orientational behaviors qualitatively dependent upon $c\cdot De$ ($De$ is the imposed shear rate times the polymer relaxation time) and $\kappa$ and quantitatively on $c$. At a small but finite $c\cdot De$, the particle spirals towards a limit cycle near the vorticity axis for all initial conditions. Upon increasing $\kappa$, the limit cycle becomes smaller. Thus, ultimately the particle undergoes a periodic motion around and at a small angle from the vorticity axis. At moderate $c\cdot De$, a particle starting near the flow-gradient plane departs it monotonically instead of spirally, as this plane (a limit cycle at smaller $c\cdot De$) obtains a saddle and an unstable node. The former is close to the flow direction. Upon further increasing $c\cdot De$, the saddle-node changes to a stable node. Therefore, depending upon the initial condition, a particle may either approach a periodic orbit near the vorticity axis or obtain a stable orientation near the flow direction. Upon further increasing $c\cdot De$, the limit cycle near the vorticity axis vanishes, and the particle aligns with the flow direction for all starting orientations.
\end{abstract}
	
	
\section{Introduction}\label{sec:Introduction}
\begin{figure}
	\centering	
	\subfloat{\includegraphics[width=0.495\textwidth]{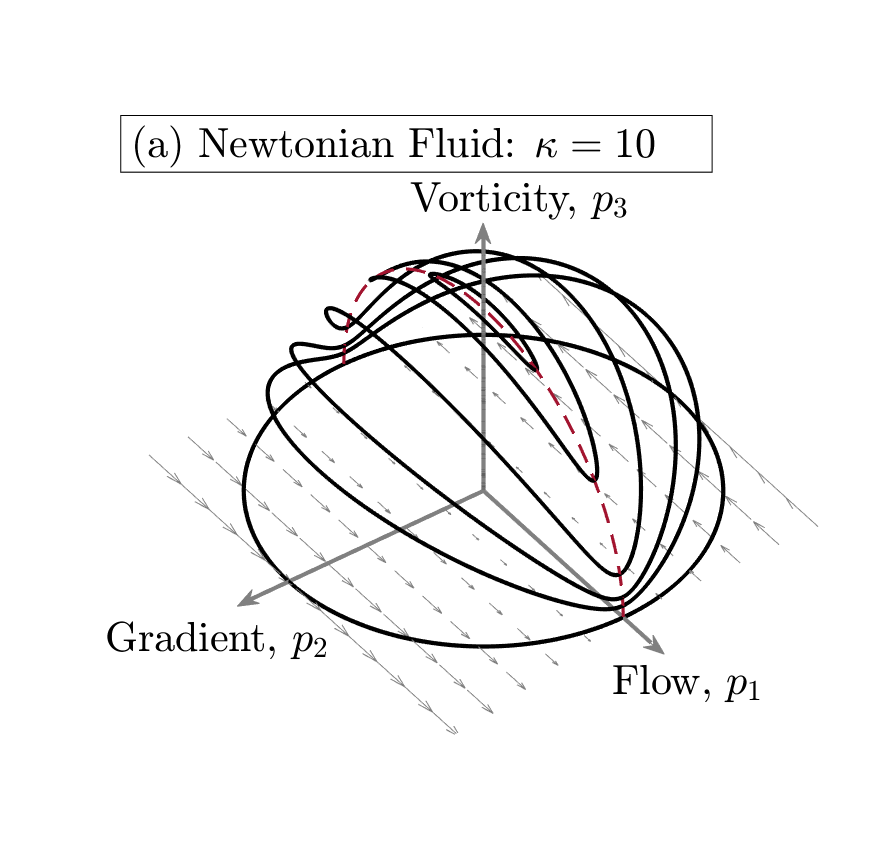}}
	\subfloat{\includegraphics[width=0.495\textwidth]{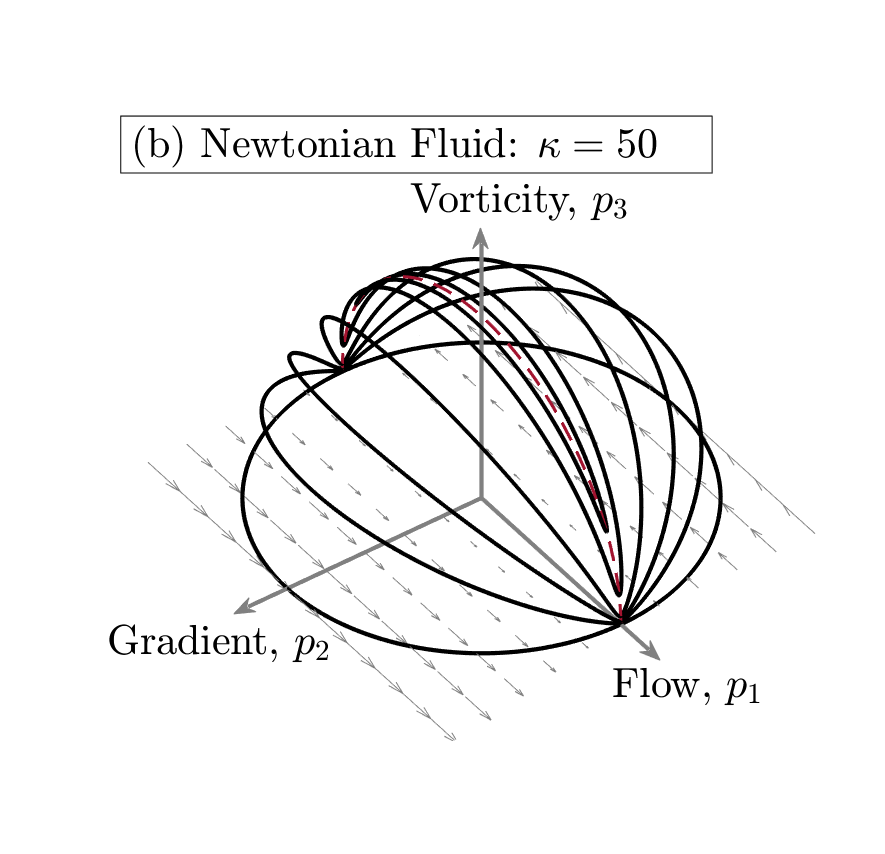}}
	\caption {\citet{jeffery1922motion} orbits or orientation trajectories in a simple shear flow of Newtonian fluid of a prolate spheroidal particle with aspect ratio, $\kappa=$ (a) 10, and (b) 50.\label{fig:TrajectoriesNewtonian}}
\end{figure}

{A particle-filled viscoelastic polymeric fluid undergoes simple shear flow in many industrial applications such as fiber spinning \citep{breitenbach2002melt,huang2003review,nakajima1994advanced,chae2008making} and roll-to-roll manufacturing of high aspect ratio, low resistance films for flexible and transparent electronics \citep{mutiso2013integrating,yin2010inkjet}. Simple shear flow occurs in the spinneret during fiber spinning and in the patterning channel during roll-to-roll manufacturing. The suspending viscoelastic fluid may include large aspect ratio particles to impart strength to the final product in fiber spinning or provide a desired anisotropy to the low resistance film. In the simple shear flow of an inertia-less Newtonian fluid, a fiber/ slender particle undergoes an initial condition-dependent periodic motion in orientational trajectories termed \cite{jeffery1922motion} orbits as shown in figure \ref{fig:TrajectoriesNewtonian} for particles with aspect ratio, $\kappa=10$ and 50. However, the interaction between the polymers in a viscoelastic fluid and the fiber breaks this degenerate periodic behavior. Previous experiments \citep{gauthier1971particle,bartram1975particle,stover1990motion,johnson1990dynamics,iso1996orientation1,gunes2008flow} indicate that depending on the $\kappa$, imposed shear rate and the properties of the viscoelastic fluid such as polymer concentration and relaxation time, a particle may exhibit various orientation dynamics. A slender particle, i.e., one with a large $\kappa$, may either spiral or monotonically drift towards the vorticity axis, align near the flow direction or settle somewhere within the flow-vorticity plane. Therefore, careful design and choice of flow parameters during the simple shear regime are essential in obtaining a final product with desired particle orientation and material strength. Theoretical studies are useful due to the many parameters required for characterizing a viscoelastic fluid. Polymers lead to new features in a viscoelastic fluid flow such as shear thinning or a finite first and second normal stress difference as compared to a Newtonian fluid flow. \cite{leal1975slow} predicts that a slender particle in a slow flow will spiral towards the vorticity axis due to the second normal stress difference in the fluid. Whereas, operating in a double limit of small polymer concentration and large Deborah number, $De$, (the product of the imposed shear rate and the polymer relaxation time) \cite{harlen1993simple} also predict the spiraling of the particle towards the vorticity axis, but due to first normal stress difference in the fluid. Neither of these theories captures any other orientation behavior observed experimentally. In this paper, using a regular perturbation expansion in polymer concentration, $c$, we develop a slender body theory that spans a range of $De$. It encapsulates the $\mathcal{O}(c)$ effect of particle-polymer interaction and qualitatively describes the rich orientation dynamics seen in previous experiments.}

{Simple shear flow is ubiquitous in industrial applications as the near wall flow can always be considered locally simple shear. Many scenarios include laminar flow between two parallel walls; in these cases, the local flow is simple shear everywhere. In many industrial scenarios, particle concentration in the suspension is dilute, and due to negligible particle-particle interaction, each particle can be considered to be suspended in an unbounded fluid. In their experiments with highly elastic fluids, \cite{iso1996orientation} observed an isolated fiber to obtain a stable orientation close to the direction of the imposed flow. They found that fibers in a moderate particle concentration suspension also obtained a highly peaked orientation distribution near the flow direction for the same parameters. Therefore, the particle-particle interaction may sometimes be ignored even with higher particle concentrations, making the studies of an isolated particle/ fiber freely rotating in a simple shear flow of a viscoelastic fluid invaluable.}

{Based on their qualitative nature, the \cite{jeffery1922motion} orbits, illustrated for particles with aspect ratio, $\kappa=10$ and 50 in figure \ref{fig:TrajectoriesNewtonian}, may be classified into log-rolling, wobbling, flipping, and tumbling. Log-rolling occurs when the particle is aligned with the axis of rotation of the imposed flow or the vorticity axis. Here the particle rotates about its major axis. Except in the log-rolling motion, the particle's angular velocity changes throughout its orbit. When initially placed in the flow-gradient plane, the particle remains in the plane. It rotates about its minor axis while tumbling from one side of the flow axis to the other. In the tumbling and flipping orbits, a large aspect ratio particle or fiber spends only an $\mathcal{O}(1/\kappa)$ proportion (non-dimensionalized with shear rate) of the Jefferey time-period of $2\pi\kappa$ away from the flow-vorticity plane indicated by dashed red lines in the plots of figure \ref{fig:TrajectoriesNewtonian}. In the flipping orbits (which are  three-dimensional extensions of the tumbling orbit when the particle is not confined to the flow-gradient plane), the particle comes within $\mathcal{O}(1/\kappa)$ of the flow direction. In these orbits, the particle's orientation rapidly flips from being aligned with the positive to the negative flow axis. During its rapid flipping phase, a particle in a flipping orbit spans a large portion of the orientation space in the gradient direction. In wobbling orbits, which are smaller in circumference than flipping orbits, the particle does not come very close to the flow direction. In these orbits, the particle gradually wobbles in its orbit around the vorticity axis.} 

\citet{gauthier1971particle} conducted experiments with $\kappa=16.1$ nylon rods in a viscoelastic fluid made with 0.03 wt. \% polyacrylamide solution in water. They conducted experiments at a shear rate of $0.53s^{-1}$ and found that a particle starting close to the flow-gradient plane spirals towards the vorticity axis as it is exposed to the Couette flow. Using a similar viscoelastic fluid and a polyethylene rod of $\kappa=9.0$  \citet{bartram1975particle} also found a similar behavior with shear rates up to $5 s^{-1}$. Upon further increase in shear rate, they found that a $\kappa=9.0$ rod released near the gradient direction initially moves within the flow-gradient plane towards an orientation near the flow direction. From here, introducing a disturbance made the particle move monotonically along the flow-vorticity plane away from the flow direction. When the particle was sufficiently close to the vorticity direction, it started to spiral toward it. \citet{bartram1975particle} observed similar behavior with a $\kappa=5.6$ rod. However, unlike the $\kappa=9.0$ rod, no disturbance was required when the particle came close to the flow direction after being placed near the gradient direction. The time period of particle rotation about the vorticity axis, that is already very large at large $\kappa$ in Newtonian fluid ($2\pi\kappa$), is further increased in experiments with viscoelastic fluids \citep{gauthier1971particle,bartram1975particle}. In the experiments where the orientation of the particle centerline was found to be spiraling towards the vorticity axis, complete alignment with the vorticity axis was not shown.

\cite{iso1996orientation1} observed rotations of different high $\kappa$ isolated fibers in a Boger fluid consisting of 1000 ppm polyisobutylene (PIB) in polybutene (PB) in a simple shear flow with different shear rates. {The polymer relaxation time and concentration, $c$, defined as the polymer-to-solvent viscosity ratio, were 3$s$ and 0.39, respectively. With a $\kappa=19$ fiber, they found the particle to be orientated very close to the vorticity axis when $De=1.5$.} Increasing $\kappa$ to 34.4, or $De$ to 3.0 or both, they found that, after initial spiraling away from the flow-gradient plane, the particle obtained a steady orientation between 5$^\circ$ and 60 $^\circ$ from the vorticity axis in the flow-vorticity plane. With $\kappa=34.4$ and $De=3.0$, they report two additional observations with no initial spiraling in contrast to other experiments at identical parameters. The authors attributed different initial orientations and fluid rheology due to slight changes in room temperature as the causes of the lack of initial spiraling.

Across their two studies \cite{iso1996orientation,iso1996orientation1} also conducted experiments in a viscoelastic liquid obtained by adding a certain amount of high molecular weight polymer polyacrylamide (PAA) to a Newtonian solvent. The shear rate in these experiments was $0.5s^{-1}$, and the fluid was slightly shear-thinning. They observed various behaviors as the amount of PAA was increased from 100ppm to 2000ppm (although the exact value of $c$ is not available, it increases with PAA amount). For 100ppm ($\kappa=14$) and 500 ppm ($\kappa=24$) solutions of PAA at a shear rate of 0.5$s^{-1}$ \citep{iso1996orientation,iso1996orientation1}, fibers either end up in a trajectory where they oscillate in a small periodic orbit close to the vorticity axis or obtain a stable orientation in the flow-vorticity plane at a particular angle from the vorticity axis similar to the Boger fluid \citep{boger1977highly} experiments at a higher shear rate of $1.0s^{-1}$ \citep{iso1996orientation1}. With 1000 ppm PAA, the $\kappa=24$ fiber obtains a stable orientation at the flow direction or 20$^\circ$ from the flow direction in the flow-vorticity plane. Irrespective of the initial condition, a $\kappa=24$ fiber in 2000 ppm PAA solution stably aligns with the flow direction. Therefore, fibers become more flow aligned with increasing elasticity or polymer concentration.

The latest available experimental results measuring the effect of viscoelasticity on the rotation of an anisotropic particle in simple shear flow are by \cite{gunes2008flow}. They considered hematite spheroidal particles with a much smaller aspect ratio, $\kappa$, between 2 and 7.5, than the previous experimental studies. In a 20\% hydroxypropylcellulose solution in water, they found $\kappa=3.8$ particles to be oriented close to vorticity and flow directions at low and large shear rates or $De$, respectively. At moderate shear rates, particles exhibited a bimodal orientation distribution. In a 2\% poly-(ethyleneoxide), flow alignment was obtained at a lower shear rate for a higher $\kappa$ or $c$. Most of the fluids they used were shear-thinning. They reported one experiment of a non-shear-thinning Boger fluid \citep{boger1977highly}, consisting of a 0.1\% polyisobutylene in polybutadiene solution, with $\kappa=3.8$. Here the particles were close to the vorticity axis at all shear rates reported.

{Due to the variety of non-Newtonian parameters needed to fully characterize a viscoelastic fluid and several physical phenomena that polymers may undergo simultaneously, it is difficult to quantitatively compare the previous experiments and identify the source of various behaviors of particle orientation. For example, temperature changes during an experiment may change the rheology of the fluid, subsequently changing the polymer's relaxation time, or multiple relaxation times may be needed to represent the fluid fully, or adding more polymers to a solution may not only increase the polymer concentration, but it may also change the relaxation time of polymers as they entangle with one another. Thus numerical and theoretical modeling of the relevant system is an important tool in isolating and understanding different physical mechanisms that can complement or inspire future experiments.}

Recently, \cite{d2014bistability} reported numerical simulations of a $\kappa=4.0$ prolate spheroidal particle in a simple shear flow of a Giesekus fluid that models polymer melts \citep{bird1987dynamics}, with large polymer concentration. They used $c=10$ and found spiraling towards the vorticity axis for $De\lesssim1$ and alignment close to the flow direction for $De\gtrsim3$. For intermediate $De$, depending on $De$, the particle obtained either one or two stable orientations between the flow and vorticity axis. They mentioned similar observations in unreported simulations with $\kappa=8$ spheroids with the transition from vorticity- to flow-aligned particle orientation occurring at a smaller $De$. Therefore, the trend of the particle's final orientation with the shear rate or $De$ and $\kappa$ between the experiments of \cite{gunes2008flow} and the numerical study of \cite{d2014bistability}, both conducted at small $\kappa$, is similar. However, it is unclear from the aforementioned numerical findings if the novel orientation dynamics are due to first or second normal stress difference, shear thinning, or synergistic effects of various non-Newtonian behaviors exhibited by the Giesekus fluid. Also, at intermediate shear rates or $De$, while the orientation behavior is bimodal, i.e., either along vorticity or flow directions, in the experiments of \cite{gunes2008flow}, the final orientations in numerical results of \cite{d2014bistability} lie between the flow and vorticity axes in the flow-vorticity plane. The latter is instead similar to some of the experimental observations of \cite{iso1996orientation} at larger $\kappa$.

A richer orientation behavior is illustrated in the previous experiments of \cite{gauthier1971particle,bartram1975particle,iso1996orientation,iso1996orientation1} at larger $\kappa$ as compared to the more recent studies of \cite{gunes2008flow,d2014bistability} at smaller $\kappa$. Numerical studies with large $\kappa$ in a viscoelastic fluid are expensive due to the large velocity and polymer conformation gradients near the particle surface. Resolving these large gradients and accurately modeling the shape of a slender particle requires very fine spatial resolution and hence smaller time steps to ensure numerical stability. Therefore, theoretical studies at large $\kappa$ are invaluable, and \cite{leal1975slow,harlen1993simple,abtahi2019jeffery} are such pre-existing examples.

Using the slender body theory of \citet{batchelor1970slender}, \citet{leal1975slow} predicts that a fiber rotating in a slow, simple shear flow of a second order fluid, will spiral towards the vorticity axis due to the second normal stress difference, $\psi_2$, of the fluid. $\psi_2$ is usually smaller than the first normal stress difference, $\psi_1$, and it is zero for Boger fluids \citep{boger1977highly}. Hence, Leal's theory predicts that a slender particle rotating in a Boger fluid undergoing a simple shear flow with a small shear rate rotates in the same manner as in a Newtonian fluid. However, the low shear rate experiments of \citet{iso1996orientation1} with a Boger fluid show spiraling towards vorticity. For a large Deborah number, $De\gg1$, small polymer concentration, $c\ll1$ and also using the slender body theory of \citet{batchelor1970slender}, \citet{harlen1993simple} predict the fiber in an Oldroyd-B fluid to spiral towards the vorticity axis, but, due to $\psi_1$ (an Oldroyd-B fluid has $\psi_2=0$). Shear-thinning, another property exhibited by polymeric fluids, does not play a role in either of these theories. \cite{abtahi2019jeffery} conducted a theoretical study on an asymptotically weakly shear thinning liquid and found that a prolate spheroid rotates in a closed periodic orbit but with a longer time period compared with the Jeffery orbit in a Newtonian fluid. {In other words, shear thinning makes a prolate spheroid's trajectory equivalent to that of a larger aspect ratio particle in a Newtonian fluid but does not qualitatively alter the topology of the trajectories.}

Spiraling towards vorticity, as indicated by the two previous theories \citep{leal1975slow,harlen1993simple} predicting a qualitative change in particle orientation trajectory, is only one of the many qualitative influences of viscoelasticity observed in the previous simple shear experiments of \cite{gauthier1971particle,bartram1975particle,iso1996orientation,iso1996orientation1}  at large particle aspect ratio, $\kappa$. In this paper, assuming a small polymer concentration, we theoretically revisit the effect of viscoelasticity on a slender fiber rotating in a simple shear flow to explain the richer qualitative behavior of a large $\kappa$ particle's orientation in a polymeric fluid observed in the experiments. We use the Oldroyd-B model to isolate the effect of elasticity from shear thinning. Also, any predicted viscoelastic effects will exclude the {impact} of the second normal stress difference and re-examine the fiber rotation in Boger fluid undergoing simple shear flow.

In the absence of fluid inertia, fluid stress at any point in the viscoelastic fluid surrounding a suspended particle is decomposed into three components: a) Particle motion induced solvent stress, i.e., the stress around the particle rotating in a Newtonian fluid, b) Elastic stress or the polymer stress, and, c) Polymer-induced solvent stress, i.e., the stress created by the perturbations in fluid velocity and pressure due to the forcing by polymer stress. Therefore, the torque acting on the particle is the sum of particle motion-induced solvent torque (MIST), elastic torque, and polymer-induced solvent torque (PIST). We consider a freely rotating or torque-free particle where the sum of the three torque components is zero.

The rest of the paper is organized as follows. In section \ref{sec:MathBackground} we discuss different torque generating mechanisms along with the mathematical formulation relevant to any freely rotating (torque-free) particle in a viscoelastic fluid. For any polymer concentration, $c$, using a generalized reciprocal theorem, we derive the formulation where PIST on any such particle can be expressed in terms of the polymer stress instead of the polymer-induced solvent stress. After section \ref{sec:MathBackground} we are concerned with viscoelastic fluid with a small polymer concentration, $c$. Therefore, in section \ref{sec:LowcFormulation} we briefly describe the regular perturbation expansion of the relevant flow variables in $c$ and the procedure to obtain the $\mathcal{O}(c)$ correction to the rotation of a torque-free particle in a low $c$ viscoelastic fluid. The formulation in section \ref{sec:MathBackground} that expresses PIST in terms of the polymer stress allows us to circumvent the numerical discretization of the partial differential equations to calculate the $\mathcal{O}(c)$ polymer-induced solvent stress. Therefore, the $\mathcal{O}(c)$ PIST (and also the elastic torque) can be evaluated using the leading order or Newtonian velocity field around the particle. In section \ref{sec:SBTVelocityTorques}, we use the matched asymptotic/ slender body theory (SBT) solution for the Newtonian flow field around a slender prolate spheroid to calculate the torques and obtain the $\mathcal{O}(c)$ correction to the \citet{jeffery1922motion} rotation rates for large aspect ratio prolate spheroids due to viscoelasticity. {In SBT, in the inner region close to the particle, owing to a large $\kappa$, the velocity field is obtained by assuming the flow to vary slowly along the length of the particle compared to its variation perpendicular to the particle surface. This solution is taken from \citet{cox1970motion}. Further from the particle surface, in the outer region, the particle center line is replaced by a line of Stokeslets and doublets. The velocity disturbance created by these are taken from \citet{batchelor1970slender}, and \citet{cox1971motion} respectively. In the SBT \citep{cox1970motion,cox1971motion,batchelor1970slender}, the inner and the outer solution approach one another in the matching reaching, i.e., in the dual limit of the radial distance from the particle centerline written in inner and outer variables approaching infinity and zero, respectively.} In section \ref{sec:MainResults} we study the dynamical system defined by these equations for different $c$ and $De$ and illustrate various orientation dynamics of the spheroid predicted by this system. We also provide a qualitative comparison of the theoretical orientation trajectories with the previous experimental observations. Finally, we conclude our findings in section \ref{sec:Conclusions}.
	
\section{Mathematical formulation and different torque generating mechanisms in viscoelastic
	fluids}\label{sec:MathBackground}
In the absence of inertia, equations governing the conservation of mass and momentum in the viscoelastic fluid surrounding a particle are,
\begin{equation}
\nabla\cdot\mathbf{u}=0,\hspace{0.2in}\nabla\cdot\boldsymbol{\sigma}=0, \label{eq:MainGovern}
\end{equation}
where $\mathbf{u}$ and $\boldsymbol{\sigma}$ are the fluid velocity vector and stress tensor field. We consider a particle with its center of mass at the origin of a simple shear flow such that it freely rotates with an angular velocity $\boldsymbol{\omega}_p$, but does not translate. Therefore, the no-slip boundary condition on particle surface and the far-field (as $|\mathbf{r}|\rightarrow\infty$) imposed flow boundary condition are,
\begin{eqnarray}
{\mathbf{u}=\boldsymbol{\omega}_p\times\mathbf{r}, \hspace{0.2in}\text{ on particle surface },\hspace{0.2in}\text{and, }
\mathbf{u}=\mathbf{r}\cdot (\nabla \mathbf{u})_\infty,  \hspace{0.2in}\text{ as }|\mathbf{r}|\rightarrow\infty,}
\end{eqnarray}
where $(\nabla \mathbf{u})_\infty$ is the velocity gradient tensor of the imposed flow. The equations are non-dimensionalized with the particle's maximum length and the inverse of the imposed shear rate as length and time scales. The fluid stress,
\begin{equation}
\boldsymbol{\sigma}=\boldsymbol{\tau}+\mathbf{\Pi},
\end{equation}
is the sum of the solvent stress, $\boldsymbol{\tau}=-p\mathbf{I}+\nabla \mathbf{u}+(\nabla \mathbf{u})^T$ and the polymeric stress, $\boldsymbol{\Pi}$. The solvent viscosity is the viscosity scale in our non-dimensionalization. In the solvent stress, $p$ is the fluid pressure. We use the Oldroyd-B model  \citep{bird1987dynamics} where the polymers are modeled as dumbells consisting of two beads attached to a linearly elastic spring. The polymeric stress is,
\begin{equation}
\boldsymbol{\Pi}=\frac{c}{De}(\boldsymbol{\Lambda}-\mathbf{I}),\label{eq:PolymerStressDefintion}
\end{equation}
where $c$ is the polymer concentration, $De$ is a non-dimensional product of the polymer relaxation time and imposed shear rate, $\boldsymbol{\Lambda}$ is the polymer conformation (defined as the average over possible polymer conformations of the outer product of the polymer end to end vector with itself)  and $\mathbf{I}$ is the identity tensor. $\boldsymbol{\Lambda}$ is affected by convection due to the fluid velocity, stretching and rotation by the velocity gradients, and relaxation of the polymer to its equilibrium orientation, $\boldsymbol{\Lambda}=\mathbf{I}$. It is governed by,
\begin{equation}
\frac{\partial \boldsymbol{\Lambda}}{\partial t}+\mathbf{u}\cdot\nabla\boldsymbol{\Lambda}=(\nabla \mathbf{u})^T\cdot\boldsymbol{\Lambda}+\boldsymbol{\Lambda}\cdot\nabla\mathbf{u}+\frac{1}{De}(\mathbf{I}-\boldsymbol{\Lambda}).\label{eq:Constitutive}
\end{equation}
Due to the absence of any non-linear term explicitly involving the velocity and pressure variables in the mass and momentum conservation equations, we can split equation \eqref{eq:MainGovern} and its boundary conditions into two parts. The first part is the same as the flow around a particle rotating with an angular velocity $\boldsymbol{\omega}_p$ in an imposed simple shear flow of a Newtonian fluid. It is governed by,
\begin{equation}
{\nabla\cdot\mathbf{u}^\text{M}=0,\hspace{0.2in}\nabla\cdot\boldsymbol{\tau}^\text{M}=0, \label{eq:MainGovernDecompose1}}
\end{equation}
{where the particle motion induced solvent stress, $\boldsymbol{\tau}^\text{M}=-p^\text{M}\mathbf{I}+\nabla \mathbf{u}^\text{M}+(\nabla \mathbf{u}^\text{M})^T$, is the sum of the pressure and rate of strain tensor in the solvent due to the particle motion in an inertia-less Newtonian fluid. These equations are subject to boundary conditions,
\begin{eqnarray}
\mathbf{u}^\text{M}=\boldsymbol{\omega}_p\times\mathbf{r}, \hspace{0.2in}\text{ on particle surface},\hspace{0.2in}\text{and, }
\mathbf{u}^\text{M}=\mathbf{r}\cdot (\nabla \mathbf{u})_\infty,  \hspace{0.2in}\text{ as }|\mathbf{r}|\rightarrow\infty.\label{eq:NewtonianBCs}
\end{eqnarray}
The second part is the balance of the divergence of the sum of the polymeric stress and $\boldsymbol{\tau}^\text{P}$,
\begin{equation}
\nabla\cdot\mathbf{u}^\text{P}=0,\hspace{0.2in}\nabla\cdot\boldsymbol{\tau}^\text{P}+\nabla\cdot\boldsymbol{\Pi}=0, \label{eq:MainGovernDecompose2}
\end{equation}
where the solvent stress generated due to the forcing by the polymeric stress is $\boldsymbol{\tau}^\text{P}=-p^\text{P}\mathbf{I}+\nabla \mathbf{u}^\text{P}+(\nabla \mathbf{u}^\text{P})^T$. $p^\text{P}$ and $(\nabla \mathbf{u}^\text{P}+(\nabla \mathbf{u}^\text{P})^T)/2$ are the modification of pressure and rate of strain tensor by the polymers. Boundary conditions for equation \eqref{eq:MainGovernDecompose2} are,
\begin{eqnarray}
\mathbf{u}^\text{P}=0, \hspace{0.2in}\text{ on particle surface},\hspace{0.2in}\text{and, }
\mathbf{u}^\text{P}=0,  \hspace{0.2in}\text{ as }|\mathbf{r}|\rightarrow\infty.\label{eq:BCNonNewtMom}
\end{eqnarray}
 The two parts are coupled via equation \eqref{eq:Constitutive}, i.e., the polymer constitutive equation, where
\begin{equation}\mathbf{u}=\mathbf{u}^\text{M}+\mathbf{u}^\text{P},\label{eq:TotalVel}\end{equation}
and the torque-free condition (equation \eqref{eq:TorqueFree}).
In this framework, the total fluid stress,
\begin{equation}
\boldsymbol{\sigma}=\boldsymbol{\tau}^\text{M}+\boldsymbol{\tau}^\text{P}+\boldsymbol{\Pi},
\end{equation}
and the total torque acting on the particle surface, $\boldsymbol{\sigma}$,
\begin{equation}
\mathbf{G}(\boldsymbol{\sigma})=\int_{\mathbf{r}_\text{p}} dS\hspace{0.1in}\mathbf{r}\times( \boldsymbol{\sigma}\cdot \mathbf{n}),\label{eq:StressFree}
\end{equation}
\begin{equation}
\mathbf{G}(\boldsymbol{\sigma})=\mathbf{G}(\boldsymbol{\tau}^\text{M})+\mathbf{G}(\boldsymbol{\tau}^\text{P})+\mathbf{G}(\boldsymbol{\Pi})=\mathbf{G}^\text{MIST}+\mathbf{G}^\text{PIST}+\mathbf{G}^\text{Elastic},
\end{equation}
are decomposed into three underlying mechanisms,
where,
\begin{align}
\mathbf{G}^\text{MIST}&=\mathbf{G}({\boldsymbol{\tau}^\text{M}})=\int_{\mathbf{r}_\text{p}} dS\hspace{0.1in}\mathbf{r}\times \boldsymbol{\tau}^\text{M}\cdot \mathbf{n},\label{eq:DiffTorques1} \\
\mathbf{G}^\text{PIST}&=\mathbf{G}({\boldsymbol{\tau}^\text{P}})=\int_{\mathbf{r}_\text{p}} dS\hspace{0.1in}\mathbf{r}\times \boldsymbol{\tau}^\text{P}\cdot \mathbf{n},\text{ and},\label{eq:DiffTorques2}\\
\mathbf{G}^\text{Elastic}&=\mathbf{G}(\boldsymbol{\Pi})=\int_{\mathbf{r}_\text{p}} dS\hspace{0.1in}\mathbf{r}\times\boldsymbol{\Pi}\cdot \mathbf{n},
\label{eq:DiffTorques3}\end{align}}
are the particle motion induced solvent, the polymer induced solvent, and the elastic or polymeric torques, respectively. The angular velocity, $\boldsymbol{\omega}_p$, of a freely rotating particle, is determined by the torque-free condition,
\begin{equation}
\boldsymbol{\omega}_p=\{\boldsymbol{\omega}_p: \mathbf{G}=\mathbf{G}^\text{MIST}+\mathbf{G}^\text{PIST}+\mathbf{G}^\text{Elastic}=0\}.\label{eq:TorqueFree}
\end{equation}

We consider the motion of a freely rotating particle in a viscoelastic fluid. Our main motivation is to study the behavior of a prolate spheroid in simple shear flow. Due to symmetry, this particle has zero net hydrodynamic force if it moves with the local velocity of the imposed flow. Its physically motivated decomposed components (particle motion induced solvent, polymer induced solvent, and elastic forces) are individually zero by symmetry. However, the force balance can be considered similar to the torque balance discussed above for determining the motion of a freely translating particle (of any shape) in a general linear flow or a particle sedimenting under gravity (where the net hydrodynamic force must balance the force due to gravity).

\subsection{Using a generalized reciprocal theorem to obtain ${\mathbf{G}^\text{PIST}}$ in terms of $\boldsymbol{\Pi}$\label{sec:Reciprocal}}
{In its original form, $\mathbf{G}^\text{PIST}=\int_{\mathbf{r}_\text{p}} dS\hspace{0.1in}\mathbf{r}\times \boldsymbol{\tau}^\text{P}\cdot \mathbf{n}$  is a function of $\boldsymbol{\tau}^\text{P}$ which in turn is driven by $\boldsymbol{\Pi}$ through equation \eqref{eq:MainGovernDecompose2}.  In this section, using the balance of the divergence of the polymeric stress ($\boldsymbol{\Pi}$) and $\boldsymbol{\tau}^\text{P}$ from equation \eqref{eq:MainGovernDecompose2} and using a generalized reciprocal theorem we derive an expression to obtain $\mathbf{G}^\text{PIST}$ directly from $\boldsymbol{\Pi}$ without the need to compute $\boldsymbol{\tau}^\text{P}$.}

The auxiliary or comparison problem in a generalized reciprocal theorem must be chosen based on the surface integral one is interested in evaluating. The effect of the torque is to rotate the particle. Hence, we consider the Stokes flow around a particle rotating in a quiescent Newtonian fluid. We define the following auxiliary Stokes problem for a `velocity' field consisting of a rank-2 pseudo-tensor $\mathbf{b}$,  `pressure' field that is a pseudovector, $\mathbf{q}$, and `fluid stress' field that is a rank-3 pseudo-tensor $\mathbf{B}$,
\begin{eqnarray}
\frac{\partial B_{ijk}}{\partial r_i}=0,\hspace{0.2in} \frac{\partial b_{ij}}{\partial r_i}=0,\hspace{0.2in} B_{ijk}=-\delta_{ij}q_k+\frac{\partial b_{jk}}{\partial r_i}+\frac{\partial b_{ik}}{\partial r_j},\label{eq:AuxEqns}
\end{eqnarray}
with boundary condition
{\begin{equation}
b_{ij}=\epsilon_{ijk}r_k, \hspace{0.2in} \text{on particle surface}, \hspace{0.2in} \text{and}, \hspace{0.2in} b_{ij}=0, \hspace{0.2in} \text{as } |\mathbf{r}|\rightarrow\infty, \label{eq:AuxBoundary}
\end{equation}}
where $\epsilon_{ijk}$ is the permutation tensor. $\mathbf{b}\cdot\boldsymbol{\omega}_\text{auxillary}$ is the velocity field around a particle rotating with an angular velocity $\boldsymbol{\omega}_\text{auxillary}$ in a quiescent Newtonian fluid.
Using the definitions of {$\boldsymbol{\tau}^\text{P}$} and $\mathbf{B}$ in terms of the respective velocities, {$\mathbf{u}^\text{P}$} and $\mathbf{b}$, incompressibility of the velocities in equations \eqref{eq:MainGovernDecompose2} and \eqref{eq:AuxEqns}, $\nabla\cdot\mathbf{B}=0$ from equation \eqref{eq:AuxEqns} and the symmetry of $B_{lki}$ and {${\tau}^\text{P}_{lk}$} about the $l$ and $k$ indices we obtain,
\begin{equation}
{{\tau}^\text{P}_{lk}\frac{\partial b_{ki}}{\partial r_l}=\frac{\partial B_{lki} {u}^\text{P}_k}{\partial r_l}.}\label{eq:StressSymmetry}
\end{equation}
Using the balance of the divergence of $\boldsymbol{\Pi}$ and {$\boldsymbol{\tau}^\text{P}$} from equation \eqref{eq:MainGovernDecompose2}, the volume integral of equation \eqref{eq:StressSymmetry} in a region bounded by the particle surface, $\mathbf{r}_\text{p}$, and a far-field spherical surface at $|\mathbf{r}|\rightarrow\infty$ is
\begin{equation}{\int_\text{Fluid} dV\hspace{0.1in}\frac{\partial}{\partial r_l}[{\tau}^\text{P}_{lk}b_{ki}-B_{lki}{u}^\text{P}_k]=-\int_\text{Fluid} dV\hspace{0.1in}b_{ki}\frac{\partial{\Pi}_{lk}}{\partial r_l}.}\label{eq:ReciprocalDerivation}\end{equation}
{Using the divergence theorem, the left side of the above equation can be written as,}
{\begin{align}\begin{split}&\int_\text{Fluid} dV\hspace{0.1in}\frac{\partial}{\partial r_l}[{\tau}^\text{P}_{lk}b_{ki}-B_{lki}{u}^\text{P}_k]=\\&
\int_{|\mathbf{r}|\rightarrow\infty} dS\hspace{0.1in}n_l[{\tau}^\text{P}_{lk}b_{ki}-B_{lki}{u}^\text{P}_k]-
\int_{\mathbf{r}_\text{p}} dS\hspace{0.1in}n_l[{\tau}^\text{P}_{lk}b_{ki}-B_{lki}{u}^\text{P}_k],
\end{split}\label{eq:ReciprocalIntermediate}\end{align}}
where the surface normal $n_l$ points into the fluid (away from particle region) on the particle surface.
The fluid velocity due to a particle that exerts a force (force dipole) on the fluid decays as $1/r$ ($1/{r}^2$) in the far field.   Hence, the velocities {$\mathbf{u}^\text{P}$} and $\mathbf{b}$ scale as $1/r$ and $1/{r}^2$, respectively, and the stresses {$\boldsymbol{\tau}^\text{P}$} and $\mathbf{B}$ scale as $1/{r}^2$ and $1/{r}^3$. Therefore, the first surface integral in equation \eqref{eq:ReciprocalIntermediate} vanishes. Using the boundary conditions for {$\mathbf{u}^\text{P}$} and $\mathbf{b}$ from equations \eqref{eq:BCNonNewtMom} and \eqref{eq:AuxBoundary} in the second surface integral in equation \eqref{eq:ReciprocalIntermediate} we obtain,
{\begin{equation}\int_\text{Fluid} dV\hspace{0.1in}\frac{\partial}{\partial r_l}({\tau}^\text{P}_{lk}b_{ki}-B_{lki}{u}^\text{P}_k)=\int_{\mathbf{r}_\text{p}}dS\hspace{0.1in}n_l{\tau}^\text{P}_{lk}\epsilon_{kim}r_m=-{G}^\text{PIST}_i.\label{eq:DivergenceResultReciprocal}\end{equation}}
Therefore, from equation \eqref{eq:ReciprocalDerivation} and \eqref{eq:DivergenceResultReciprocal}, for a particle of any shape,
\begin{equation}
\boldsymbol{G}^\text{PIST}=\int_\text{Fluid} dV\hspace{0.1in}(\nabla\cdot{\boldsymbol{\Pi}})\cdot\mathbf{b}.\label{eq:PISTReciprocal}
\end{equation}
This completes the derivation expressing $\boldsymbol{G}^\text{PIST}$ in terms of the polymer stress $\boldsymbol{\Pi}$ and a quasi-steady 2-tensor field $\mathbf{b}$ that is dependent on the particle shape and is a solution of the auxiliary Stokes problem defined by equations \eqref{eq:AuxEqns} and \eqref{eq:AuxBoundary}. 

In a linear imposed flow such as a simple shear, the undisturbed polymer stress, $\boldsymbol{\Pi}^U$, i.e., the polymer stress without the particle, is spatially constant. Hence,
\begin{equation}
\boldsymbol{G}^\text{PIST}=\int_\text{Fluid} dV\hspace{0.1in}(\nabla\cdot{(\boldsymbol{\Pi}-\boldsymbol{\Pi}^U)})\cdot\mathbf{b}.\label{eq:PISTOriginal}
\end{equation}
Using the chain rule and divergence theorem,
\begin{equation}
{G}^\text{PIST}_i=\int_{|\mathbf{r}|\rightarrow\infty} dS\hspace{0.01in}({\Pi}_{lk}-{\Pi}^{U}_{lk}) b_{ki} n_l-\int_{\mathbf{r}_\text{p}} dS\hspace{0.01in}({\Pi}_{lk}-{\Pi}^{U}_{lk}) b_{ki}n_l-\int_\text{Fluid} dV\hspace{0.01in}({\Pi}_{lk}-{\Pi}^{U}_{lk})\frac{\partial b_{ki}}{\partial r_l}.\label{eq:GPistReciprocal}
\end{equation}
The disturbance of the polymer stress created by the particle, ${\Pi}_{lk}-{\Pi}^{U}_{lk}$, also decays as $1/{r}^2$ in the far-field since it is forced by the disturbance to the far-field velocity gradients. In the case of Oldroyd-B fluids or other dumbbell models such as FENE-P and Giesekus \citep{bird1987dynamics}, this far-field scaling of  ${\Pi}_{lk}-{\Pi}^{U}_{lk}$ is ascertained by linearizing the polymer constitutive equation (for example \eqref{eq:PolymerStressDefintion} and \eqref{eq:Constitutive} for the Oldroyd-B model) about the far-field velocity and polymer conformation to obtain a governing equation for ${\Pi}_{lk}-{\Pi}^{U}_{lk}$. Therefore, the first surface integral in the above equation vanishes, and using the surface boundary condition of equation \eqref{eq:AuxBoundary} leads to,
\begin{equation}
{G}^\text{PIST}_i=-\int_{\mathbf{r}_\text{p}} dS\hspace{0.1in}({\Pi}_{lk}-{\Pi}^{U}_{lk}) \epsilon_{kif}r_fn_l-\int_\text{Fluid} dV\hspace{0.1in}({\Pi}_{lk}-{\Pi}^{U}_{lk})\frac{\partial b_{ki}}{\partial r_l},
\end{equation}
and using ${G}^\text{Elastic}_i=\int_{\mathbf{r}_\text{p}} dS\hspace{0.1in}{\Pi}_{lk} \epsilon_{kif}r_fn_l$,
\begin{equation}
{G}^\text{PIST}_i+{G}^\text{Elastic}_i=\int_{\mathbf{r}_\text{p}} dS\hspace{0.1in}{\Pi}^{U}_{lk} \epsilon_{kif}r_fn_l-\int_\text{Fluid} dV\hspace{0.1in}({\Pi}_{lk}-{\Pi}^{U}_{lk})\frac{\partial b_{ki}}{\partial r_l}.\label{eq:TotalTorque}
\end{equation}
{The first term on the RHS is the torque on a particle about its center of mass due to (constant) undisturbed polymer stress acting on its surface. It is zero by symmetry for a constant density fore-aft and axisymmetric particle such as a slender prolate spheroid, and hence for such a particle,}
\begin{eqnarray}
&{\boldsymbol{G}^\text{PIST}}+{\boldsymbol{G}^\text{Elastic}}=-\int_\text{Fluid} dV\hspace{0.1in}(\boldsymbol{\Pi}-\boldsymbol{\Pi}^{U}):\nabla\mathbf{b},\text{ and,}\label{eq:TotalPolymericStress}\\
&{\boldsymbol{G}^\text{Elastic}}=-\int_\text{Fluid}dV\hspace{0.1in}\nabla\cdot((\boldsymbol{\Pi}-\boldsymbol{\Pi}^{U})\cdot\mathbf{b}).
\end{eqnarray}
Equation \eqref{eq:TotalTorque} (or \eqref{eq:TotalPolymericStress} for a fore-aft and axisymmetric particle) represents the total torque acting on the particle due to the presence of the polymers (including effects from both the polymeric stress and the polymer induced solvent stress), as a function of polymer stress, $\boldsymbol{\Pi}$, and its undisturbed value, $\boldsymbol{\Pi}^U$. For the Oldroyd-B fluid in a simple shear flow with 1, 2, and 3 as flow, gradient, and vorticity directions, respectively, in Cartesian coordinates, ${\Pi}^{U}_{ij}=c(2De\delta_{i1}\delta_{j1}+(\delta_{i2}\delta_{j1}+\delta_{i1}\delta_{j2}))$.
\section{Regular perturbation expansion for small polymer concentration}\label{sec:LowcFormulation}
The leading order solution in a regular perturbation expansion for $c\ll1$ corresponds to a freely rotating particle in  simple shear flow of a Newtonian fluid. At this order the stresses and hence the respective torques arising due to the polymers, i.e. ${\boldsymbol{G}^\text{PIST}}$ and ${\boldsymbol{G}^\text{Elastic}}$ are zero and the particle rotates with an angular velocity that satisfies ${\boldsymbol{G}^\text{MIST}}=0$ (equation \eqref{eq:TorqueFree}). This is simply the \cite{jeffery1922motion} rotation. At the leading order the polymer configuration is driven by the leading order velocity field (equation \eqref{eq:Constitutive}) and this leads to an $\mathcal{O}(c)$ polymer stress (equation \eqref{eq:PolymerStressDefintion}). Therefore, the torques ${\boldsymbol{G}^\text{PIST}}$ and ${\boldsymbol{G}^\text{Elastic}}$ are $\mathcal{O}(c)$ (equations \eqref{eq:PISTOriginal}, \eqref{eq:DiffTorques1}, \eqref{eq:DiffTorques2}, \eqref{eq:DiffTorques3}). Hence, the particle rotation must be modified at $\mathcal{O}(c)$ such that the sum all three torques ${\boldsymbol{G}^\text{MIST}}$, ${\boldsymbol{G}^\text{PIST}}$ and ${\boldsymbol{G}^\text{Elastic}}$ is zero at $\mathcal{O}(c)$. The regular perturbation expansion of the relevant flow variables is,
{\begin{eqnarray}
\mathbf{u}^\text{M}=&{\mathbf{u}^\text{M}}^{(0)}+c{\mathbf{u}^\text{M}}^{(1)}+\mathcal{O}(c^2),\\
p^\text{M}=&{p^\text{M}}^{(0)}+c{p^\text{M}}^{(1)}+\mathcal{O}(c^2),\\ \boldsymbol{\tau}^\text{M}=&{\boldsymbol{\tau}^\text{M}}^{(0)}+c{\boldsymbol{\tau}^\text{M}}^{(1)}+\mathcal{O}(c^2),\\
\mathbf{u}^\text{P}=&c{\mathbf{u}^\text{P}}^{(1)}+\mathcal{O}(c^2),\\p^\text{P}=&c{p^\text{P}}^{(1)}+\mathcal{O}(c^2),\\\boldsymbol{\tau}^\text{P}=&c{\boldsymbol{\tau}^\text{P}}^{(1)}+\mathcal{O}(c^2),\\ \boldsymbol{\Lambda}=&\boldsymbol{\Lambda}^{(0)}+c\boldsymbol{\Lambda}^{(1)}+\mathcal{O}(c^2),\\\boldsymbol{\Pi}=&c\boldsymbol{\Pi}^{(1)}+\mathcal{O}(c^2),\\
\boldsymbol{\omega}_p=&{\boldsymbol{\omega}^{(0)}_p}+c{\boldsymbol{\omega}^{(1)}_p}+\mathcal{O}(c^2),\\ \mathbf{G}^\text{MIST}=&c{\mathbf{G}^\text{MIST}}^{(1)}+\mathcal{O}(c^2),\\\mathbf{G}^\text{PIST}=&c{\mathbf{G}^\text{PIST}}^{(1)}+\mathcal{O}(c^2),\\\mathbf{G}^\text{Elastic}=&c{\mathbf{G}^\text{Elastic}}^{(1)}+\mathcal{O}(c^2).
\end{eqnarray}}
{In an inertia-less Newtonian fluid undergoing simple shear a particle rotating at an angular velocity $\boldsymbol{\omega}^{(0)}_p$ generates the velocity field ${\mathbf{u}^\text{M}}^{(0)}$.
As mentioned earlier, the leading order angular velocity, ${\boldsymbol{\omega}^{(0)}_p}$, is the \cite{jeffery1922motion} rotation. It allows the leading order torque free condition,
$ \int_{\mathbf{r}_\text{p}} dS\hspace{0.1in}\mathbf{r}\times {\boldsymbol{\tau}^\text{M}}^{(0)}\cdot \mathbf{n}=0$ to be satisfied. The leading order polymer constitutive equation is
\begin{equation}
\frac{\partial \boldsymbol{\Lambda}^{(0)}}{\partial t}+{\mathbf{u}^\text{M}}^{(0)}\cdot\nabla\boldsymbol{\Lambda}^{(0)}=(\nabla {\mathbf{u}^\text{M}}^{(0)})^T\cdot\boldsymbol{\Lambda}^{(0)}+\boldsymbol{\Lambda}^{(0)}\cdot\nabla{\mathbf{u}^\text{M}}^{(0)}+\frac{1}{De}(\mathbf{I}-\boldsymbol{\Lambda}^{(0)}).\label{eq:ConstitutiveEqn0}
\end{equation}
Solving this equation, one obtains the $\mathcal{O}(c)$ polymer stress,
\begin{eqnarray}
\boldsymbol{\Pi}^{(1)}=\frac{1}{De}(\boldsymbol{\Lambda}^{(0)}-\mathbf{I}),\label{eq:OrdercPolyStress}
\end{eqnarray}
and hence the $\mathcal{O}(c)$ polymer induced solvent and elastic torques,
\begin{eqnarray}
{\mathbf{G}^\text{PIST}}^{(1)}=&\int_\text{Fluid} dV\hspace{0.2in}\nabla\cdot (\boldsymbol{\Pi}^{(1)}-\boldsymbol{\Pi}^{U}/c) \cdot \mathbf{b},\\
{\mathbf{G}^\text{Elastic}}^{(1)}=&\int_{\mathbf{r}_\text{p}} dS\hspace{0.1in}\mathbf{r}\times\boldsymbol{\Pi}^{(1)}\cdot \mathbf{n}.\label{eq:TorqueExpansion1}
\end{eqnarray}
The $\mathcal{O}(c)$ angular velocity, ${\boldsymbol{\omega}^{(1)}_p}$ is the one that allows the $\mathcal{O}(c)$ torque-free condition to be satisfied,
\begin{equation}
{\mathbf{G}^\text{MIST}}^{(1)}=\int_{\mathbf{r}_\text{p}} dS\hspace{0.1in}\mathbf{r}\times {\boldsymbol{\tau}^\text{M}}^{(1)}\cdot \mathbf{n}=-{\mathbf{G}^\text{PIST}}^{(1)}-{\mathbf{G}^\text{Elastic}}^{(1)},\label{eq:OrdercTorqueFree}
\end{equation}
where ${\boldsymbol{\tau}^\text{M}}^{(1)}=-{p^\text{M}}^{(1)}\mathbf{I}+\nabla {\mathbf{u}^\text{M}}^{(1)}+(\nabla {\mathbf{u}^\text{M}}^{(1)})^T$ is obtained from the $\mathcal{O}(c)$ equations,
\begin{equation}
\nabla\cdot{\mathbf{u}^\text{M}}^{(1)}=0,\hspace{0.2in}\nabla\cdot{\boldsymbol{\tau}^\text{M}}^{(1)}=0,\label{eq:Ordercequations}
\end{equation}
subject to the velocity boundary conditions,
\begin{eqnarray}
{\mathbf{u}^\text{M}}^{(1)}={\boldsymbol{\omega}^{(1)}_p}\times\mathbf{r}, \hspace{0.2in}\text{ on particle surface},\hspace{0.2in}\text{and, }
{\mathbf{u}^\text{M}}^{(1)}=0,  \hspace{0.2in}\text{ as }|\mathbf{r}|\rightarrow\infty.\label{eq:OrdercBCs}
\end{eqnarray}
Equations \eqref{eq:Ordercequations} and \eqref{eq:OrdercBCs} represent the rotation of the particle in a quiescent Newtonian fluid i.e. a Stokes flow. Due to the linearity of the Stokes flow, ${\mathbf{G}^\text{MIST}}^{(1)}$ is a linear function of ${\boldsymbol{\omega}^{(1)}_p}$. The latter can be viewed as the additional angular velocity of the particle that generates a large enough viscous torque, ${\mathbf{G}^\text{MIST}}^{(1)}$, to balance the sum of polymer induced solvent and elastic torque.}

The formulation in this section is valid for any polymer constitutive model and particle shape. By using the formulation of $\mathbf{G}^\text{PIST}$ in equation \eqref{eq:PISTOriginal} to express its $\mathcal{O}(c)$ value in equation \eqref{eq:TorqueExpansion1}, we have avoided dealing with the  $\mathcal{O}(c)$ equation for the balance of the divergence of the polymer stress and the polymer induced solvent stress (obtained by regularly expanding equation \eqref{eq:MainGovernDecompose2} in $c$). Otherwise, obtaining $\mathbf{G}^\text{PIST}$ would have required a numerical solution via discretization of the governing partial differential equations. 

\section{Rotation of a fiber due to simple shear flow in viscoelastic fluid with small polymer concentration, $c$}\label{sec:SBTVelocityTorques}
In this section, we calculate torques from the various physical mechanisms discussed in section \ref{sec:MathBackground} on a large aspect ratio prolate spheroid (considered a slender fiber) freely rotating in a simple shear flow of a viscoelastic fluid with a small polymer concentration, $c$, using the procedure indicated in section \ref{sec:LowcFormulation}. These torques are then balanced to obtain the $\mathcal{O}(c)$ correction to the particle's rotation rate due to the presence of the polymers. The Jeffery orbit period of a slender fiber with aspect ratio $\kappa$ is $2\pi\kappa$ and the proportion of time spent outside $|p_2|>\mathcal{O}(1/\kappa)$ is only $\mathcal{O}(1/\kappa)$, where $p_2=0$ defines the flow-vorticity plane. Therefore, a slender fiber suspended in a Newtonian fluid spends most of its \cite{jeffery1922motion} orbit close to the flow-vorticity plane, where the particle rotation rate is very small. Hence, most of the elasticity influence arises when the particle is in this orientation, and polymer conformation when the fiber is close to the flow-vorticity plane is quasi-steady. Thus,the  polymer constitutive equation \eqref{eq:ConstitutiveEqn0} is simplified to,
\begin{equation}{
\frac{\partial \boldsymbol{\Lambda}^{(0)}}{\partial t}\approx 0\rightarrow {\mathbf{u}^\text{M}}^{(0)}\cdot\nabla\boldsymbol{\Lambda}^{(0)}\approx(\nabla {\mathbf{u}^\text{M}}^{(0)})^T\cdot\boldsymbol{\Lambda}^{(0)}+\boldsymbol{\Lambda}^{(0)}\cdot\nabla{\mathbf{u}^\text{M}}^{(0)}+\frac{1}{De}(\mathbf{I}-\boldsymbol{\Lambda}^{(0)}).}\label{eq:ConstitutiveEqn0Use}
\end{equation}
The Stokes flow solution of a Newtonian fluid around a large aspect ratio or a slender prolate spheroid is analytically solvable via a matched asymptotic expansion in $1/\kappa$ called slender body theory (SBT). In SBT, the fluid velocity close to the particle or in the {inner region} is considered quasi-two-dimensional. It is solved by ignoring the end effects and treating the slender particle as an infinite cylinder. In terms of the radial distance from the centerline (non-dimensionalized with the major radius of the particle) $\rho$, the inner region is defined by $\rho\ll 1$. Away from the particle in the outer region, defined by $\rho\gg1/\kappa$ ($1/\kappa$ is the minor radius), the particle is assumed to be a line of point forces \citep{batchelor1970slender,cox1970motion} and force doublets \citep{cox1971motion}. These singularity solutions are used to represent the velocity and pressure fields. For a large enough $\kappa$ (which is required for SBT to be valid), an intermediate or {matching region} exists which is defined by $1/\kappa\ll\rho\ll 1$. In this region, the $\rho\rightarrow\infty$ asymptote of the flow in the inner region and $\rho\rightarrow0$ asymptote of the flow in the outer region are matched. See previous SBT calculations in \citep{cox1970motion,cox1971motion,batchelor1970slender} for more details. In the rest of this section, we will use these SBT results to obtain the effect of viscoelasticity on the rotation of a slender prolate spheroid. In contrast to a fiber with blunt ends, such as a cylinder, a slender prolate spheroid is more convenient for analysis due to the absence of localized forces at the ends.

\subsection{Flow of Newtonian fluid around a slender fiber and particle rotation rates}\label{sec:SBTFlow}
{Due to the relative velocity between the particle's centerline and the imposed flow, the flow disturbance at $\mathcal{O}(1/\log(\kappa))$ and higher orders in $1/\log(\kappa)$ can be considered to be generated by a line of point forces or force Stokeslets located at the particle centerline. This flow is considered by the general slender body theory (SBT) of \cite{cox1970motion} up to $\mathcal{O}(1/\log(\kappa)^2)$. An equivalent theory by \cite{batchelor1970slender} is valid at all orders in $1/\log(\kappa)$ for a prolate spheroid. For quantitative accuracy, it is advantageous to use the Stokeslet distribution, defined as $\boldsymbol{h}^{(0)}$ below, from the SBT of \cite{batchelor1970slender} instead of \cite{cox1970motion}. For a prolate spheroidal particle fixed in the flow-vorticity plane of a simple shear flow or rotating about its centerline in a quiescent fluid, the force Stokeslet from these two theories is zero. The flow driven by the local velocity gradients, which acts at $\mathcal{O}(1/\kappa^2)$, is dominant in these cases. Considering the velocity gradients in the far-field relative to the particle, \citet{cox1971motion} provides the solution for this flow. It is a combination of flows driven by $\mathcal{O}(1/\kappa^2)$ force Stokeslets ($\boldsymbol{f}^{(0)}$) and doublets ($\boldsymbol{\mathcal{G}}^{(0)}$).}

{The Newtonian or the leading order (in $c$) outer flow generated by a fiber freely rotating with an angular velocity $\boldsymbol{\omega}^{(0)}_p$ in an imposed shear flow of Newtonian fluid with 1, 2, and 3 as the flow, gradient and vorticity directions, when in orientation $\mathbf{p}$,
\begin{equation}
\mathbf{p}=\begin{bmatrix}
p_1&p_2&p_3
\end{bmatrix}^T,
\end{equation}
close to the flow-vorticity plane is obtained by superposition of flows from \cite{batchelor1970slender}, and \cite{cox1971motion} with the assumption $p_2\ll1$ and $\kappa\gg1$.
The outer flow in the presence of the fiber is
{\begin{equation}
\mathbf{u}_\text{out}^{\text{M}{(0)}}(\mathbf{r};\boldsymbol{\omega}^{(0)}_p,\mathbf{p})=\mathbf{r}\cdot (\nabla \mathbf{u})_\infty+{\mathbf{u}_\text{out}^{\text{M}{(0)}}}'(\mathbf{r};\boldsymbol{\omega}^{(0)}_p,\mathbf{p}).\label{eq:Outer1}
\end{equation}}
The disturbance velocity field,
\begin{eqnarray}
&{\mathbf{u}_\text{out}^{\text{M}{(0)}}}'(\mathbf{r};\boldsymbol{\omega}^{(0)}_p,\mathbf{p})=\frac{1}{8\pi}\int_{-1}^1d\lambda\hspace{0.1in}\Big\{\frac{\boldsymbol{h}^{(0)}(\lambda;\boldsymbol{\omega}^{(0)}_p,\mathbf{p})}{\log(2\kappa)-1.5}+\frac{\mathbf{f}^{(0)}(\lambda;\mathbf{p})}{\kappa^2}\Big\}\cdot\Big\{\frac{\mathbf{I}}{|\mathbf{r}-\lambda\mathbf{p}|}+\frac{(\mathbf{r}-\lambda \mathbf{p})(\mathbf{r}-\lambda \mathbf{p})}{|\mathbf{r}-\lambda\mathbf{p}|^3}\Big\}+\nonumber\\&\frac{1}{8\pi}\int_{-1}^1d\lambda\hspace{0.1in}\frac{1}{\kappa^2}\boldsymbol{\mathcal{G}}^{(0)}(\lambda;\boldsymbol{\omega}^{(0)}_p,\mathbf{p}):\nabla\Big\{\frac{\mathbf{I}}{|\mathbf{r}-\lambda\mathbf{p}|}+\frac{(\mathbf{r}-\lambda \mathbf{p})(\mathbf{r}-\lambda \mathbf{p})}{|\mathbf{r}-\lambda\mathbf{p}|^3}\Big\},\label{eq:Outer2}
\end{eqnarray}
is the sum of flows due to the point forces distributions $\boldsymbol{h}^{(0)}$ and $\mathbf{f}^{(0)}$, and the force doublet distribution $\boldsymbol{\mathcal{G}}^{(0)}$ discussed above. The different source distributions are,
\begin{eqnarray}
&\boldsymbol{h^{(0)}}(\lambda;\boldsymbol{\omega}^{(0)}_p,\mathbf{p})=-4\pi\lambda [\mathbf{p}\cdot(\nabla \mathbf{u})_\infty-\boldsymbol{\omega}^{(0)}_p\times\mathbf{p}]\cdot\Big(\frac{1}{2}\mathbf{pp}+(\mathbf{I}-\mathbf{p}\mathbf{p})\frac{\log(2\kappa)-1.5}{\log(2\kappa)-0.5}\Big),\nonumber\\
&\mathbf{f}^{(0)}(\lambda;\mathbf{p})=-4\pi\lambda \begin{bmatrix}0&p_1&0\end{bmatrix}^T\Big(1-\frac{1}{\log(2\kappa)-0.5}\Big)+\mathcal{O}(p_2),\label{eq:Cox2ForceStrengths}\\
&\boldsymbol{\mathcal{G}}^{(0)}(\lambda;\boldsymbol{\omega}^{(0)}_p,\mathbf{p})=2\pi (1-\lambda^2)\Bigg(\begin{bmatrix}
0&1+\frac{p_3^2}{2}&0\\
\frac{p_3^2}{2}&0&-\frac{1}{2}p_1p_3\\
0&-\frac{1}{2}p_1p_3&0
\end{bmatrix}+(\boldsymbol{\omega}^{(0)}_p\cdot\mathbf{p})\boldsymbol{\epsilon}\cdot\mathbf{p}\Bigg)+\mathcal{O}(p_2)\nonumber.\label{eq:Sources}
\end{eqnarray}
A torque-free spheroid with aspect ratio, $\kappa$ rotating in a simple shear flow has the exact result for the temporal evolution of the orientation vector \citep{jeffery1922motion,kim2013microhydrodynamics},
\begin{equation}
\dot{\mathbf{p}}^{(0)}=\boldsymbol{\omega}^{(0)}_p\times\mathbf{p}=\boldsymbol{\omega}_\infty\times\mathbf{p}+\frac{\kappa^2+1}{\kappa^2-1}(\mathbf{E}_\infty\cdot\mathbf{p})\cdot(\mathbf{I}-\mathbf{pp}),\label{eq:JeffRot}
\end{equation}
where $\boldsymbol{\omega}_\infty$ and $\mathbf{E}_\infty$ are the vorticity vector and strain rate tensor of the imposed simple shear,
\begin{equation}
\boldsymbol{\omega}_\infty=-\frac{1}{2}\begin{bmatrix}
0&0&1
\end{bmatrix}^T,\hspace{0.2in}\mathbf{E}_\infty=\frac{1}{2}\begin{bmatrix}0&1&0\\1&0&0\\0&0&0\end{bmatrix}.
\end{equation}
For a slender spheroid close to the flow-vorticity plane,
\begin{eqnarray}
\boldsymbol{\omega}^{(0)}_p\cdot\mathbf{p}=-\frac{p_3}{2}+\mathcal{O}(p_2).\label{eq:omegadotp}
\end{eqnarray}
Using $p_3=1$ in the above equation, we obtain the log-rolling velocity (when the particle is oriented with the vorticity axis) of the particle in a Newtonian fluid and find that in this orientation, a slender particle rotates at the angular velocity of the fluid, i.e., half of the shear rate. $\boldsymbol{\omega}^{(0)}_p\cdot\mathbf{p}$ from equation \eqref{eq:omegadotp} and $\boldsymbol{\omega}^{(0)}_p\times\mathbf{p}$ from equation \eqref{eq:JeffRot} are used to obtain $\boldsymbol{\mathcal{G}}^{(0)}$ and $\boldsymbol{h^{(0)}}$, respectively, in equation \eqref{eq:Cox2ForceStrengths}.}

{The flow due to $\boldsymbol{h}^{(0)}$ is taken from \cite{batchelor1970slender}. It is obtained by an expansion in $1/\log(\kappa)$ for a slender particle. But for a slender prolate spheroid, this expansion terminates such that the flow due to $\boldsymbol{h}^{(0)}$ captures the disturbance created by the prolate spheroid at all orders in $1/\log(\kappa)$ when expressed as in equation \eqref{eq:Outer2} and \eqref{eq:Sources}. The flow generated at the next order in $\kappa$ arises at $\mathcal{O}(\kappa^{-2})$ and is generated by $\mathbf{f}^{(0)}$ and $\boldsymbol{\mathcal{G}}^{(0)}$. It is taken from \cite{cox1971motion} where only the $\mathcal{O}(\kappa^{-2})$ flow is available. Thus, the disturbance velocity in the outer region given by equation \eqref{eq:Outer2} has an overall error of $\mathcal{O}(\kappa^{-3})$. If only $\boldsymbol{h}^{(0)}$ is considered, the error is of $\mathcal{O}(\kappa^{-2})$. The flow generated by $\boldsymbol{h}^{(0)}$ is proportional to $p_2$ and therefore, no flow is produced by $\boldsymbol{h}^{(0)}$ when the particle is aligned in the flow-vorticity plane. In this plane $\mathbf{f}^{(0)}$ and $\boldsymbol{\mathcal{G}}^{(0)}$ capture the (highest)  $\mathcal{O}(\kappa^{-2})$ disturbance created by the particle. Accounting for the flow generated by $\mathbf{f}^{(0)}$ and $\boldsymbol{\mathcal{G}}^{(0)}$ allows us to consider the influence of elasticity within the flow-vorticity plane.}

As mentioned just before equation \eqref{eq:ConstitutiveEqn0Use} we expect most of the changes in the particle's rotation rate due to elasticity to arise when the particle is near the flow-vorticity plane, i.e., when $|p_2|\le\mathcal{O}(1/\kappa)$. Therefore, in $\boldsymbol{h}^{(0)}$ we only consider the flow at $\mathcal{O}(p_2)$ i.e. a flow of $\mathcal{O}(p_2/\log(\kappa))$ with an error of $\mathcal{O}(p_2^2/\log(\kappa))$. Since the primary purpose of using the $\mathbf{f}^{(0)}$ and $\boldsymbol{\mathcal{G}}^{(0)}$ flow is to capture the finite effect of elasticity when the particle is in the flow-vorticity plane, we use $\mathbf{f}^{(0)}$ and $\boldsymbol{\mathcal{G}}^{(0)}$ with $p_2=0$ (as expressed in equation \eqref{eq:Cox2ForceStrengths}) which leads to a flow of   $\mathcal{O}(1/\kappa^2)$ with an error of  $\mathcal{O}(p_2/\kappa^2)$. When $|p_2|/\log(\kappa)$ is more than $1/\kappa^2$, $\boldsymbol{h}^{(0)}$ driven flow dominates. In the $|p_2|\le\mathcal{O}(1/\kappa)$ regime considered, the errors in $\boldsymbol{h}^{(0)}$ driven flow, i.e., $\mathcal{O}(p_2^2/\log(\kappa))$, are always smaller than the flow generated by   $\mathbf{f}^{(0)}$ and $\boldsymbol{\mathcal{G}}^{(0)}$, i.e., $\mathcal{O}(1/\kappa^2)$. As $p_2$ approaches zero, the $\boldsymbol{h}^{(0)}$ driven flow and the associated errors fall rapidly to zero making $\mathbf{f}^{(0)}$ and $\boldsymbol{\mathcal{G}}^{(0)}$ driven flow at $\mathcal{O}(1/\kappa^2)$ the dominating one. Hence, the error terms arising from either of the $\boldsymbol{h}^{(0)}$ or $\mathbf{f}^{(0)}$ and $\boldsymbol{\mathcal{G}}^{(0)}$ driven flow are always lower than the actual flow when $|p_2|\le\mathcal{O}(1/\kappa)$. For $|p_2|>\mathcal{O}(1/\kappa)$, the Newtonian rotation rate of the fiber dominates over the changes due to elasticity, and we consider the exact \cite{jeffery1922motion} rotation to account for it.

{We have considered spheroidal particles in our theory. However, in experiments with slender particles \citep{gauthier1971particle,bartram1975particle,iso1996orientation1,iso1996orientation} it is convenient to fabricate cylindrical particles. The forces generated by the blunt ends of a slender cylinder lead to an additional torque of $\mathcal{O}(1/\kappa^2)$ \citep{cox1971motion} rendering the $\mathcal{O}(1/\kappa^2)$ flow generated by $\mathbf{f}^{(0)}$ and $\boldsymbol{\mathcal{G}}^{(0)}$ inaccurate. Instead of being valid at all orders in $1/\log(\kappa)$, the $\boldsymbol{h}^{(0)}$ generated flow may be considered up to order $1/\log(\kappa)^3$ \citep{batchelor1970slender}. Taking the $\mathcal{O}(p_2)$ terms with this velocity field can still allow us to consider the flow accurately up to $\mathcal{O}(p_2/\log(\kappa))$. Thus the upcoming theoretical development for the orientation dynamics of a slender prolate spheroid leading up to equation \eqref{eq:FullOrbitEquationNewtExact} can still be used for a slender cylinder while ignoring the $\mathbf{f}^{(0)}$ and $\boldsymbol{\mathcal{G}}^{(0)}$ flow and using only the appropriate terms up to $1/\log(\kappa)^3$ instead of the factor $1/(2\log(2\kappa)-3)$ appearing in the following text. Most of the features described in section \ref{sec:MainResults} while analyzing the theoretical prediction of the influence of viscoelasticity on the orientation of a slender spheroid will be qualitatively valid for a slender cylinder or a general slender particle.}

{The Newtonian flow at $\mathcal{O}(c)$ is due to a fiber rotating with the perturbed angular velocity, $\boldsymbol{\omega}^{(1)}_p$, in a quiescent fluid, i.e., equations \eqref{eq:Ordercequations}-\eqref{eq:OrdercBCs}. In the outer region, the fluid velocity is,
\begin{align}
{{\mathbf{u}_\text{out}^\text{M}}^{(1)}}&(\mathbf{r};\boldsymbol{\omega}^{(1)}_p,\mathbf{p})=\frac{1}{8\pi}\int_{-1}^1d\lambda\hspace{0.1in}\frac{\boldsymbol{h}^{(1)}(\lambda;\boldsymbol{\omega}^{(1)}_p,\mathbf{p})}{\log(2\kappa)-0.5}\cdot\Big\{\frac{\mathbf{I}}{|\mathbf{r}-\lambda\mathbf{p}|}+\frac{(\mathbf{r}-\lambda \mathbf{p})(\mathbf{r}-\lambda \mathbf{p})}{|\mathbf{r}-\lambda\mathbf{p}|^3}\Big\}+\nonumber\\&\frac{1}{8\pi}\int_{-1}^1d\lambda\hspace{0.1in}\frac{1}{\kappa^2}\boldsymbol{\mathcal{G}}^{(1)}(\lambda;\boldsymbol{\omega}^{(1)}_p,\mathbf{p}):\nabla\Big\{\frac{\mathbf{I}}{|\mathbf{r}-\lambda\mathbf{p}|}+\frac{(\mathbf{r}-\lambda \mathbf{p})(\mathbf{r}-\lambda \mathbf{p})}{|\mathbf{r}-\lambda\mathbf{p}|^3}\Big\},\label{eq:OrdercOuterVelocity}
\end{align}
where,
\begin{equation}
\boldsymbol{h^{(1)}}(\lambda;\boldsymbol{\omega}^{(1)}_p,\mathbf{p})=4\pi\lambda \boldsymbol{\omega}^{(1)}_p\times\mathbf{p},\hspace{0.2in}
\boldsymbol{\mathcal{G}}^{(1)}(\lambda;\boldsymbol{\omega}^{(1)}_p,\mathbf{p})=-2\pi (1-\lambda^2)(\boldsymbol{\omega}^{(1)}_p\cdot\mathbf{p})\boldsymbol{\epsilon}\cdot\mathbf{p}.\label{eq:Order1ForceandDoublet}\end{equation}
The torque generated by this flow is,
\begin{align}\begin{split}
&{\mathbf{G}^\text{MIST}}^{(1)}(\boldsymbol{\omega}^{(1)}_p,\mathbf{p})=\\&\int_{-1}^1d\lambda\hspace{0.1in}\Big(\frac{1}{\kappa^2}\boldsymbol{\mathbf{\epsilon}}:\boldsymbol{\mathcal{G}}^{(1)}-\frac{\lambda\mathbf{p}\times\boldsymbol{h}^{(1)}}{\log(2\kappa)-0.5}\Big)=\frac{8\pi}{3}\Big(\frac{(\boldsymbol{\omega}^{(1)}_p\cdot\mathbf{p})\mathbf{p}}{\kappa^2}-\frac{\boldsymbol{\omega}^{(1)}_p\cdot(\mathbf{I}-\mathbf{pp})}{\log(2\kappa)-0.5}\Big)
\end{split}\end{align}
Taking the cross product of this equation with orientation vector $\mathbf{p}$ leads to the $\mathcal{O}(c)$ rotation rate,
\begin{equation}
\dot{\mathbf{p}}^{(1)}=\boldsymbol{\omega}^{(1)}_p\times\mathbf{p}=-\frac{3}{8\pi}({\log(2\kappa)-0.5})({\mathbf{G}^\text{MIST}}^{(1)}(\boldsymbol{\omega}^{(1)}_p,\mathbf{p})\times\mathbf{p}),
\end{equation}
Using the torque balance at $\mathcal{O}(c)$ from equation \eqref{eq:OrdercTorqueFree} we obtain,
\begin{equation}
\dot{\mathbf{p}}^{(1)}=\dot{\mathbf{p}}^{(1)}_\text{PIST}+\dot{\mathbf{p}}^{(1)}_\text{Elastic},\label{eq:TotalRotation}
\end{equation}
where
\begin{equation}
\dot{\mathbf{p}}^{(1)}_\text{PIST}=\frac{3}{8\pi}({\log(2\kappa)-0.5})({\mathbf{G}^\text{PIST}}^{(1)}\times\mathbf{p}), \label{eq:PISTRotation}
\end{equation}
is the effect of polymer-induced solvent stresses on the particle rotation rate and,
\begin{equation}
\dot{\mathbf{p}}^{(1)}_\text{Elastic}=\frac{3}{8\pi}({\log(2\kappa)-0.5})({\mathbf{G}^\text{Elastic}}^{(1)}\times\mathbf{p}),\label{eq:ElasticRotation}
\end{equation}
is the effect of the elastic stress on the particle rotation rate.}

\subsection{Rotation due to polymer induced solvent stress }\label{sec:PISTRotation}
{From equations \eqref{eq:PolymerStressDefintion}, \eqref{eq:OrdercPolyStress}, \eqref{eq:TorqueExpansion1} and \eqref{eq:PISTRotation}, the rotation rate due to the polymer induced solvent stress is,
\begin{equation}
\dot{\mathbf{p}}^{(1)}_\text{PIST}=-\frac{3}{8\pi De}({\log(2\kappa)-0.5})\mathbf{p}\times\int_\text{Fluid} dV\hspace{0.1in}(\nabla\cdot({\boldsymbol{\Lambda}^{(0)}}-{\boldsymbol{\Lambda}^U}))\cdot\mathbf{b}.\label{eq:MainPISTRotation}
\end{equation}
We remind the reader that here $\mathbf{b}$ is the auxillary `velocity' field used in the reciprocal theorem for deriving the polymer induced solvent torque (equation \eqref{eq:PISTReciprocal} or \eqref{eq:TorqueExpansion1}). $\mathbf{b}\cdot\boldsymbol{\omega}$ corresponds to the fluid velocity around a fiber rotating with an angular velocity $\boldsymbol{\omega}$ in a quiescent fluid. Therefore,
\begin{equation}
\mathbf{b}=\nabla_{\boldsymbol{\omega}^{(1)}_p} {\mathbf{u}^\text{M}}^{(1)}\label{eq:auxvelandunewt1},
\end{equation}
where ${\mathbf{u}^\text{M}}^{(1)}$ is the solution to equations \eqref{eq:Ordercequations} and \eqref{eq:OrdercBCs}. The volume integral in equation \eqref{eq:MainPISTRotation} is approximated from the outer region of slender body theory, where the particle is replaced with a line of point forces and doublets. We find the integral from the outer region to converge, and the contribution from the inner region is expected to be small due to the smaller volume of that region. Therefore,
\begin{equation}
\dot{\mathbf{p}}^{(1)}_\text{PIST}\approx-\frac{3}{8\pi De}({\log(2\kappa)-0.5})\mathbf{p}\times\int d^3\mathbf{r}\hspace{0.1in}\nabla\cdot({\boldsymbol{\Lambda}_\text{out}^{(0)}}-{\boldsymbol{\Lambda}^U})\cdot\mathbf{b}_\text{out},\label{eq:PISTOuter}
\end{equation}
where the volume integral is taken over the entire space, ${\boldsymbol{\Lambda}_\text{out}^{(0)}}$ is the polymer conformation in the outer region and from equation \eqref{eq:OrdercOuterVelocity}, \eqref{eq:Order1ForceandDoublet} and \eqref{eq:auxvelandunewt1},
\begin{eqnarray}
\mathbf{b}_\text{out}=&\frac{1}{8\pi}\int_{-1}^1d\lambda\hspace{0.1in}\frac{4\pi\lambda\mathbf{p}\cdot\boldsymbol{\epsilon}}{\log(2\kappa)-0.5}\cdot\Big(\frac{\mathbf{I}}{|\mathbf{r}-\lambda\mathbf{p}|}+\frac{(\mathbf{r}-\lambda \mathbf{p})(\mathbf{r}-\lambda \mathbf{p})}{|\mathbf{r}-\lambda\mathbf{p}|^3}\Big)+\mathcal{O}(\kappa^{-2}).
\end{eqnarray}
It is straightforward to show,
\begin{equation}
\dot{\mathbf{p}}^{(1)}_\text{PIST}\approx\frac{3}{2}(\mathbf{I}-\mathbf{p}\mathbf{p})\cdot\int_{-1}^1d\lambda\hspace{0.1in}\lambda\mathbf{u}_\text{out}^\text{polymer}(\lambda\mathbf{p})\label{eq:PISTlikeHarlen},
\end{equation}
where,
\begin{equation}
\mathbf{u}_\text{out}^\text{polymer}(\lambda\mathbf{p})=\frac{1}{8\pi}\int d^3\mathbf{r}\hspace{0.1in}\frac{1}{De}\nabla\cdot({\boldsymbol{\Lambda}_\text{out}^{(0)}}-{\boldsymbol{\Lambda}^U})\cdot\Big(\frac{\mathbf{I}}{|\mathbf{r}-\lambda\mathbf{p}|}+\frac{(\mathbf{r}-\lambda \mathbf{p})(\mathbf{r}-\lambda \mathbf{p})}{|\mathbf{r}-\lambda\mathbf{p}|^3}\Big)\label{eq:VelocityfromPolymer},
\end{equation}
is the velocity field evaluated on the particle center-line that is produced by polymeric force, $\frac{1}{De}\nabla\cdot({\boldsymbol{\Lambda}_\text{out}^{(0)}}-{\boldsymbol{\Lambda}^U})$, acting in the outer region.  This form of the rotation rate, equations \eqref{eq:PISTlikeHarlen} and \eqref{eq:VelocityfromPolymer}, was considered by \citep{harlen1993simple}. However, they did not account for all the relevant terms in calculating ${\boldsymbol{\Lambda}_\text{out}^{(0)}}$ as we show below.}

{As discussed in section \ref{sec:LowcFormulation} and shown by equations \eqref{eq:ConstitutiveEqn0} and \eqref{eq:ConstitutiveEqn0Use}, the polymer conformation, ${\boldsymbol{\Lambda}^{(0)}}$ depends on the leading order Newtonian velocity. In the outer region, the velocity disturbance created by the particle is $\mathcal{O}(\max[p_2/\log(\kappa),1/\kappa^2])$ smaller than the velocity of the imposed simple shear (equations \eqref{eq:Outer1}, \eqref{eq:Outer2} and \eqref{eq:Cox2ForceStrengths}). Thus, we linearize equation \eqref{eq:ConstitutiveEqn0Use} in the outer region about the imposed flow field, $\mathbf{r}\cdot (\nabla \mathbf{u})_\infty$, and obtain the governing equation for the disturbance in the polymer conformation from its undisturbed value,
\begin{equation}
{\boldsymbol{\Lambda}_\text{out}^{(0)}}'=\boldsymbol{\Lambda}_\text{out}^{(0)}-\boldsymbol{\Lambda}_\text{out}^U,
\end{equation}
to be
\begin{align}\begin{split}
\Big(\mathbf{r}\cdot(\nabla \mathbf{u})_\infty\cdot\nabla+\frac{1}{De}\Big){\boldsymbol{\Lambda}_\text{out}^{(0)}}'&-(\nabla \mathbf{u})_\infty^T\cdot{\boldsymbol{\Lambda}_\text{out}^{(0)}}'-{\boldsymbol{\Lambda}_\text{out}^{(0)}}'\cdot(\nabla \mathbf{u})_\infty=\\&(\nabla {\mathbf{u}_\text{out}^{\text{M}{(0)}}}')^T\cdot\boldsymbol{\Lambda}^U+\boldsymbol{\Lambda}^U\cdot\nabla{\mathbf{u}_\text{out}^{\text{M}{(0)}}}'.
\end{split}\label{eq:ConstitutiveLinear}
\end{align}
It is more convenient to solve \eqref{eq:ConstitutiveLinear} in Fourier space,
\begin{align}\begin{split}
\Big(-k_1\frac{\partial}{\partial k_2}+\frac{1}{De}\Big){\hat{\boldsymbol{\Lambda}}_\text{out}^{(0)}}'&-(\nabla \mathbf{u})_\infty^T\cdot{\hat{\boldsymbol{\Lambda}}_\text{out}^{(0)}}'-{\hat{\boldsymbol{\Lambda}}_\text{out}^{(0)}}'\cdot(\nabla \mathbf{u})_\infty\\&=i({{\hat{\mathbf{u}}_\text{out}^\text{M}}^{(0)}}'\mathbf{k}\cdot\boldsymbol{\Lambda}^U+\boldsymbol{\Lambda}^U\cdot\mathbf{k}{{\hat{\mathbf{u}}_\text{out}^\text{M}}^{(0)}}'),\label{eq:ConstitutiveLinearFourier}
\end{split}\end{align}
where ${\hat{\boldsymbol{\Lambda}}_\text{out}^{(0)}}'=\mathcal{F}({{\boldsymbol{\Lambda}}_\text{out}^{(0)}}')$ and ${{\hat{\mathbf{u}}_\text{out}^\text{M}}^{(0)}}'=\mathcal{F}({{{\mathbf{u}}_\text{out}^\text{M}}^{(0)}}')$ are the Fourier transforms of the disturbance of polymer conformation and fluid velocity in the outer region.
The rotation rate in equation \eqref{eq:PISTlikeHarlen} is expressed as
\begin{equation}
\dot{\mathbf{p}}^{(1)}_\text{PIST}\approx\frac{3}{2}(\mathbf{I}-\mathbf{p}\mathbf{p})\cdot\int_{-1}^1d\lambda\int d^3\mathbf{k}\hspace{0.1in}\lambda\exp(2\pi i\lambda\mathbf{k}\cdot\mathbf{p})\hat{\mathbf{u}}_\text{out}^\text{polymer},\label{Eq:PISTRotation}
\end{equation}
where from the convolution theorem,
\begin{equation}
\hat{\mathbf{u}}_\text{out}^\text{polymer}=i\frac{1}{De}\mathbf{k}\cdot{\hat{\boldsymbol{\Lambda}}_\text{out}^{(0)}}'(\mathbf{k};\mathbf{p})\cdot\hat{\mathbf{J}}(\mathbf{k}), \text{ with }
\hat{\mathbf{J}}(\mathbf{k})=\mathcal{F}\Big(\frac{1}{8\pi}\Big(\frac{\mathbf{I}}{r}-\frac{\mathbf{rr}}{r^3}\Big)\Big)=\frac{1}{k^2}\Big(\mathbf{I}-\frac{\mathbf{kk}}{k^2}\Big).
\end{equation}}

{We solve the polymer constitutive equation \eqref{eq:ConstitutiveLinearFourier} in Fourier space using the method of characteristics to obtain ${\hat{\boldsymbol{\Lambda}}_\text{out}^{(0)}}'$. The characteristics are the streamlines of the imposed shear flow (in the Fourier space) and so only integration along the $k_2$ direction is needed. The limits of the integral are $\infty \text{sgn}(k_1)$ and $k_2$ and the boundary condition is ${\hat{\boldsymbol{\Lambda}}_\text{out}^{(0)}}'(\infty \text{sgn}(k_1))=0$. We use computer algebra for this calculation. ${\hat{\boldsymbol{\Lambda}}_\text{out}^{(0)}}'$ hence obtained allows us to evaluate the integral in equation \eqref{Eq:PISTRotation}. The expression used for the disturbance in the Newtonian velocity field, ${\mathbf{u}_\text{out}^{\text{M}{(0)}}}'$,  has errors of $\mathcal{O}(\text{max}[p_2^2/\log(\kappa),p_2/\kappa^2,1/\kappa^3])$. Due to the linearization of the constitutive equation, the error in evaluation of rotation rate is $\mathcal{O}(\epsilon_{\dot{p}})$, where,
\begin{equation}\epsilon_{\dot{p}}=\text{max}\Bigg[\frac{p_2^2}{\log(\kappa)},\frac{p_2}{\kappa^2},\frac{1}{\kappa^3}\Bigg].\label{eq:ErrorInpdot}\end{equation}
Due to fortuitous canceling of terms, a simple expression for the 2nd (gradient direction) component of the rotation rate is obtained,
\begin{equation}
\dot{{p}}^{(1)}_\text{2,PIST}=-\frac{De p_1^2 p_2}{2\log(2\kappa)-3}+\mathcal{O}(\epsilon_{\dot{p}}),\label{eq:p2Pist}
\end{equation}
valid at all $De$. Equivalently, from the outer region integral of equation \eqref{eq:TorqueExpansion1}, i.e.,
\begin{equation}
{\mathbf{G}^\text{PIST}}^{(1)}\approx \frac{1}{De}\int d^3\mathbf{r}\hspace{0.1in}\nabla\cdot({\boldsymbol{\Lambda}_\text{out}^{(0)}}-{\boldsymbol{\Lambda}^U})\cdot\mathbf{b}_\text{out},
\end{equation}
we obtain the 1st and 3rd components of the polymer induced solvent torque,
\begin{equation}
\frac{{{G}_1^\text{PIST}}^{(1)}}{p_3}=-\frac{{{G}_3^\text{PIST}}^{(1)}}{p_1}=\frac{4De^2 p_1^2 p_2}{3(\log(2\kappa)-0.5)(\log(2\kappa)-1.5)}+\mathcal{O}\Bigg(\frac{\epsilon_{\dot{p}}}{\log(\kappa)}\Bigg).\label{eq:GPIST}
\end{equation}
Since, $\dot{\mathbf{p}}^{(1)}_\text{PIST}={3}/{(8\pi)}({\log(2\kappa)-0.5})({\mathbf{G}^\text{PIST}}^{(1)}\times\mathbf{p})$ (equation \eqref{eq:PISTRotation}) we may obtain $\dot{{p}}^{(1)}_\text{2,PIST}=-De{p_1^2 p_2}/({2\log(2\kappa)-3})$.
However, the equations for determining the torque component in the gradient direction, ${{G}_2^\text{PIST}}^{(1)}$ or rotation rates in flow and vorticity direction, $\dot{{p}}^{(1)}_\text{1,PIST}$ and $\dot{{p}}^{(3)}_\text{1,PIST}$ are not tractable for a general $De$. Therefore, to obtain these components we consider small and large $De$ limits separately in sections \ref{sec:LargeDePIST} and \ref{sec:SmallDePIST}.}

We find the particle rotation rate due to polymer induced solvent stress in gradient direction, $\dot{{p}}^{(1)}_\text{2, PIST}$ to be dependent on the first normal stress difference of the Oldroyd-B fluid. This is because $\dot{\mathbf{p}}^{(1)}_\text{PIST}$ in equation \eqref{Eq:PISTRotation} is directly proportional to ${\hat{\boldsymbol{\Lambda}}_\text{out}^{(0)}}'$ that is in turn dependent on $\boldsymbol{\Lambda}^U$ via equation \eqref{eq:ConstitutiveLinearFourier}. This first normal stress difference dependence is further confirmed by repeating the above calculation by artificially setting ${\Lambda}^U_{ij}=De^2\delta_{i1}\delta_{j1}$, such that the only non-zero non-Newtonian property of the fluid is a finite first normal stress difference equivalent to that of an Oldroyd-B fluid. Similarly, we find the two remaining rotation rate components in the limit of large $De$ calculation of section \ref{sec:LargeDePIST} to arise from the first normal stress difference of the Oldroyd-B fluid. The rotation rate for a second order fluid ($\mathcal{O}(De)$ rotation rate in the small $De$ calculation of section \ref{sec:SmallDePIST}) is also due to the first normal stress difference. An Oldroyd-B fluid has no second normal stress difference. In addition to its appropriate modeling of the simple shear flow of a dilute polymeric liquid, the Oldroyd-B model the advantages over other models such as FENE-P or Giesekus \citep{bird1987dynamics} of simplicity and hence better analytical tractability. As mentioned in section \ref{sec:Introduction}, the second normal stress difference for most polymeric fluids is much smaller than the first normal stress difference. Hence, the effect of the first normal stress difference as ascertained from the Oldroyd-B model is likely to be the most important contribution in determining the influence of polymers on the orientational dynamics of a prolate spheroid in simple shear flow.

\subsubsection{Large $De$}\label{sec:LargeDePIST}
In the large $De$ regime, the relaxation of the disturbance in polymer conformation in the outer region is much slower than its convection and stretching by the imposed velocity field. Therefore, when $De\gg1$, equation \eqref{eq:ConstitutiveLinearFourier} simplifies to,
\begin{equation}
-k_1\frac{\partial}{\partial k_2}{\hat{\boldsymbol{\Lambda}}_\text{out}^{(0)}}'-(\nabla \mathbf{u})_\infty^T\cdot{\hat{\boldsymbol{\Lambda}}_\text{out}^{(0)}}'-{\hat{\boldsymbol{\Lambda}}_\text{out}^{(0)}}'\cdot(\nabla \mathbf{u})_\infty=i({{\hat{\mathbf{u}}_\text{out}^\text{M}}^{(0)}}'\mathbf{k}\cdot\boldsymbol{\Lambda}^U+\boldsymbol{\Lambda}^U\cdot\mathbf{k}{{\hat{\mathbf{u}}_\text{out}^\text{M}}^{(0)}}').\label{eq:LinearizedConstitutiveLarge}
\end{equation}
This equation is solved in a similar way as equation \eqref{eq:ConstitutiveLinearFourier} described earlier. Using ${\hat{\boldsymbol{\Lambda}}_\text{out}^{(0)}}'$ obtained from this equation we find the particle rotation rate due to polymer induced solvent stresses at large $De$ to be
\begin{equation}{
\lim\limits_{De\gg 1}\dot{\mathbf{p}}^{(1)}_\text{PIST}=De\begin{bmatrix}
-p_1p_3^2 \alpha_{De\gg 1}\\-\frac{p_1^2 p_2}{2\log(2\kappa)-3}\\p_1^2p_3\alpha_{De\gg 1}
\end{bmatrix}, \text{ with }\alpha_{De\gg 1}=\frac{-4}{\kappa^2}+\frac{3\pi}{4\log(2\kappa)-6}p_2\tan^{-1}\frac{p_1}{|p_3|}.}\label{eq:PistRotLargeDe}
\end{equation}
The large $De$ approximation can be viewed as the leading $\mathcal{O}(De)$ term in an expansion in powers of $1/De$.  Thus the error due to the expansion in $1/De$ is $\mathcal{O}(1)$ in the 1 and 3 components of $\lim\limits_{De\gg 1}\dot{\mathbf{p}}^{(1)}_\text{PIST}$. There is no error due to expansion in $1/De$ for $\dot{{p}}^{(1)}_\text{2,PIST}$ as same value is obtained via $De\gg 1$ approximation as that for a general $De$ in equation \eqref{eq:p2Pist}. As mentioned earlier, $\dot{{p}}^{(1)}_\text{2,PIST}$ has an error of $\mathcal{O}(\epsilon_{\dot{p}})$ (equation \eqref{eq:ErrorInpdot}). $\dot{{p}}^{(1)}_\text{1,PIST}$ and $\dot{{p}}^{(1)}_\text{3,PIST}$ have an error of $\mathcal{O}(De\epsilon_{\dot{p}})$.
To the best of our knowledge, the only previous theoretical study concerning the rotation of a slender particle in a viscoelastic fluid at high $De$  was conducted by \cite{harlen1993simple}. However they did not account for the stretching of the conformation disturbance, ${\hat{\boldsymbol{\Lambda}}_\text{out}^{(0)}}'$, by the mean velocity gradients i.e. they omitted $(\nabla \mathbf{u})_\infty^T\cdot{\hat{\boldsymbol{\Lambda}}_\text{out}^{(0)}}'+{\hat{\boldsymbol{\Lambda}}_\text{out}^{(0)}}'\cdot(\nabla \mathbf{u})_\infty$ term in equation \eqref{eq:LinearizedConstitutiveLarge}.

\subsubsection{Small $De$}\label{sec:SmallDePIST}
When $De\ll1$, a solution of equation \eqref{eq:ConstitutiveLinear} or \eqref{eq:ConstitutiveLinearFourier} is obtained via a regular perturbation expansion of ${\hat{\boldsymbol{\Lambda}}_\text{out}^{(0)}}'$ in the powers of $De$
\begin{equation}
{{\hat{\boldsymbol{\Lambda}}_\text{out}^{(0)}}'}=\Sigma_{n=0}^\infty De^n {{\hat{\boldsymbol{\Lambda}}_\text{out}^{(0)}}'}^{(n)}.\label{eq:ConfigPertExpansion}
\end{equation}
The first two terms in this expansion are
{\begin{align}\begin{split}
&{{\hat{\boldsymbol{\Lambda}}_\text{out}^{(0)}}'}^{(0)}=\mathbf{0},\\
&{{\hat{\boldsymbol{\Lambda}}_\text{out}^{(0)}}'}^{(1)}=i(\mathbf{k}{\hat{\mathbf{u}}^\text{M}}^{(0)}+{\hat{\mathbf{u}}^\text{M}}^{(0)}\mathbf{k}).\label{eq:PolyConfigFirstOrder}\end{split}
\end{align}
The polymer conformation at higher order in $De$ is obtained from the following equations
\begin{align}\begin{split}
{\hat{\boldsymbol{\Lambda}}_\text{out}^{(0')}}^{(2)}=&(\nabla \mathbf{u})_\infty^T\cdot{\hat{\boldsymbol{\Lambda}}_\text{out}^{(0')}}^{(1)}+{\hat{\boldsymbol{\Lambda}}_\text{out}^{(0')}}^{(1)}\cdot(\nabla \mathbf{u})_\infty\\&+i({\hat{\mathbf{u}}^\text{M}}^{(0)}\mathbf{k}\cdot{\boldsymbol{\Lambda}^U}^{(1)}+{\boldsymbol{\Lambda}^U}^{(1)}\cdot\mathbf{k}{\hat{\mathbf{u}}^\text{M}}^{(0)})+k_1\frac{\partial}{\partial k_2}{\hat{\boldsymbol{\Lambda}}_\text{out}^{(0')}}^{(1)},\\
{\hat{\boldsymbol{\Lambda}}_\text{out}^{(0')}}^{(3)}=&(\nabla \mathbf{u})_\infty^T\cdot{\hat{\boldsymbol{\Lambda}}_\text{out}^{(0')}}^{(2)}+{\hat{\boldsymbol{\Lambda}}_\text{out}^{(0')}}^{(2)}\cdot(\nabla \mathbf{u})_\infty\\&+i({\hat{\mathbf{u}}^\text{M}}^{(0)}\mathbf{k}\cdot{\boldsymbol{\Lambda}^U}^{(2)}+{\boldsymbol{\Lambda}^U}^{(2)}\cdot\mathbf{k}{\hat{\mathbf{u}}^\text{M}}^{(0)})+k_1\frac{\partial}{\partial k_2}{\hat{\boldsymbol{\Lambda}}_\text{out}^{(0')}}^{(2)},\\
{\hat{\boldsymbol{\Lambda}}_\text{out}^{(0')}}^{(n)}=&(\nabla \mathbf{u})_\infty^T\cdot{\hat{\boldsymbol{\Lambda}}_\text{out}^{(0')}}^{(n-1)}+{\hat{\boldsymbol{\Lambda}}_\text{out}^{(0')}}^{(n-1)}\cdot(\nabla \mathbf{u})_\infty+k_1\frac{\partial}{\partial k_2}{\hat{\boldsymbol{\Lambda}}_\text{out}^{(0')}}^{(n-1)},n\ge 4.\end{split}\label{eq:ConstitutiveFourierSmallDe}
\end{align}}
Therefore, in the limit of low $De$, $\dot{\mathbf{p}}^{(1)}_\text{PIST}$ can be obtained up to arbitrary order in $De$,
\begin{align}\begin{split}
\lim\limits_{De\ll 1}\dot{\mathbf{p}}^{(1)}_\text{PIST}=&-\frac{3}{8\pi }({\log(2\kappa)-0.5})\mathbf{p}\times\Sigma_{n=1}^\infty De^{n-1} \int_\text{Fluid} dV\hspace{0.1in}(\nabla\cdot({{{\boldsymbol{\Lambda}}_\text{out}^{(0)}}'}^{(n)}))\cdot\mathbf{b}.
\end{split}\label{eq:PISTExpansionSmall}
\end{align}
We will discuss the contribution of the $n=1$ term, i.e. $\mathcal{O}(1)$ term in $De$ expansion of $\dot{\mathbf{p}}^{(1)}_\text{PIST}$ in the above equation in section \ref{sec:NetEffect}, along with the similar term in the expansion of $\dot{\mathbf{p}}^{(1)}_\text{Elastic}$. Higher order contributions to  $\dot{\mathbf{p}}^{(1)}_\text{PIST}$ can be evaluated in Fourier space. From equations \eqref{eq:ConfigPertExpansion}-\eqref{eq:ConstitutiveFourierSmallDe}, using expansion of ${{\boldsymbol{\Lambda}}_\text{out}^{(0)}}'$ up to $\mathcal{O}(De^{8})$ (even higher orders can be easily evaluated), we obtain,
\begin{align}\begin{split}
\lim\limits_{De\ll 1} \dot{\mathbf{p}}^{(1)}_\text{PIST}=&-\frac{3}{8\pi }({\log(2\kappa)-0.5})\mathbf{p}\times \int_\text{Fluid} dV\hspace{0.1in}(\nabla\cdot({{{\boldsymbol{\Lambda}}_\text{out}^{(0)}}'}^{(1)}))\cdot\mathbf{b}\\&+De\begin{bmatrix}
-p_1p_3^2 \alpha_{De\ll 1}\\-\frac{p_1^2 p_2}{2\log(2\kappa)-3}\\p_1^2p_3\alpha_{De\ll 1}
\end{bmatrix},
\end{split}\label{eq:SmallDePIST}\end{align}
where,
\begin{equation}
\alpha_{De\ll 1}=-\frac{4}{\kappa^2}+\frac{ p_1p_2 }{8\log(2\kappa)-12} \Big(5 De-\frac{3}{2}(2+p_3^2)De^3+\frac{13}{10}(8+p_3^2(4+3p_3^2))De^5\Big).\label{eq:SmallDealfa}\end{equation}
As expected, we obtain the same expression for $\dot{{p}}^{(1)}_\text{2,PIST}$ from this perturbation expansion in $De$ valid at small $De$ as was obtained directly for a general $De$ in equation \eqref{eq:p2Pist}. The error in the rotation rate in equation \eqref{eq:SmallDePIST} is $\mathcal{O}(De\epsilon_{\dot{p}})$ for all the components ($\epsilon_{\dot{p}}$ is given in equation \eqref{eq:ErrorInpdot}).

\subsection{Rotation due to elastic stress }\label{sec:ElasticRotation}
The elastic torque, ${\mathbf{G}^\text{Elastic}}^{(1)}$, is due to the polymer stress on the particle surface (equations \eqref{eq:OrdercPolyStress} and \eqref{eq:TorqueExpansion1}). This is evaluated through the polymer constitutive equations on the particle surface written in the frame of reference moving with the particle surface. Due to the absence of polymer convection relative to the particle on the latter's surface, the quasi-steady constitutive equation \eqref{eq:ConstitutiveEqn0Use} simplifies to a set of coupled algebraic equations,
\begin{equation}{
(\nabla{\mathbf{u}_\text{in}^\text{M}}^{(0)}(\mathbf{r}_p;\boldsymbol{\omega}^{(0)}_p,\mathbf{p}))^T\cdot\boldsymbol{\Lambda}^{(0)}_p+\boldsymbol{\Lambda}^{(0)}_p\cdot\nabla{\mathbf{u}_\text{in}^\text{M}}^{(0)}(\mathbf{r}_p;\boldsymbol{\omega}^{(0)}_p,\mathbf{p})+\frac{1}{De}(\mathbf{I}-\boldsymbol{\Lambda}^{(0)}_p)=0,}
\label{eq:SurfaceConstitutive}\end{equation}
at each point on the surface. Here $\mathbf{r}_p$ is the position vector of a point on the surface. {$\nabla{\mathbf{u}_\text{in}^\text{M}}^{(0)}(\mathbf{r}_p;\boldsymbol{\omega}^{(0)}_p,\mathbf{p})$ and $\boldsymbol{\Lambda}^{(0)}_p$ represent the surface velocity gradient of the inner velocity field and the surface polymer conformation respectively at $\mathbf{r}_p$.} The inner velocity field is obtained from \citet{cox1970motion} and \cite{cox1971motion}. These velocity fields include the effects of point force Stokeslets and doublets that correspond to the outer velocity field accurate up to $\mathcal{O}(p_2/\log(\kappa)^2)$ and $\mathcal{O}(1/\kappa^2)$. Solving equation \eqref{eq:SurfaceConstitutive} via an asymptotic expansion in $1/\kappa$ and ignoring higher order terms leads to the elastic torque,
\begin{align}\begin{split}
{\mathbf{G}}^{(1)}_\text{Elastic}&=\frac{8\pi c}{3\kappa^2}\begin{bmatrix}
p_1 p_3\\ 0\\-p_1^2
\end{bmatrix}+\mathcal{O}(\max[p_2\kappa^{-2},\kappa^{-3}]).\end{split}\label{eq:ElasticTorque}
\end{align}
and the corresponding rotation rate,
\begin{align}\begin{split}
\dot{\mathbf{p}}^{(1)}_\text{Elastic}&=-\frac{3}{8\pi}({\log(2\kappa)-0.5})\mathbf{p}\times\int dS\hspace{0.1in}\Big(\mathbf{r}\times\frac{1}{De}(\boldsymbol{\Lambda}^{(0)}_p-\mathbf{I})\cdot \mathbf{n}\Big)\\&=\frac{{\log(2\kappa)-0.5}}{\kappa^2}p_1\begin{bmatrix}
0\\ 1\\0
\end{bmatrix}+\mathcal{O}(\log(\kappa)\max[p_2\kappa^{-2},\kappa^{-3}]).\end{split}\label{eq:ElasticRotationRate}
\end{align}
Therefore, for all $De$, in limits of large $\kappa$ and small $p_2$, the $\mathcal{O}(c)$ elastic torque is independent of the polymer relaxation time, $De$.

\subsection{Net rotation due to the polymers}\label{sec:NetEffect}
Combining the results of previous two sections, we obtain the net change in the fiber's rotation rate due to polymers.

For $De\ll1$, we can obtain the $\mathcal{O}(1)$ term in the $De$ expansion of $\dot{\mathbf{p}}^{(1)}_\text{PIST}$ and $\dot{\mathbf{p}}^{(1)}_\text{Elastic}$ for all particle orientations, without resorting to the outer region approximation for the former. The first two terms in the polymer configuration, everywhere (inner and outer region) is similar to that mentioned earlier in equation \eqref{eq:PolyConfigFirstOrder}, i.e.,
\begin{equation}
{{{{\boldsymbol{\Lambda}}^{(0)}}}=\mathbf{I}+De(\nabla {\mathbf{u}^\text{M}}^{(0)}+(\nabla {\mathbf{u}^\text{M}}^{(0)})^T)+\mathcal{O}(De^2).}\label{eq:PolyConfigFirstOrderEverywhere}
\end{equation}
From the combined effect of polymers from equations \eqref{eq:TorqueExpansion1}, \eqref{eq:TotalRotation}, \eqref{eq:PISTRotation} and \eqref{eq:ElasticRotation}, along with using the value of ${{{\boldsymbol{\Lambda}}^{(0)}}}$ from equation \eqref{eq:PolyConfigFirstOrderEverywhere}, divergence theorem, the leading order momentum equation {(i.e. $\nabla {p^\text{P}}^{(0)}=\nabla\cdot[\nabla {\mathbf{u}^\text{P}}^{(1)}+(\nabla {\mathbf{u}^\text{P}}^{(1)})^T]$)}, and the auxillary Stokes problem (equations \eqref{eq:AuxEqns} and \eqref{eq:AuxBoundary}), we obtain,
\begin{align}
\begin{split}
&\lim\limits_{De\ll 1}{\dot{\mathbf{p}}^{(1)}}_\text{Elastic}+{\dot{\mathbf{p}}^{(1)}}_\text{PIST}\\&
=\int_{\mathbf{r}_\text{p}} dS\hspace{0.1in}\mathbf{r}\times[(\nabla {\mathbf{u}^\text{M}}^{(0)}+(\nabla {\mathbf{u}^\text{M}}^{(0)})^T-{p^\text{M}}^{(0)}\mathbf{I})\cdot \mathbf{n}]+\mathcal{O}(De)\\&= \int_{\mathbf{r}_\text{p}} dS\hspace{0.1in}\mathbf{r}\times {\boldsymbol{\tau}^\text{M}}^{(0)}+\mathcal{O}(De)\cdot \mathbf{n}=\mathcal{O}(De).
\end{split}
\end{align}
The $\mathcal{O}(1)$ term in this sum is equivalent to the rotation due to the leading order torque (in $c$) acting on the particle. This is the torque on a freely rotating particle in a Newtonian fluid and is hence zero. Thus, for a freely rotating particle for $De\ll1$, the $\mathcal{O}(1)$ rotation rates (in the $De$ expansion) due to the torque generated by the elastic and polymer induced solvent stress cancel each other and lead to no change in the net particle rotation. The change in the particle rotation due to the polymers arises at $\mathcal{O}(De)$. Close to the flow-vorticity plane, the rotation rate due to the elastic torque was shown to be independent of $De$ for all $De$ in equation \eqref{eq:ElasticRotationRate}, which we have just shown balances the equivalent rotation rate due to polymer induced solvent torque for $De\ll1$. Thus the $\mathcal{O}(De)$ and higher effect of polymers on the particle's rotation rate arises entirely due to the polymer induced solvent torque (equation \eqref{eq:SmallDePIST}) and is given by,
\begin{equation}
\lim\limits_{De\ll 1}(\dot{\mathbf{p}}^{(1)}_\text{Elastic}+\dot{\mathbf{p}}^{(1)}_\text{PIST})=De\begin{bmatrix}
-p_1p_3^2 \alpha_{De\ll 1}\\-\frac{p_1^2 p_2}{2\log(2\kappa)-3}\\p_1^2p_3\alpha_{De\ll 1}
\end{bmatrix},\label{eq:SmallDeRotation}
\end{equation}
where, $\alpha_{De\ll 1}$ is given in equation \eqref{eq:SmallDealfa} and the errors are $\mathcal{O}(De\epsilon_{\dot{p}})$ for all the components (where $\epsilon_{\dot{p}}$ is given in equation \eqref{eq:ErrorInpdot}).

{For $De\gg 1$, the total effect of viscoelasticity on the particle rotation also arises mainly from the polymer induced solvent stress. But, unlike the $De\ll 1$ regime, here the reason is that the effect of the elastic stress is $\mathcal{O}(De)$ smaller. The net rotation rate due to the polymers is,
\begin{equation}
\lim\limits_{De\gg 1}(\dot{\mathbf{p}}^{(1)}_\text{Elastic}+\dot{\mathbf{p}}^{(1)}_\text{PIST})=De\begin{bmatrix}
-p_1p_3^2 \alpha_{De\gg 1}\\-\frac{p_1^2 p_2}{2\log(2\kappa)-3}\\p_1^2p_3\alpha_{De\gg 1}
\end{bmatrix}.\label{eq:LargeDeRotation}
\end{equation}
where $\alpha_{De\gg 1}$ is given by equation \eqref{eq:PistRotLargeDe}. The error in $\dot{{p}}^{(1)}_\text{1,PIST}$ and $\dot{{p}}^{(1)}_\text{3,PIST}$ is of $\mathcal{O}(De\epsilon_{\dot{p}})$ (equation \eqref{eq:ErrorInpdot}). Neglecting the elastic torque leads to an additional error of $\mathcal{O}(\log(\kappa)/\kappa^2)$ in the second component and $\mathcal{O}(\log(\kappa)\max[p_2\kappa^{-2},\kappa^{-3}])$ in the first and third component of the net rotation rate due to the polymers (equation \eqref{eq:ElasticRotationRate}).}

\subsection{Equations of motion of a freely rotating fiber in low $c$ viscoelastic fluid at all $De$}\label{sec:SBTRotationRates}
Equations \eqref{eq:SmallDeRotation} and \eqref{eq:LargeDeRotation} encompass our main result for the effect of viscoelasticity on the rotation of a particle suspended in simple shear flow at large and small $De$, respectively. In the limit of large $De$ and for a second order fluid ($\mathcal{O}(De)$ rotation rate in small $De$ limit), the effect of viscoelasticity is due to the first normal stress difference of the fluid as discussed in section \ref{sec:PISTRotation}. The governing equation for the orientation dynamics of a slender prolate spheroid that includes the viscoelastic effects near the flow-vorticity plane is,
\begin{eqnarray}
\dot{\mathbf{p}}=\frac{d\mathbf{p}}{d t}=\begin{bmatrix}
\frac{p_2}{2}+\frac{\kappa^2-1}{\kappa^2+1}\Big(\frac{p_2}{2}-p_1^2p_2\Big)-cDep_1p_3^2 \alpha\\
- \frac{p_1}{2}+\frac{\kappa^2-1}{\kappa^2+1}\Big(\frac{p_1}{2}-p_1p_2^2\Big)-cDe\frac{p_1^2 p_2}{2\log(2\kappa)-3}\\
-\frac{\kappa^2-1}{\kappa^2+1}p_1 p_2 p_3+cDep_1^2p_3\alpha
\end{bmatrix},\label{eq:FullOrbitEquationNewtExact}
\end{eqnarray}
where in the large $De$ limit $\alpha$ is $\alpha_{De\gg 1}$ (equation \eqref{eq:PistRotLargeDe}). In the small $De$ limit we consider $\alpha_{De\ll 1}$ (equation \eqref{eq:SmallDealfa}) up to $\mathcal{O}(De)$ and interpolating between these two limits of $De$, we obtain the following uniformly valid approximation for $\alpha$ at all $De$,
\begin{equation}
\alpha=-\frac{4}{\kappa^2}+\frac{p_2}{4\log(2\kappa)-6}\frac{2.5p_1De}{1+2.5p_1De/(3\pi\tan^{-1}\frac{p_1}{|p_3|})},\label{eq:FullOrbitalpha}
\end{equation}
Equation \eqref{eq:FullOrbitEquationNewtExact} introduces no errors in the Newtonian rotation rate of a prolate spheroidal particle. The slender body theory provides a good approximation for the Newtonian velocity field for $\kappa\gtrsim10$.
The errors in these equations due to the viscoelastic terms are of $\mathcal{O}(cDe\epsilon_{\dot{p}})$ for all the components ($\epsilon_{\dot{p}}$ is given in equation \eqref{eq:ErrorInpdot}) in the $De\ll1$ limit. For $De\gg1$, the errors are of $\mathcal{O}(cDe\epsilon_{\dot{p}})$ in the 1 and 3 component and $\mathcal{O}(c\max[\log(\kappa)/\kappa^2,\epsilon_{\dot{p}}])$ in the 2 component. Before analyzing the influence of polymers on particle orientation as suggested by this equation we compare our theory at low $De$ with that of \cite{leal1975slow}.

\subsection{Comparison of second order fluid result with \cite{leal1975slow}}
\cite{leal1975slow} considered the motion of a fiber in a second order fluid and found,
\begin{eqnarray}
\dot{p}_2^\text{Leal}=&- p_1p_2 (p_2+Vp_1 (1-2p_2^2))\\
\dot{p}_3^\text{Leal}=&- p_1 p_2 p_3(1-2V p_1p_3),
\end{eqnarray}
where, $V=-\frac{3 \lambda \gamma}{16 \log(\kappa)} M_1(1+2\epsilon_1)$,
$\lambda=\Phi_3 U/\mu l$ is the polymer relaxation time ($\mu$ is the zero shear rate viscosity, $l$ the particle half length and $U$ a characteristic velocity scale) and $\epsilon_1=\Phi_2/\Phi_3$,
\begin{eqnarray}
\Phi_2=-\lim\limits_{\gamma\rightarrow 0}\frac{\sigma_{11}-\sigma_{22}}{2\gamma^2}, \hspace{0.2in}\text{and}\hspace{0.2in}\Phi_3=\lim\limits_{\gamma\rightarrow 0}\frac{\sigma_{11}-\sigma_{33}}{\gamma^2},
\end{eqnarray}
$\sigma_{11}$, $\sigma_{22}$, $\sigma_{33}$ and $\gamma$ are the first, second and third normal stresses and the shear rate. and $M_1$ is a positive number depending upon the shape of the particle. In non-dimensional terms $\lambda=2De$, $V=\frac{-3De}{8\log(\kappa)}M_1(1+2\epsilon_1)$. $(1+2\epsilon_1)$ and hence $V$ in Leal's theory is proportional to the second normal stress difference. Up to $\mathcal{O}(De p_2)$ Leal's results are therefore,
\begin{eqnarray}
\dot{p}_2^\text{Leal}=-p_1p_2^2+\frac{3De}{8\log(\kappa)}M_1 (1+2\epsilon_1)p_1^2p_2 +\mathcal{O}(De^2,p_2^2De)\\
\dot{p}_3^\text{Leal}=-p_1p_2p_3+\mathcal{O}(De^2,p_2^2De).
\end{eqnarray}
By taking only up to $\mathcal{O}(De)$ terms from equation \eqref{eq:FullOrbitEquationNewtExact}, our results up to this order for a prolate spheroidal particle in a second order fluid  are,
\begin{eqnarray}
\dot{p}_2=-p_1p_2^2-\frac{cDe}{2\log(2\kappa)-3}p_1^2p_2 +\mathcal{O}(cDe^2,cp_2^2De)\\
\dot{p}_3=-p_1p_2p_3+\mathcal{O}(cDe^2,cp_2^2De).
\end{eqnarray}
Hence, our theory's first viscoelastic effects at low $De$ have the same functional dependence of rotation rates on the particle orientation as that of Leal's second order fluid theory. However, our theory predicts these effects to arise from the first normal stress difference. In contrast, Leal's theory predicts these to emerge from the second normal stress difference. According to our theory, a prolate spheroid rotating in a simple shear flow of a Boger fluid that has zero second normal stress difference \citep{magda1991second}, will exhibit a different rotational motion at a finite but small $c\cdot De$ as compared to $c=0$ (Newtonian fluid). Specifically, a particle spirals towards the stable limit cycle close to the vorticity, as discussed in the next section. However, Leal's theory predicts Jeffery rotations or no effect of viscoelasticity. \citet{brunn1977slow} considered the motion of rigid particles in second order fluid. While \citet{brunn1977slow} does not calculate the rotation rates for a rod-like particle such as prolate spheroid, for a dumbbell representing two spheres joined by a thin, rigid rod, he also obtains the same functional dependence of rotations rates on the particle orientation. However, instead of the factor $M_1(1+2\epsilon_1)$, \citet{brunn1977slow} obtains $(1+4\epsilon_1)$ so that the rotation rate of a dumbbell in a second order fluid with zero second normal stress difference is finite.

\section{Analysis of particle orientation dynamics}\label{sec:MainResults}
\begin{figure}
	\centering
	\includegraphics[width=0.75\linewidth]{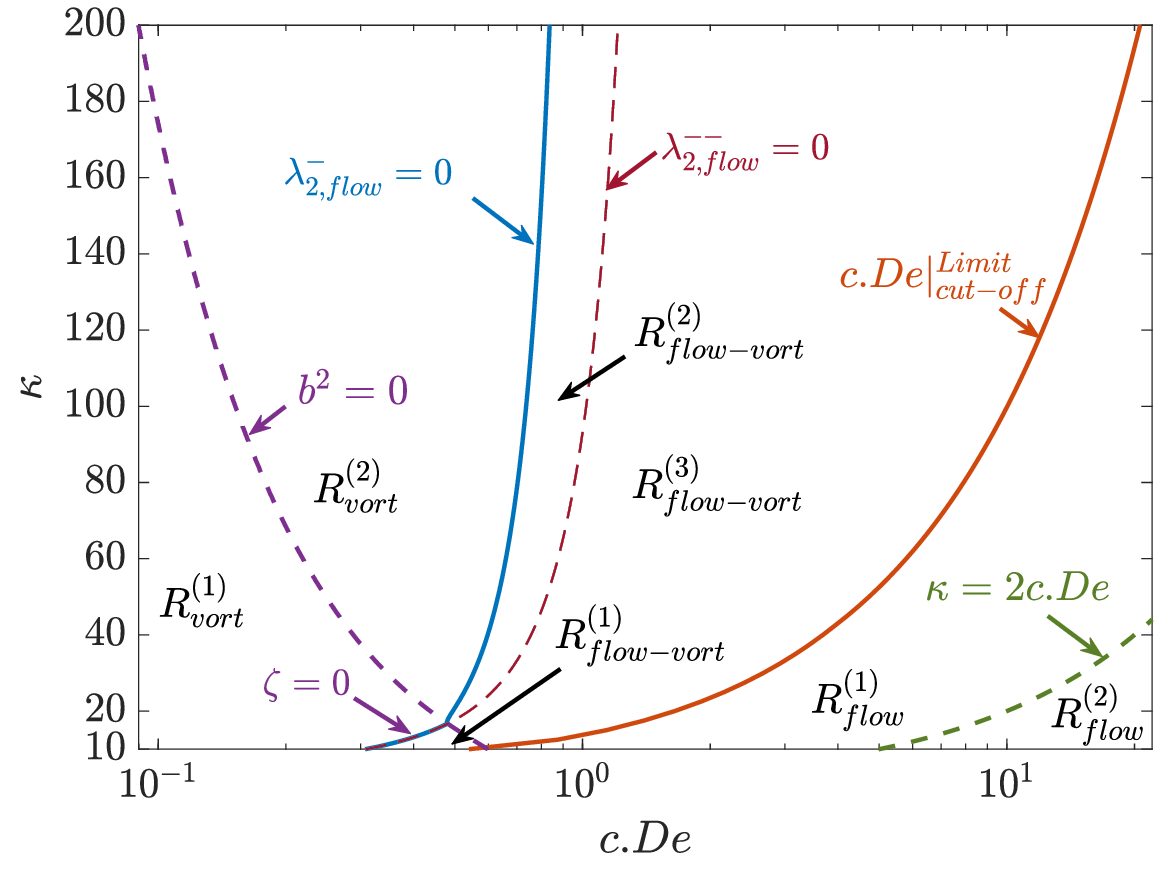}
	\caption{$\kappa-c\cdot De$ parameter space, for $c=0.005$, divided into regions with different qualitative behavior of particle's orientation dynamics. $b^2$ and $\zeta$ are given in equations \eqref{eq:bsquare} and \eqref{eq:bsqrandzeta}} respectively. $\lambda_{2,\text{flow}}^{0-}$ and $\lambda_{2,\text{flow}}^{0--}$ are in equation \eqref{eq:EigenFlow}. Procedure for numerically obtaining $c\cdot De|_\text{cut-off}^\text{Limit}$ is described in section \ref{sec:StableLC}.
	\label{fig:PhasecDekapa}
\end{figure}

Before a more detailed analytical and numerical treatment of equation \eqref{eq:FullOrbitEquationNewtExact}, we qualitatively describe the changes in orientation dynamics introduced by viscoelasticity as given by this equation. We obtain three primary regions in $c\cdot De- \kappa$ space where viscoelasticity leads to different final fiber orientation behaviors. These regions, $R_{vort}=R_{vort}^{(1)}\cup R_{vort}^{(2)}$, $R_{flow}=R_{flow}^{(1)}\cup R_{flow}^{(2)}$, and, $R_{flow-vort}=R_{flow-vort}^{(1)}\cup R_{flow-vort}^{(2)}\cup R_{flow-vort}^{(3)}$, separated by the curves $\lambda_{2,flow}^{-}=0$ and $c\cdot De=c\cdot De|_{cut-off}^{Limit}$ are shown in figure \ref{fig:PhasecDekapa} for $c=0.005$. In $R_{vort}$, i.e. the region to the left of the curve $\lambda_{2, flow}^{-}=0$ in figure \ref{fig:PhasecDekapa}, irrespective of the initial orientation, the particle is attracted towards a stable limit cycle near the vorticity axis where the particle revolves in a small periodic orbit around the vorticity axis. In $R_{flow}$, i.e. the region to the right of the curve $c\cdot De=c\cdot De|_{cut-off}^{Limit}$ in figure \ref{fig:PhasecDekapa}, a particle aligns close to the flow direction for all initial orientations as a stable fixed point at this location is the only stable attractor in the orientation space. This flow alignment behavior is similar to the particle trajectories observed by \cite{iso1996orientation} at large elasticity (large $c$). In $R_{flow-vort}$, defined as the region between $\lambda_{2,flow}^{-}=0$ and $c\cdot De=c\cdot De|_{cut-off}^{Limit}$ in figure \ref{fig:PhasecDekapa}, depending upon the initial orientation, the particle can either obtain a final orientation within the flow-gradient plane (it may either obtain a stable orientation or rotate within the plane) or rotate in a small periodic orbit around the vorticity axis.

There are subdivisions of the regions $R_{vort}$, $R_{flow-vort}$ and $R_{flow}$ based on the behavior of the particle's orientation trajectory near the flow-gradient plane and the vorticity axis. In $R_{vort}^{(1)}$ (figure \ref{fig:PhasecDekapa}), a particle starting near the flow-gradient plane spirals away from the plane and towards the limit cycle near the vorticity axis. In contrast, in $R_{vort}^{(2)}$ (figure \ref{fig:PhasecDekapa}), once a particle comes close to the flow direction, it drifts along the flow-vorticity plane in a monotonic fashion. $R_{vort}^{(1)}$ trajectories are reminiscent of the observations of \cite{gauthier1971particle} at low shear rates (low $De$) and also some of the particle trajectories of \cite{iso1996orientation1,iso1996orientation} at low to medium elasticity (small to medium $c$). The trajectories in $R_{vort}^{(2)}$, on the other hand, are similar to that observed by \cite{bartram1975particle} at larger shear rates or $De$ (than the earlier low shear rate experiments with the same fluid reported by the same laboratory in \cite{gauthier1971particle}). Within the region $R_{flow}$ shown in figure \ref{fig:PhasecDekapa}, trajectories in $R_{flow}^{(1)}$  and $R_{flow}^{(2)}$ have different behavior near the vorticity axis. In the former, to the left of the curve $\kappa=c\cdot De$ in figure \ref{fig:PhasecDekapa}, the vorticity axis is an unstable spiral. Thus the particle leaves the region close to the vorticity axis in a spiral motion. In the latter, the vorticity axis is an unstable node, and the particle leaves the vicinity of the vorticity axis in a monotonic fashion.  The trajectories near the vorticity axis in also in $R_{vort}$ and $R_{flow-vort}$ are similar to those in $R_{flow}^{(1)}$.  As mentioned above, in the region of multiple final particle orientations, $R_{flow-vort}$, in addition to a stable limit cycle/ periodic orbit close to the vorticity axis, a stable attractor exists in the flow-gradient plane. $R_{flow-vort}$ can be further subdivided into three sub-regions. In $R_{flow-vort}^{(1)}$ (figure \ref{fig:PhasecDekapa}), the particle spirals toward the flow-gradient plane and then tumbles within the plane. In this case, the particle slows down close to the flow direction before speeding up again as it departs from this orientation. In $R_{flow-vort}^{(2)}$ and $R_{flow-vort}^{(3)}$ (figure \ref{fig:PhasecDekapa}), there is a stable fixed point very close to the flow direction. Hence a particle starting close to the flow-gradient plane ends up being flow-aligned. In $R_{flow-vort}^{(2)}$, there is an unstable node within the flow-gradient plane (further away from the flow direction than the stable fixed point). In contrast, in $R_{flow-vort}^{(3)}$, the unstable node is replaced with a saddle point with its stable manifold perpendicular to the flow-gradient plane. Therefore, the basin of attraction of the stable fixed point (with the stable limit cycle being the other stable attractor) is larger for $R_{flow-vort}^{(3)}$ than for $R_{flow-vort}^{(2)}$.

The above is a qualitative description of all the different behaviors exhibited by the dynamical system of equation \eqref{eq:FullOrbitEquationNewtExact}. However, the predictions for regions of large $c\cdot De$ (such as $R_{flow}^{(2)}$) are speculative. This is because the first normal stress difference in the polymer stress is proportional to $c\cdot De$ and our regular perturbation theory is based on the polymer stress being smaller than the Newtonian stress. We nevertheless include these regions here to provide a complete description of the dynamical system and later numerical studies of the original governing equations described in section \ref{sec:MathBackground} may be useful in determining the quantitative and qualitative validation of the theory considered here.

In the rest of this section we will derive the boundaries that determine the aforementioned divisions of $c\cdot De- \kappa$ space and provide a more detailed analytical and numerical treatment of equation \eqref{eq:FullOrbitEquationNewtExact}. $b^2=0$ and $\kappa=2c\cdot De$ boundaries shown in figure \ref{fig:PhasecDekapa} will be determined in sections \ref{sec:LogRollingandTumbling} and \ref{sec:NearVorticity} respectively. $\lambda_{2,\text{flow}}^{0-}=0$, $\lambda_{2,\text{flow}}^{0--}=0$ and $\zeta=0$ boundaries are derived in section \ref{sec:NearFlow}. $c\cdot De|_\text{cut-off}^\text{Limit}$, is numerically obtained (the other boundaries of figure \ref{fig:PhasecDekapa} are analytical) in section \ref{sec:StableLC}. To numerically integrate the orientation trajectory we transform equation \eqref{eq:FullOrbitEquationNewtExact} into $\theta-\phi$ coordinates, where,
\begin{eqnarray}
\mathbf{p}=\begin{bmatrix}
p_1\\p_2\\p_3
\end{bmatrix}=\begin{bmatrix}
\sin(\theta)\sin(\phi)\\
\sin(\theta)\cos(\phi)\\
\cos(\theta)
\end{bmatrix}\rightarrow&\begin{split}
&\frac{d \theta}{d t}= -\frac{1}{\sin(\theta)}\frac{\partial p_3}{\partial t}, \\
&\frac{d \phi}{d t}=\frac{1}{\sin^2(\theta)}\Big(\frac{\partial p_2}{\partial t}\cos(\phi)-\frac{\partial p_1}{\partial t}\sin(\phi)\Big)
\end{split}. \label{eq:OrienttoAngles}
\end{eqnarray}
Integrating this $\theta-\phi$ system instead of the equivalent equation \eqref{eq:FullOrbitEquationNewtExact} directly for $\mathbf{p}$ numerically preserves $||\mathbf{p}||_2=1$ constraint.

The boundaries in the $c\cdot De-\kappa$ space shown in figure \ref{fig:PhasecDekapa} are obtained for $c=0.005$. The primary $c$ dependence of the various boundaries is in the form of $c\cdot De$ and similar partitioning of the $c\cdot De- \kappa$ space is found with different values of $c$, albeit with small quantitative changes. The boundaries associated with $b^2=0$ (equation \eqref{eq:bsquare}), and $\kappa=2cDe$ only depend on $c\cdot De$ for a given $\kappa$. We find (section \ref{sec:StableLC}) the $c\cdot De|_\text{cut-off}^\text{Limit}$ boundary to be insensitive to changes in $c$ at constant $c\cdot De$ for $c\lessapprox0.1$. However, increasing $c$ moves the $\lambda_{2,\text{flow}}^{0-}=0$, $\lambda_{2,\text{flow}}^{0--}=0$ and $\zeta=0$ boundaries in the $c\cdot De-\kappa$ kappa space to the right or larger $c\cdot De$ (not shown) even at small $c$. The $\lambda_{2,\text{flow}}^{0--}=0$ curve moves more than $\lambda_{2,\text{flow}}^{0-}=0$ upon increasing $c$. Thus, in the $c\cdot De-\kappa$ space, upon increasing $c$, $R_{vort}^{(1)}$, $R_{flow}^{(1)}$ and $R_{flow}^{(2)}$ are unchanged, $R_{vort}^{(2)}$ and $R_{flow-vort}^{(2)}$ are enlarged, and, $R_{flow-vort}^{(1)}$ and $R_{flow-vort}^{(3)}$ are reduced in size.

\subsection{Low dimensional orbits: Log-rolling and tumbling}\label{sec:LogRollingandTumbling}
We begin our analysis with two convenient orientational states: a) Log-rolling with the particle aligned with the vorticity axis, and, b) Tumbling in the flow-gradient plane ($p_3=0$). Due to symmetry, and as suggested by equations \eqref{eq:FullOrbitEquationNewtExact}, viscoelasticity does not change the particle's orientation from the log-rolling state. Also, in the log-rolling state the theory predicts that the polymer induced solvent and elastic torques due to viscoelasticity, i.e. ${{G}_3^\text{PIST}}^{(1)}$ from equation \eqref{eq:GPIST} and ${{G}_3^\text{Elastic}}^{(1)}$ from equation \eqref{eq:ElasticTorque}, are both zero. Thus, the polymers do not change the log-rolling angular velocity of the particle at $\mathcal{O} (c)$.

In the flow-gradient plane, close to the flow direction $p_1\approx 1$, the equation governing the particle orientation is,
\begin{equation}
 \dot{p}_2\approx-p_2^2-\frac{1}{\kappa^2}-cDe\frac{p_2}{2\log(2\kappa)-3}.\label{eq:p2dotnearflow}
\end{equation}
Therefore, we notice that viscoelasticity ($c\cdot De>0$) reduces the rotation rate of the particle as compared to the Newtonian value ($c\cdot De=0$). This equation has an analytical solution,
\begin{equation}
p_2=-b\tan({(t-t_0)b})\frac{cDe}{4\log(2\kappa)-6},\label{eq:p2solution}
\end{equation}
where,
\begin{equation}
b^2={\frac{1}{\kappa^2}-\Big(\frac{cDe}{4\log(2\kappa)-6}\Big)^2}.\label{eq:bsquare}
\end{equation}
At small values of $c\cdot De$ when $b^2>0$, the solution is periodic with a time-period $2\pi/b$. Upon increasing $c\cdot De$, $b=0$ when $cDe_\text{crit}=(4\log(2\kappa)-6)/\kappa$ and an infinite period bifurcation occurs. Further increasing $c\cdot De$ lead to $b^2<0$ and two fixed points appear at $[p_{2,\text{flow}}^{0-},0]$ and $[p_{2,\text{flow}}^{0--},0]$ within the flow-gradient plane, where
 \begin{equation}
 p_{2,\text{flow}}^{0-}=-\frac{cDe}{4\log(2\kappa)-6}+\sqrt{-b^2},\hspace{0.2in} p_{2,\text{flow}}^{0--}=-\frac{cDe}{4\log(2\kappa)-6}-\sqrt{-b^2}.\label{eq:FixedPointsNearFlow}
 \end{equation}
There are two additional fixed points at $-p_{2,\text{flow}}^{0-}$ and $-p_{2,\text{flow}}^{0--}$ that can be mathematically obtained by repeating the above analysis near $p_1=-1$. In the sense of Jeffery orbits, the fixed points are downstream of the flow-vorticity plane and very close to the flow direction when $\kappa$ is large, and $c\cdot De$ is small. $p_{2,\text{flow}}^{0--}$ is further downstream. Within the flow-gradient plane, particle trajectories approach $\pm p_{2,\text{flow}}^{0-}$, while they depart $\pm p_{2,\text{flow}}^{0--}$. Therefore, in the $b^2<0$ regime, a particle placed in the flow-gradient plane approaches a steady state orientation $p_2=\pm p_{2,\text{flow}}^{0-}$. We will later observe in section \ref{sec:NearFlow} that in the orientation space near the flow-direction off the flow-gradient plane, i.e., for a finite $0<p_3\ll1$, the trajectories may either approach or leave these two fixed points along the vorticity direction.

\subsubsection{More accurate location of fixed points on flow-gradient plane in $b^2<0$ regime}
\begin{figure}
	\centering
	\includegraphics[width=0.45\linewidth]{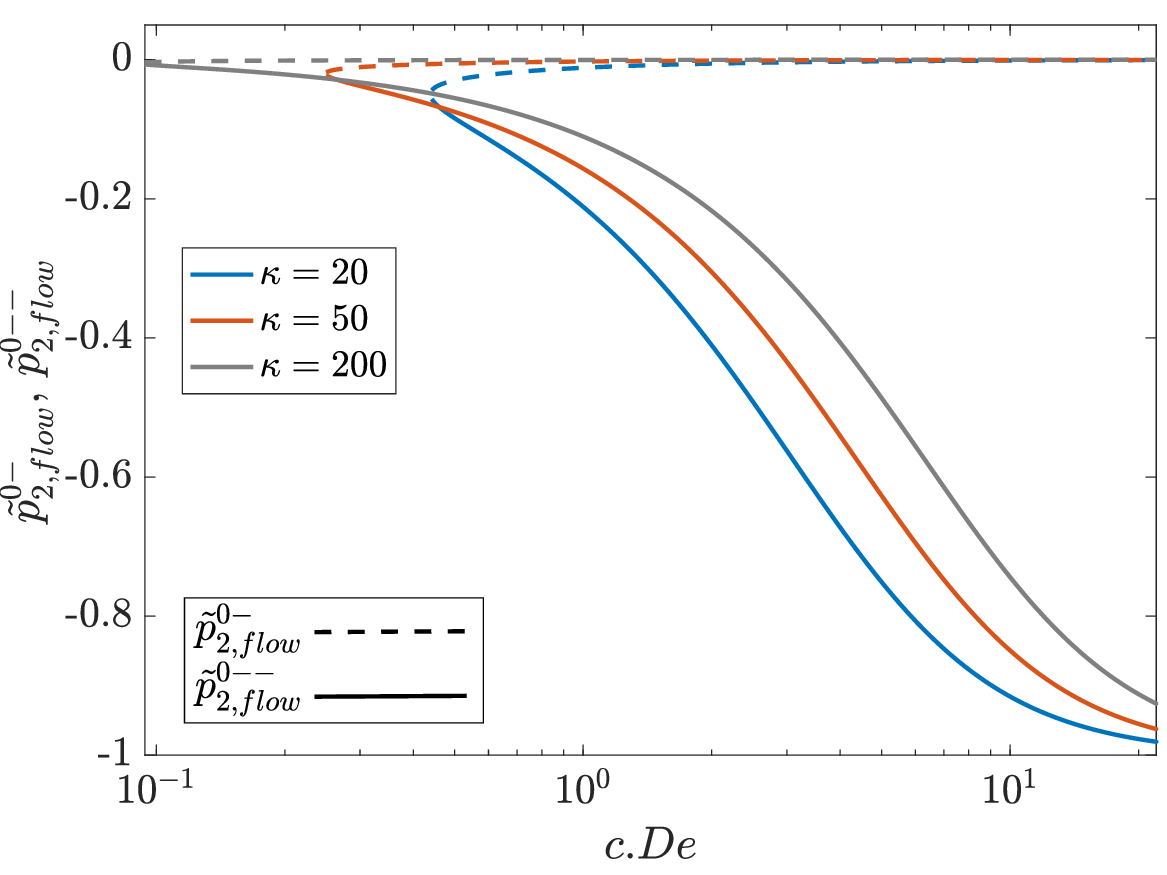}
	\caption{Variation of the fixed points' location on the flow-gradient plane with $c\cdot De$ for various $\kappa$ in the $\tilde{b^2}\approx{b^2}<0$ regime. }
	\label{fig:accuratefixpointsonfg}
\end{figure}

Above, we analyzed the equations near the flow direction under the assumption $p_1\approx1$ and found two fixed points in the $b^2<0$ regime at orientations given by equation (\ref{eq:FixedPointsNearFlow}). These expressions are only valid when the fixed points are near the flow direction. However, as  $c\cdot De$ is increased at a given $\kappa$, the fixed points separate from each other in the flow gradient plane. The fixed point at $p_{2,\text{flow}}^{-}$ moves closer to the flow direction, while $p_{2,\text{flow}}^{--}$ moves away from the flow direction. To better estimate the latter's location we relax the assumption of $p_1\approx 1$ in the expression for ${dp_2}/{dt}$ and obtain the improved expressions,
\begin{align}\begin{split}
\tilde{p}_{2,\text{flow}}^{0-}&=-\sqrt{\frac{-f^2/\kappa^2+{c^2De^2}/2-fc\cdot De\sqrt{-\tilde{b}^2}}{{c^2De^2+f^2}}},\\\tilde{p}_{2,\text{flow}}^{0--}&=-\sqrt{\frac{-f^2/\kappa^2+{c^2De^2}/2+fc\cdot De\sqrt{-\tilde{b}^2}}{{c^2De^2+f^2}}},\label{eq:FixedPointsNearFlowAccurate}
\end{split}\end{align}
where
$\tilde{b}^2={{(1+\kappa^2)}/{\kappa^4}-({cDe}/{(2f)})^2}\approx b^2$ and  $f=2\log(2\kappa)-3$.
From figure \ref{fig:accuratefixpointsonfg}, as $c\cdot De$ or $\kappa$ are varied, we observe the qualitative behaviors of $\tilde{p}_{2,\text{flow}}^{0-}$ and $\tilde{p}_{2,\text{flow}}^{0--}$ are the same as those of  ${p}_{2,\text{flow}}^{0-}$ and ${p}_{2,\text{flow}}^{0--}$ mentioned above. $\tilde{p}_{2,\text{flow}}^{0-}$ is shown with dashed lines and $\tilde{p}_{2,\text{flow}}^{0--}$ with solid lines in the region $\tilde{b^2}\approx{b^2}<0$. At the beginning of the dashed and solid lines, $\tilde{b^2}\approx{b^2}=0$, for a given $\kappa$, $\tilde{p}_{2,\text{flow}}^{0-}=\tilde{p}_{2,\text{flow}}^{0--}$, because these fixed points arise out of an infinite period bifurcation \citep{strogatz2018nonlinear}, as mentioned earlier.

Therefore, while polymers have no influence on the log-rolling motion, they slow down or stop the tumbling motion of a prolate spheroidal particle within the flow-gradient plane. We now move on to consider three-dimensional orbits by first observing the particle's orientation trajectories close to the vorticity axis in section \ref{sec:NearVorticity} and then the trajectories near the flow-gradient plane in section \ref{sec:NearFlow}.

\subsection{Effect of viscoelasticity near the vorticity direction, $p_3\approx 1$}\label{sec:NearVorticity}
Equation \eqref{eq:FullOrbitEquationNewtExact} has a fixed point on the vorticity axis, \begin{equation}\mathbf{p}^0_\text{vort}=\begin{bmatrix}
0&0&1
\end{bmatrix}.\end{equation}
Due to the constraint $||\mathbf{p}||_2=1$, we consider the linear stability of the $p_1-p_2$ dynamical system using $p_3=\sqrt{1-p_1^2-p_2^2}$. In the $p_1-p_2$ coordinate system, at the fixed point $\begin{bmatrix}
0& 0
\end{bmatrix}$, the eigenvalues are,
\begin{equation} \lambda_{1,\text{vort}}=2cDe/\kappa^2-1/\kappa\sqrt{4c^2De^2/\kappa^2-1}, \lambda_{2,\text{vort}}=2cDe/\kappa^2+1/\kappa\sqrt{4c^2De^2/\kappa^2-1},\label{eq:EigVort}
\end{equation}
with the corresponding eigenvectors
\begin{equation}
\mathbf{v}_{1,\text{vort}}=\begin{bmatrix}
\lambda_{1,\text{vort}}&1
\end{bmatrix},
\mathbf{v}_{2,\text{vort}}=\begin{bmatrix}
1&\lambda_{2,\text{vort}}
\end{bmatrix}.
\end{equation}
The fixed point at the vorticity axis undergoes a Hopf bifurcation at $c\cdot De=0$ and is unstable for all finite $c\cdot De$.
The vorticity axis is a center (imaginary eigenvalues) when $c\cdot De=0$. In the presence of viscoelasticity, $c\cdot De\ne 0$, it becomes a hyperbolic fixed point (eigenvalues with non-zero real part). Therefore, the linear stability analysis provides qualitative insight into effect of viscoelasticity on the behavior of fiber orientation governed by the full nonlinear equation \eqref{eq:FullOrbitEquationNewtExact} near the vorticity axis when $c\cdot De\ne 0$. This follows from the Hartman-Grobman theorem (see \citet{guckenheimer2013nonlinear}), which guarantees the homeomorphism between the linearized and full non-linear system near hyperbolic fixed points, while also preserving the time parametrization. The instability of the vorticity axis arises from the polymer conformation driven by the force doublet and $\mathcal{O}(1/\kappa^2)$ Stokeslet discussed in section \ref{sec:SBTFlow}. When $0<cDe<\kappa/2$ a small perturbation leads to a particle departing the vorticity axis in a spiral fashion (as the vorticity fixed point is an unstable spiral). However, for $cDe>\kappa/2$ a particle departs the vorticity axis monotonically (as the vorticity fixed point is an unstable fixed point).

\begin{figure}
	\centering	
	\subfloat{\includegraphics[width=0.49\textwidth]{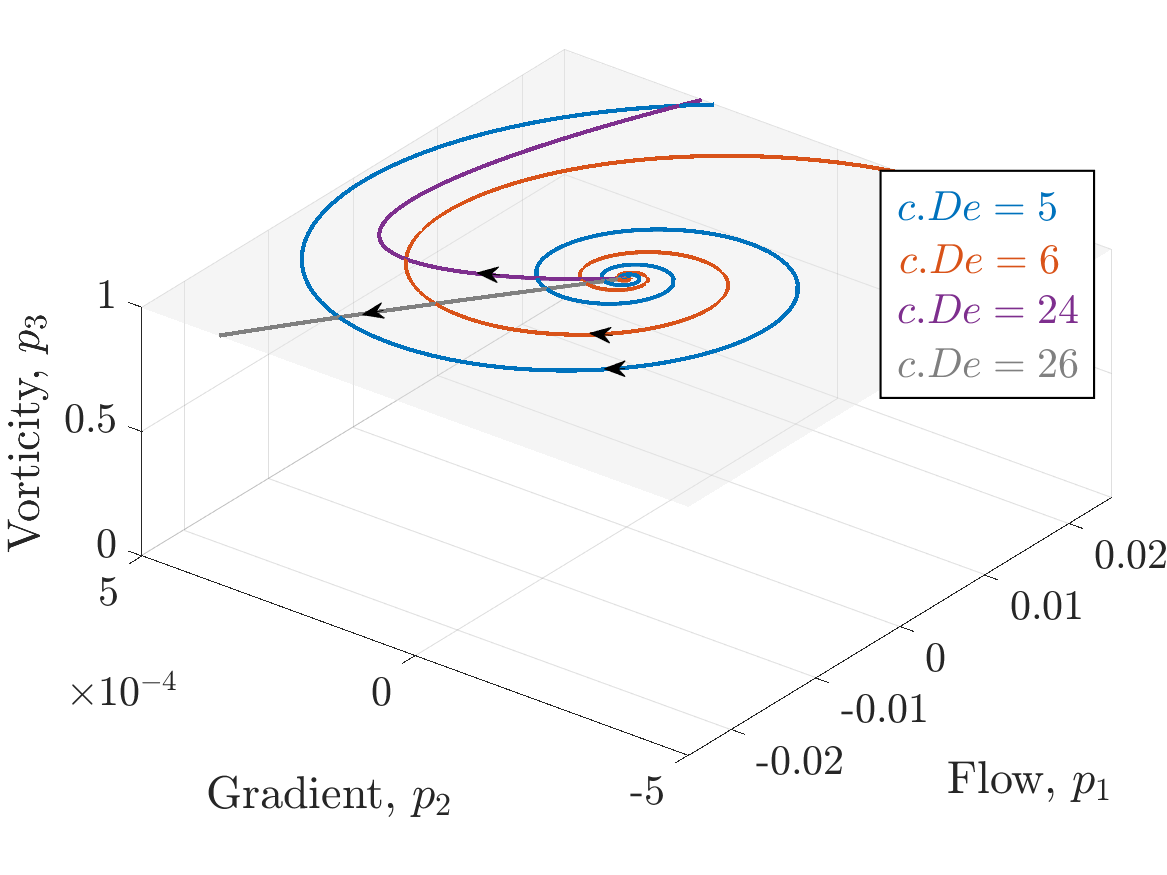}}\hfill
	\subfloat{\includegraphics[width=0.49\textwidth]{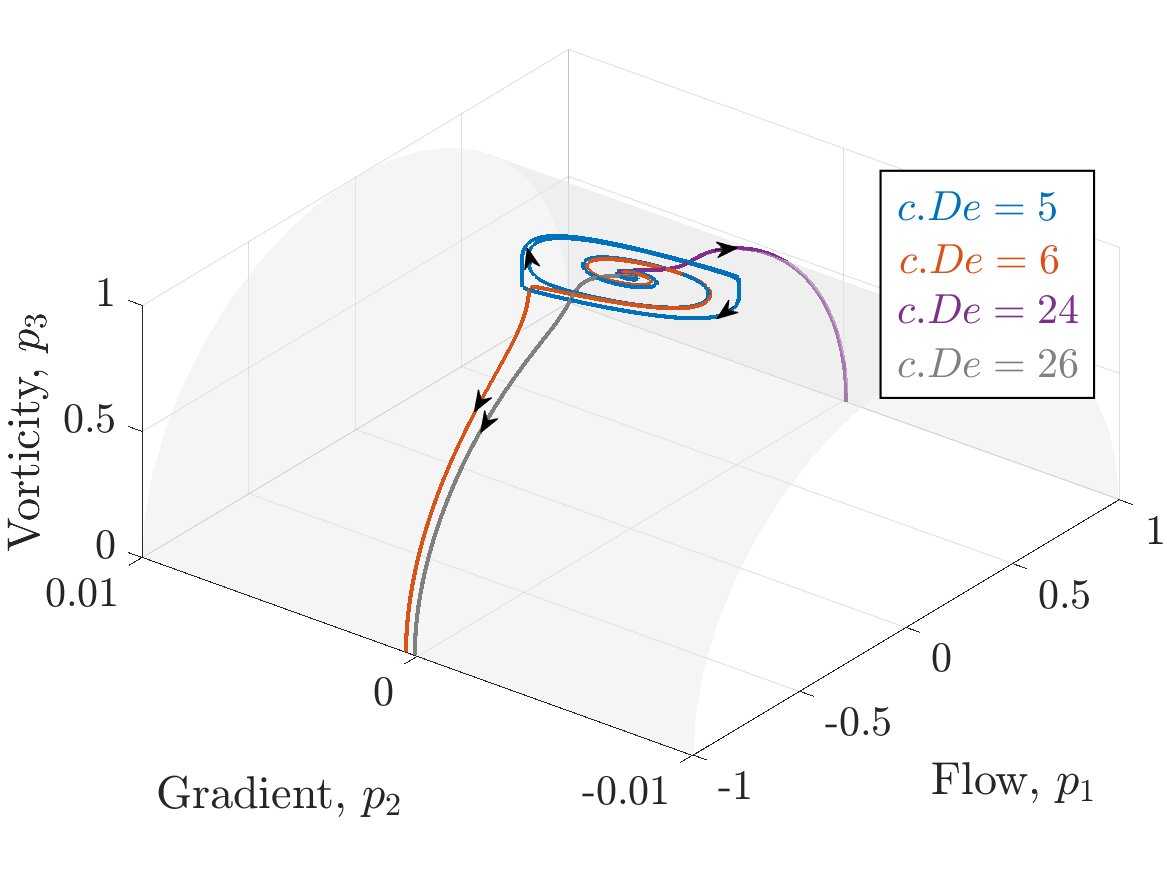}}
	\caption {Various orientation behaviors near the vorticity axis: trajectories of particle orientation starting very close to the vorticity axis at different $c\cdot De$ in $R_{flow}^{(1)}$ ($c De=6$, 24), $R_{flow}^{(2)}$ ($c De=26$), and $R_{flow-vort}^{(3)}$ ($c De=5$) at $\kappa=50$. All trajectories start at the same point. Left panel is same as right panel (showing complete particle trajectory) but zoomed near the vorticity axis ($p_3=1$). The gray surface is the unit sphere i.e. the orientation space. \label{fig:TrajectoriesR1R6R7}}
\end{figure}

The linear stability analysis is confirmed by the full numerical integration of equation \eqref{eq:FullOrbitEquationNewtExact} for $\kappa=50$ at $c\cdot De=24$ (also for $c\cdot De=5$ and 6) vs. $c\cdot De=26$ in the left panel of figure \ref{fig:TrajectoriesR1R6R7} zoomed near the vorticity axis. At $\kappa=50$, $c\cdot De=26$ is a point in $R_{flow}^{(2)}$, $c\cdot De=24$ and 6 are points in $R_{flow}^{(1)}$ and $c\cdot De=5$ is a point in $R_{flow-vort}^{(3)}$. In $R_{flow}^{(2)}$, the particle drifts out of the vorticity axis monotonically, and in the rest, it spirals out. We will discuss the dynamical system's features of the right panel of figure \ref{fig:TrajectoriesR1R6R7} in more detail later in section \ref{sec:StableLC} where we will find the unstable spiral at the vorticity axes to be surrounded by a stable limit cycle up to a cut-off $c\cdot De$. Hence, the Hopf bifurcation occurring at $c\cdot De=0$ at the vorticity axis is supercritical \citep{strogatz2018nonlinear}. In $R_{flow-vort}^{(3)}$ ($c\cdot De=5$), the particle starting very close to the vorticity axis spirals outwards into the stable limit cycle, where it then moves periodically around the vorticity axis. The limit cycle does not exist in $R_{flow}^{(1)}$ and $R_{flow}^{(2)}$, so that, after the particle either spirals or monotonically drifts out of the vorticity axis, it ends up in the flow aligned state as it approaches the stable fixed point ($\pm\tilde{p}_{2,\text{flow}}^{0-}$) near the flow direction in the flow-gradient plane for $c\cdot De=6$, 24 and 25.

The identification of the stable limit cycle near the vorticity axis is made possible by including the flow generated by force dipoles per unit length (flow from \cite{cox1971motion}) along the fiber in addition to the flow generated by the force per unit length (flow from \cite{batchelor1970slender}). If this dipole generated flow is neglected from our theory the stable limit cycle will not be predicted and instead when the theory predicts particle to be repelled from the flow-gradient plane it will approach the vorticity axis. A previous slender body theory for a second order fluid by \citet{ferec2021rigid}, relying on only the flow generated by the force per unit length, predicts the fiber to align with the vorticity axis. When the axis of symmetry of an axisymmetric slender fiber is aligned in the flow-vorticity plane of the imposed flow no force per unit length is exerted by the fluid (Newtonian or polymeric). This is  because in the flow-vorticity aligned state the imposed flow has no variation along the fiber axis. However, according to equations (29) and (35) of \citet{ferec2021rigid} when a fiber is in the flow-vorticity plane, the force per unit length is nonzero and proportional to the polymer relaxation time. This implies an error in the expression for the tension force in equation (29) of \citet{ferec2021rigid}.

Figure 3a  of \cite{wang2019numerical} clearly shows that a $\kappa=4.0$ spheroid started  away from the vorticity axis in a plane Couette flow at $De=0.1$ reaches a closed orbit around the vorticity axis, instead of approaching the axis. Figures 2a and 2b of \cite{d2014bistability} at $De=1.0$ and 2.0 respectively for a $\kappa=4.0$ spheroid in an unbounded simple shear flow show drift towards the vorticity axis, but do not show the particle approaching the axis. Although these studies were conducted at high polymer concentration, $c$, and small $\kappa$, outside the formal range of validity of our theory, they indicate the possibility of a stable limit cycle around the vorticity axis. Furthermore, in the second order fluid regime with $De=0.1$, careful observation of figure 8a of the theoretical investigation of \cite{wang2020dynamics} shows a $\kappa=3.0$ spheroid approaching a stable limit cycle, instead of reaching the vorticity axis of an unbounded parabolic slit flow ($u=1-y^2$). Lastly, while the boundary element formulation aided study of  \cite{phan2002viscoelastic} in an Oldroyd-B fuid shows the $\kappa=2.0$ particle drifting towards the vorticity axis, the simulation (see figures 7 and 8 of their paper) stops before we can conclude if it will approach the axis or a stable limit cycle.

\subsection{Effect of viscoelasticity near the flow-gradient plane and flow direction}\label{sec:NearFlow}
We described the effect of viscoelasticity on the particle's motion near the flow direction within the flow-gradient plane (FGP) in section \ref{sec:LogRollingandTumbling}. Here, we consider the particle motion near (but not exactly on) the FGP. The analysis of section \ref{sec:LogRollingandTumbling} for the $p_2$ (gradient) direction is valid even outside the FGP. Near the flow direction, $p_1\approx1$,  the orientation dynamics and solution in the $p_2$ direction are given by equation \eqref{eq:p2dotnearflow} and \eqref{eq:p2solution}. The orientation dynamics in the $p_3$ (vorticity) direction for  $p_1\approx 1$ is governed by the simplified equation (from equation \eqref{eq:FullOrbitEquationNewtExact}),
\begin{equation} \dot{p}_3\approx-p_2p_3+cDep_3\Big(-\frac{4}{\kappa^2}+\frac{p_2}{8\log(2\kappa)-12}\frac{15\pi^2De}{3\pi^2+5De}\Big)+\mathcal{O}(p_3^2).\label{eq:p3dotnearflow}
\end{equation}
Its closed form solution is,
\begin{equation}
p_3=C\exp(cDe\zeta (t-t_0))\cos({(t-t_0)b})^{-1+cDe\gamma/(8\log(2\kappa)-12)}\label{eq:ClosedFormNearFlow}
\end{equation}
where $b^2$ is given in equation \eqref{eq:bsquare} and,
\begin{equation}
\gamma=\frac{15\pi^2De}{3\pi^2+5De},\hspace{0.1in}
\zeta=\frac{1}{4\log(2\kappa)-6}-cDe\frac{\gamma}{2(4\log(2\kappa)-6)^2}-\frac{4}{\kappa^2}.\label{eq:bsqrandzeta}
\end{equation}

\begin{figure}
	\centering	
	\subfloat{\includegraphics[width=0.49\textwidth]{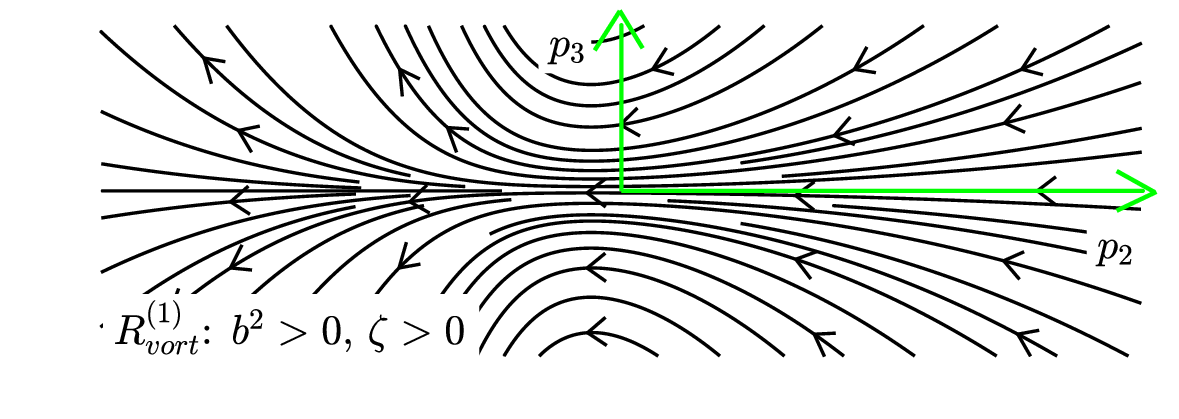}}\hfill
	\subfloat{\includegraphics[width=0.49\textwidth]{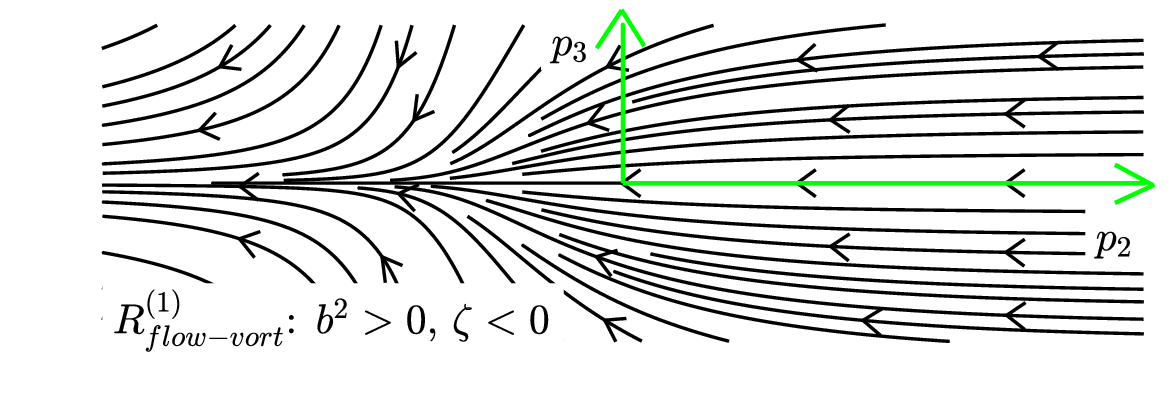}}
	\caption {{Phase portraits of equations \eqref{eq:p2dotnearflow} and \eqref{eq:p3dotnearflow} in the $b^2>0$ regime in the gradient ($p_2$)- vorticity ($p_3$) plane with $\zeta>0$ (left) and $\zeta<0$ (right). When $b^2>0$ and $\zeta>0$, i.e., in the left panel representing $R_{vort}^{(1)}$ close to the flow gradient plane ($p_3=0$), the $p_3$ component of the phase velocity changes sign from negative to positive along with an increase in magnitude downstream of the $p_2=0$ plane indicating a departure of a particle from the flow gradient plane at a rate higher than it approaches the plane. However, when $b^2>0$ and $\zeta<0$, i.e., in the right panel representing $R_{flow-vort}^{(1)}$  the $p_3$ component of the phase velocity is negative for all $p_2$ indicating an approach towards the flow gradient plane. Since $p_2$ never approaches zero on the flow gradient plane for $b^2>0$ it is an unstable and stable limit cycle for $R_{vort}^{(1)}$ ($\zeta>0$, left panel) and $R_{flow-vort}^{(1)}$ ($\zeta<0$, right panel) respectively.} \label{fig:PhasePotraitNearFlow}}
\end{figure}

When $b$ is real-valued, i.e. $b^2>0$ (equation \eqref{eq:bsquare}), there are no fixed points on the FGP. Thus, the particle undergoes periodic motion in the FGP, as discussed in section \ref{sec:LogRollingandTumbling}. For a particle perturbed slightly away from the FGP, from equation \eqref{eq:ClosedFormNearFlow} we note that the sign of $\zeta$ determines whether the particle will be attracted to or repelled from the FGP ($p_3=0$). $\zeta>0$ represents the region $R_{vort}^{(1)}$ where the particle spirals away from the FGP and  $\zeta<0$ represents the region $R_{flow-vort}^{(1)}$ where the particle spirals into the FGP and continues tumbling within the plane (albeit with a larger orbit period $2\pi/b$ than in Newtonian case). In the $b^2>0$ regime, the phase portrait of the dynamical system of equation \eqref{eq:FullOrbitEquationNewtExact} (through the simplified equations \eqref{eq:p2dotnearflow} and \eqref{eq:p3dotnearflow}) projected in $p_2-p_3$ plane near $p_1=1$ is shown in figure \ref{fig:PhasePotraitNearFlow} for $\zeta>0$ and $\zeta<0$. The phase flow in the $p_2-p_3$ plane in the case of a Newtonian fluid ($c\cdot De=0$) is similar to that in the region $R_{vort}^{(1)}$ (left panel of figure \ref{fig:PhasePotraitNearFlow}), but it is symmetric about $p_2=0$ (not shown). Therefore a particle in a Newtonian fluid continues in a particular (periodic) Jeffery orbit before and after $p_2=0$. At small but finite $c\cdot De$, i.e. in the region $R_{vort}^{(1)}$, represented by the left panel of figure \ref{fig:PhasePotraitNearFlow}, this symmetry about $p_2=0$ is broken and a particle comes out of the $p_2=0$ plane with a larger $\dot{p}_3$ velocity than it enters the plane. This explains the drift towards the vorticity axis (greater $p_3$). In the region $R_{flow-vort}^{(1)}$ represented by the right panel of figure \ref{fig:PhasePotraitNearFlow} we can observe that the phase flow points towards $p_3=0$, indicating migration of a particle towards the FGP. The FGP is a stable limit cycle in $R_{flow-vort}^{(1)}$ and an unstable limit cycle in $R_{vort}^{(1)}$.

\begin{figure}
	\centering	
	\subfloat{\includegraphics[width=0.6\textwidth]{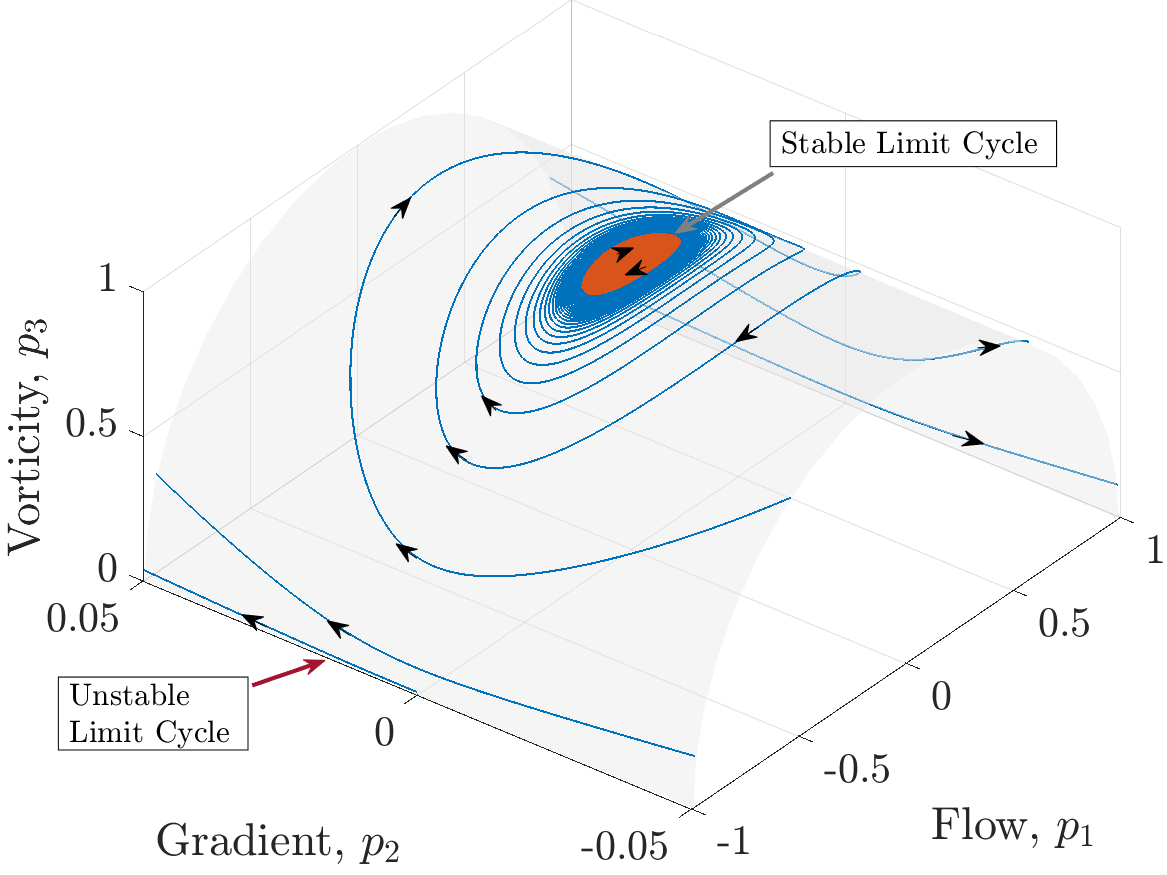}\label{fig:TrajcscDept1kappa50}}
	\caption {{In $R_{vort}^{(1)}$ (shown here for $c=0.005$, $\kappa=50, c\cdot De=0.01$), trajectories starting near the flow-gradient plane (exemplified here with the blue trajectory) spiral out of the plane. Globally they approach the same stable limit cycle as the trajectories starting near the vorticity axis (exemplified here with the orange trajectory). The blue trajectory starting near the flow direction spans a larger portion of phase space in $p_2$, but we show the region close to the flow-vorticity plane to highlight the limit cycle.}  \label{fig:TrajectoriesR2}}
\end{figure}

\begin{figure}
	\centering	
	\subfloat{\includegraphics[width=0.6\textwidth]{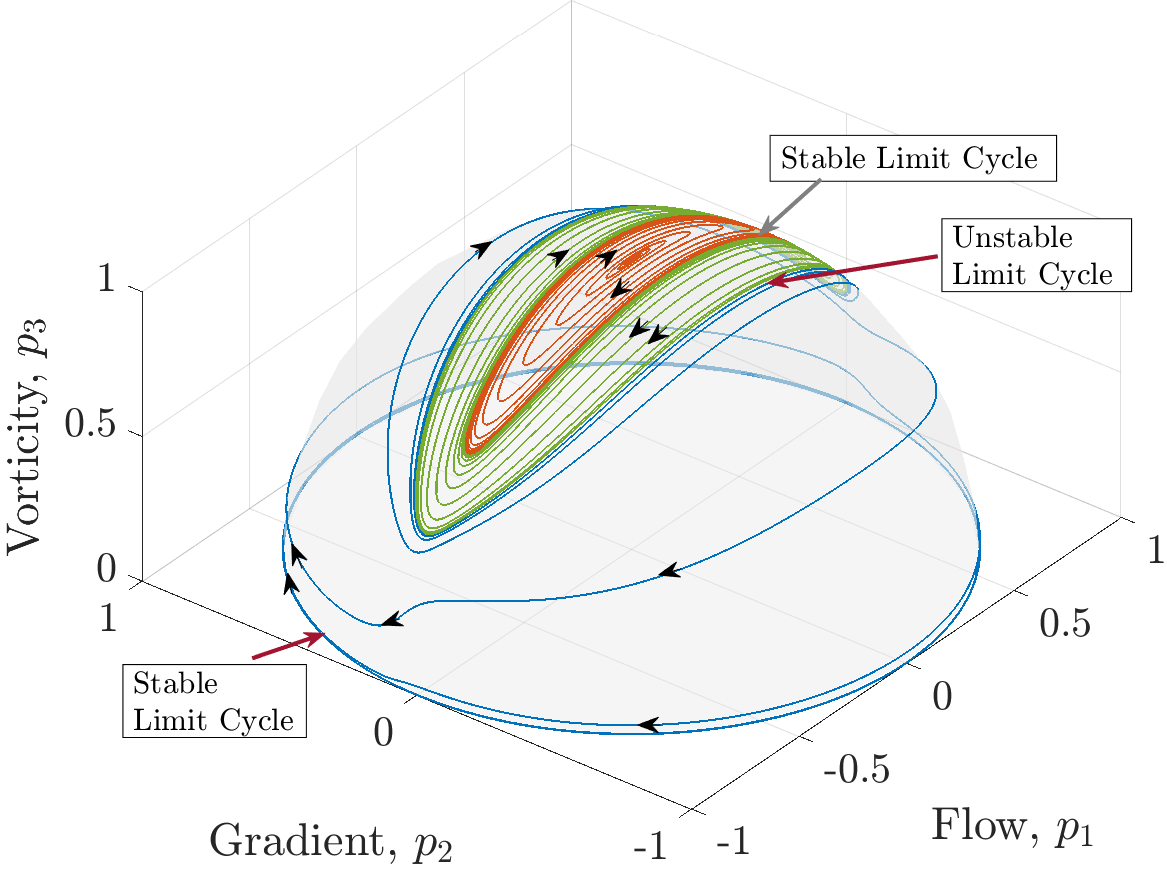}\label{fig:TrajcscDept48kappa10}}
	\caption{{In $R_{flow-vort}^{(1)}$ (shown here for $c=0.005$, $\kappa=10, c\cdot De=0.48$), trajectories of particle orientation starting near the flow-gradient plane (blue) spiral into the plane. Globally they emanate from an unstable limit cycle- the boundary between blue and green trajectories (that are started very close to each other in this numerical integration). There is a stable limit cycle above this unstable limit cycle at the boundary between green and orange trajectories.} \label{fig:TrajectoriesR3}}
\end{figure}

The numerically integrated trajectories for parameters chosen in $R_{vort}^{(1)}$ and $R_{flow-vort}^{(1)}$ for $c=0.005$ are shown in figures \ref{fig:TrajectoriesR2} and \ref{fig:TrajectoriesR3} respectively.  The prediction of the spiral exit of the orientation trajectories from the FGP in $R_{vort}^{(1)}$ and the spiral approach towards it in $R_{flow-vort}^{(1)}$ mentioned above is confirmed from these figures. The spiraling rate in $R_{vort}^{(1)}$ increases with $c\cdot De$ and $\kappa$ (not shown).

As $b^2$ is reduced by increasing $c\cdot De$ or $\kappa$, the time period, $2\pi/b$, increases. The effect of viscoelasticity (increasing $c\cdot De$) to increase the time period is consistent with the experimental observations of \citet{gauthier1971particle} and \citet{bartram1975particle}.  As mentioned earlier in section \ref{sec:LogRollingandTumbling}, when $b^2=0$, due to an infinite period bifurcation \citep{strogatz2018nonlinear} two fixed points emerge on the FGP, and we discuss the trajectories off the FGP in the $b^2<0$ regime next.

\subsubsection*{${b^2<0}$: Monotonic behavior near flow-gradient plane}
\begin{figure}
	\centering
	\subfloat{\includegraphics[width=0.49\textwidth]{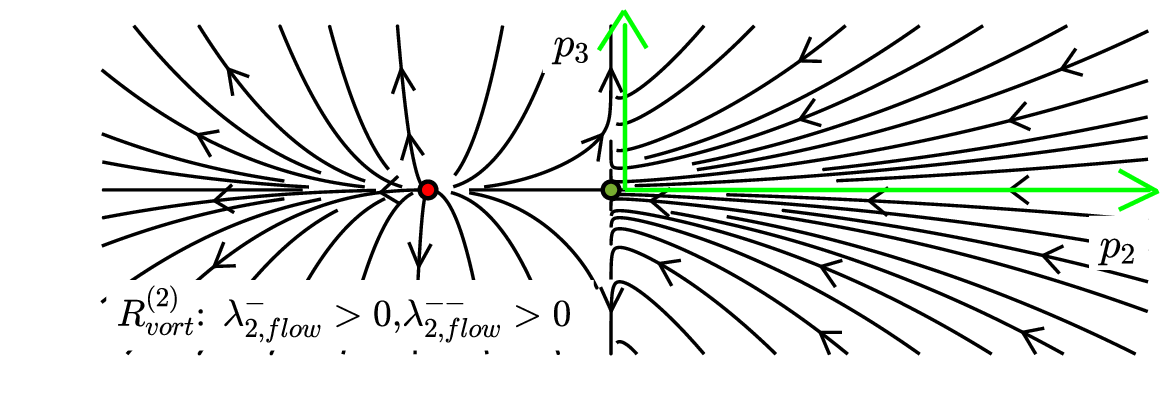}}\\
	\subfloat{\includegraphics[width=0.49\textwidth]{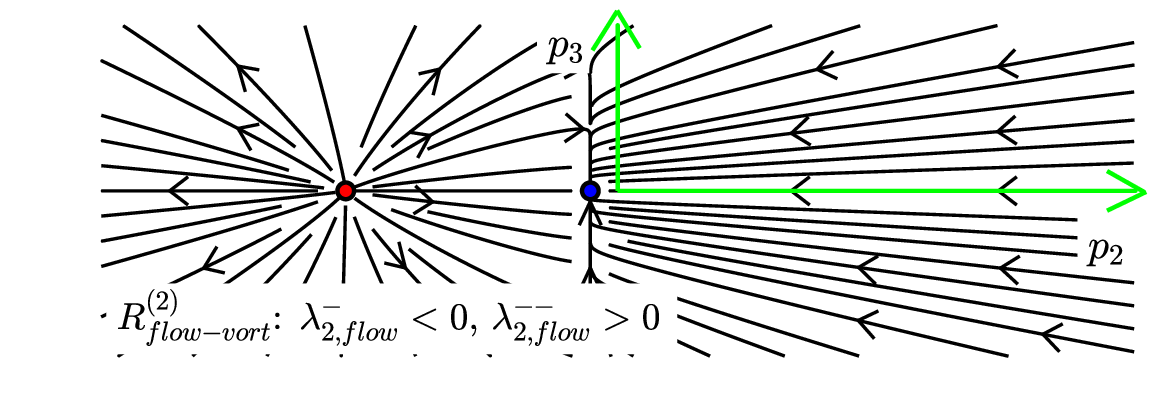}}\hfill
	\subfloat{\includegraphics[width=0.49\textwidth]{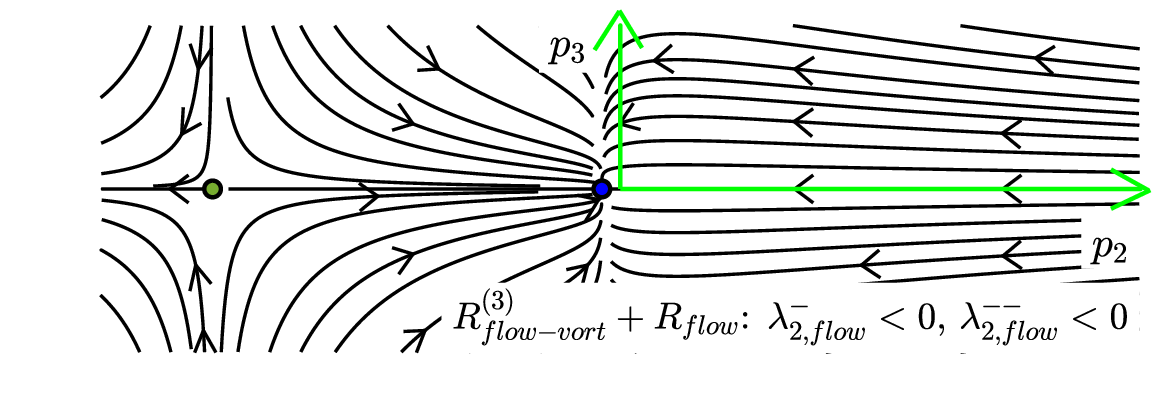}}
	\caption {{Phase portraits of the system of equations \eqref{eq:p2dotnearflow} and \eqref{eq:p3dotnearflow} in the $b^2<0$ regime. In this regime, two fixed points exist on the flow-gradient plane ($p_3=0$) close to the flow direction. Both the fixed points are downstream of the flow-vorticity plane ($p_2=0$). An unstable (red marker) and a saddle (green marker) node with its unstable manifold along the $p_3$ axis in the top panel ($R_{vort}^{(2)}$) indicates a monotonic drift of the particle away from the flow gradient plane. A stable fixed point (blue marker) near the flow direction ($p_2\approx 0, p_3=0$) in the bottom left ($R_{flow-vort}^{(2)}$) and right ($R_{flow-vort}^{(3)}$) panels indicate that particles with starting orientation near the flow gradient plane may align near the flow direction. The presence of an unstable point in $R_{flow-vort}^{(2)}$ (bottom left) instead of a saddle point with its stable manifold perpendicular to the flow gradient plane in $R_{flow-vort}^{(2)}$ (bottom right) in addition to the stable point in these cases indicates a lower proportion of trajectories leading to flow alignment in $R_{flow-vort}^{(2)}$ than in $R_{flow-vort}^{(3)}$. \label{fig:PhasePotraitNearFlow2}}} 
\end{figure}

Equation \eqref{eq:p2dotnearflow} has two fixed points at $p_{2,\text{flow}}^{0-}$ and $p_{2,\text{flow}}^{0--}$ from equation \eqref{eq:FixedPointsNearFlow} (or their more accurate values in equation \eqref{eq:FixedPointsNearFlowAccurate}). For the dynamical system in the $p_2-p_3$ plane near the flow direction, defined by the system of equations \eqref{eq:p2dotnearflow} and \eqref{eq:p3dotnearflow}, these fixed points are $[p_{2,\text{flow}}^{0-},0]$ and $[p_{2,\text{flow}}^{0--},0]$. The eigenvectors at each of the fixed points are the same, i.e.
\begin{equation}
\mathbf{v}_{1,\text{flow}}=\begin{bmatrix}
1&0
\end{bmatrix},
\mathbf{v}_{2,\text{flow}}=\begin{bmatrix}
0&1
\end{bmatrix}.
\end{equation}
The corresponding eigenvalues at $[p_{2,\text{flow}}^{0-},0]$ and $[p_{2,\text{flow}}^{0--},0]$ are $\lambda_{i,\text{flow}}^{0-}$ and $\lambda_{i,\text{flow}}^{0--}$, where $i\in[1,2]$,
\begin{align}\begin{split}
&\lambda_{1,\text{flow}}^{0-}=-2\sqrt{-b^2},\hspace{0.2in}
\lambda_{2,\text{flow}}^{0-}=\zeta c De -(\zeta+4/\kappa^2) (4\log(2\kappa)-6)\sqrt{-b^2}, \\&
\lambda_{1,\text{flow}}^{0--}=2\sqrt{-b^2},\hspace{0.2in}\lambda_{2,\text{flow}}^{0--}=\zeta c De +(\zeta+4/\kappa^2) (4\log(2\kappa)-6)\sqrt{-b^2}.\label{eq:EigenFlow}
\end{split}\end{align}
Both the fixed points are hyperbolic in the $b^2<0$ regime. Thus similar to the fixed point at the vorticity axis using the Hartman-Grobman theorem, the linear stability of the fixed points allows qualitative insight into the complete non-linear behavior of particle orientation. Both eigenvalues of both the fixed points are real in the $b^2<0$ regime. The first eigenvalues ($\lambda_{1,\text{flow}}^{0-}$ and $\lambda_{1,\text{flow}}^{0--}$) of both the fixed points do not change sign in $b^2<0$ regime i.e. $\lambda_{1,\text{flow}}^{0-}<0$ and $\lambda_{1,\text{flow}}^{0--}>0$. But, the second eigenvalues ($\lambda_{1,\text{flow}}^{0-}$ and $\lambda_{1,\text{flow}}^{0--}$), that are initially positive in the $b^2<0$ regime, become negative at different parameter ($c$, $De$ and $\kappa$) values. $\lambda_{2,\text{flow}}^{0-}$ becomes negative first. $\lambda_{2,\text{flow}}^{0-}=0$ is the bifurcation boundary where $[p_{2,\text{flow}}^{0-},0]$  changes from a saddle point (stable manifold along gradient direction and unstable along vorticity) to a stable fixed point. $\lambda_{2,\text{flow}}^{0--}=0$ is the bifurcation boundary where $[p_{2,\text{flow}}^{0--},0]$  changes from an unstable node to a saddle point (stable manifold along vorticity direction and unstable along gradient). Therefore, three new regimes arise within the $b^2<0$ region that are labeled in figure \ref{fig:PhasecDekapa} as (1) $R_{vort}^{(2)}$  ($\lambda_{1,\text{flow}}^{0-}>0$ and $\lambda_{1,\text{flow}}^{0--}>0$), (2) $R_{flow-vort}^{(2)}$ ($\lambda_{1,\text{flow}}^{0-}<0$ and $\lambda_{1,\text{flow}}^{0--}>0$), and (3) $R_{flow-vort}^{(3)}+R_{flow}$ ($\lambda_{1,\text{flow}}^{0-}<0$ and $\lambda_{1,\text{flow}}^{0--}<0$). The phase flow close to the fixed points in the gradient-vorticity plane for these three cases is shown in figure \ref{fig:PhasePotraitNearFlow2}. We can observe the phase flow approaching $[p_{2,\text{flow}}^{0-},0]$ and leaving $[p_{2,\text{flow}}^{0--},0]$ along the gradient direction ($p_2$) for all three cases. This implies that a particle with an initial orientation close to the flow-gradient plane (FGP) will approach the flow direction (specifically the fixed point $[p_{2,\text{flow}}^{0-},0]$) while slowing down. For the parameters within $R_{vort}^{(2)}$, the particle will then monotonically drift away from the FGP. This is similar to the experimental observation of \cite{bartram1975particle} discussed in section \ref{sec:Introduction}. For a given $\kappa$ and $c$, at larger $c\cdot De$ than $R_{vort}^{(2)}$, within the regions $R_{flow-vort}^{(2)}$ and $R_{flow-vort}^{(3)}$, the particle starting near the flow-gradient plane (and for all starting orientations in $R_{flow}$) achieves a stable orientation near the flow direction, similar to the experiments of \cite{iso1996orientation}.

\begin{figure}
	\centering	
	\subfloat{\includegraphics[width=0.6\textwidth]{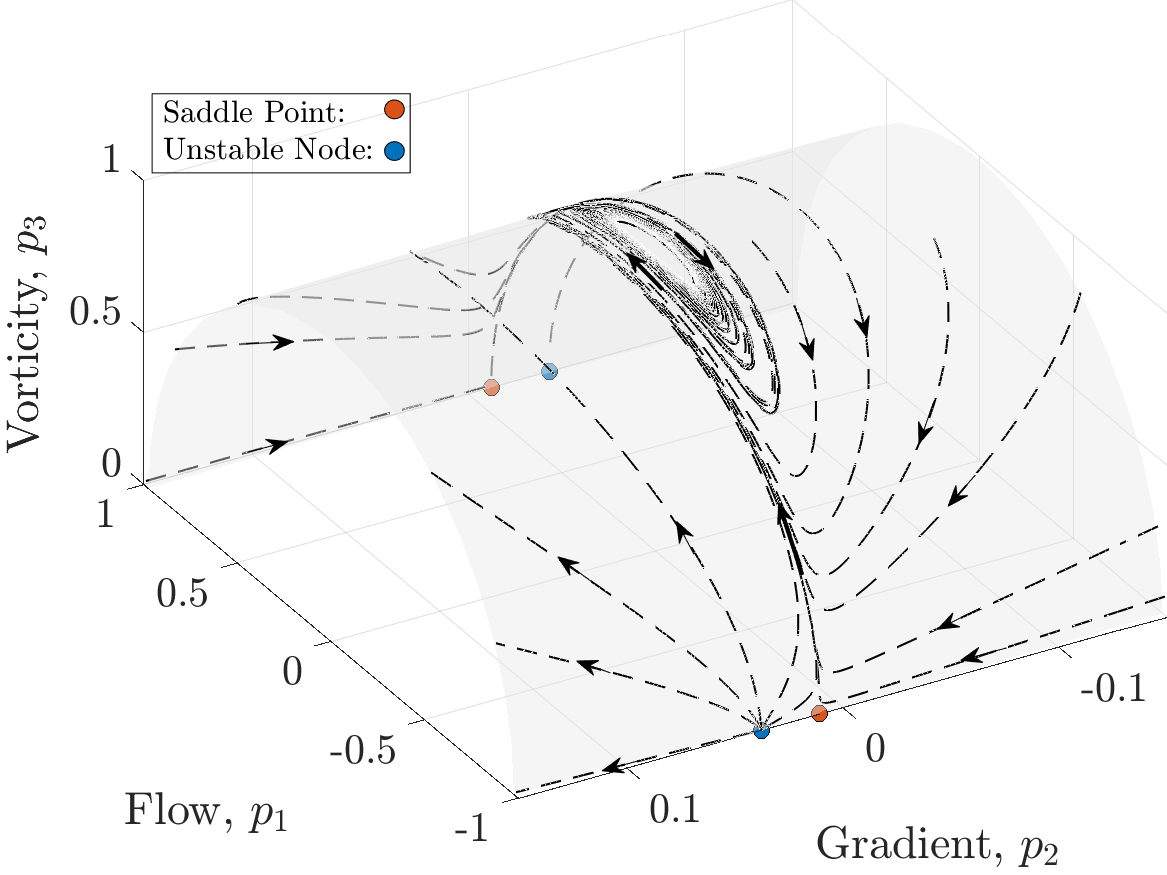}}\hfill
	\caption {In $R_{vort}^{(2)}$ (shown here for $c=0.005$, $\kappa=50, c\cdot De=0.3$), a particle's orientation trajectories approach the stable limit cycle around and near the vorticity axis. In contrast to $R_{vort}^{(1)}$ (figure \ref{fig:TrajectoriesR2}) where the trajectories spiral away from the flow-gradient plane, here they leave the flow-gradient plane monotonically. We show the region close to the flow-vorticity plane as this is where the different attractors lie. The gray surface is the unit sphere, i.e., the orientation space. \label{fig:TrajectoriesR4}}
\end{figure}

\begin{figure}
	\centering	
	\subfloat{\includegraphics[width=0.6\textwidth]{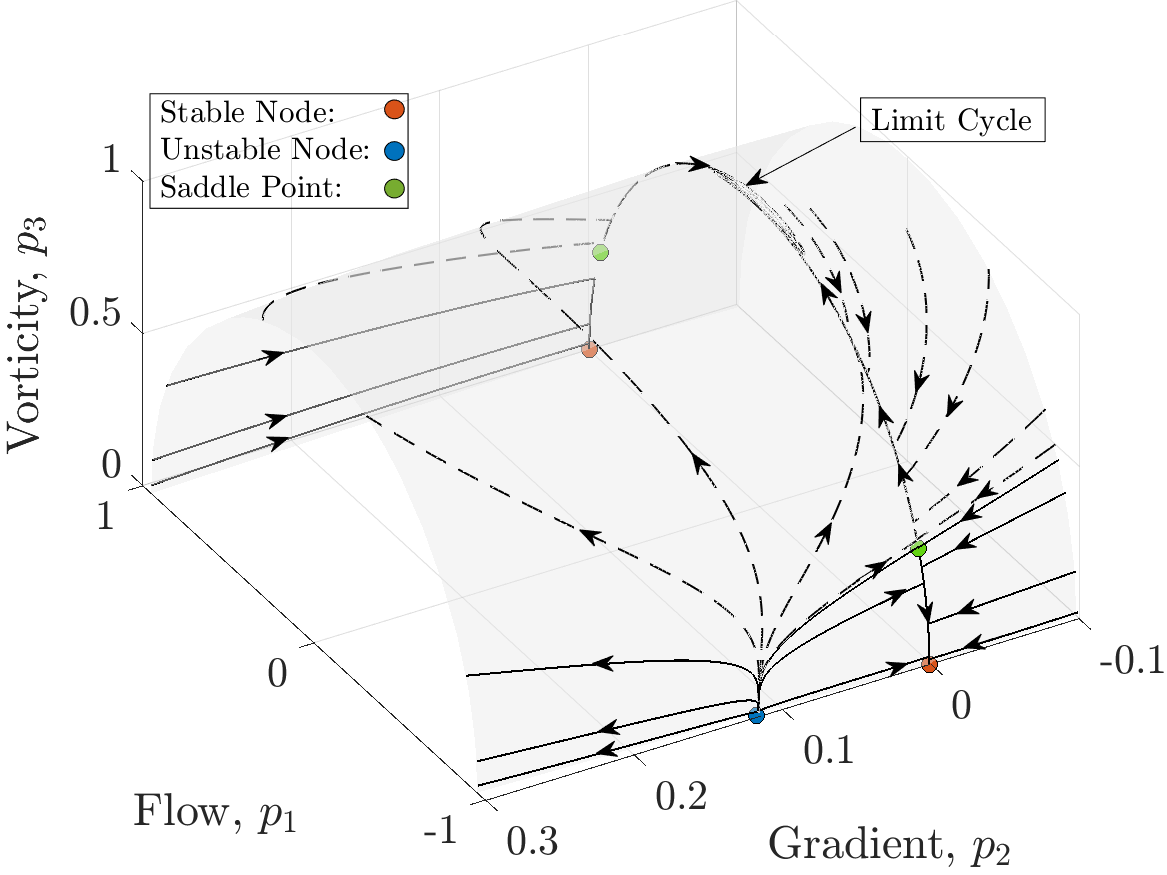}}
	\caption {{In $R_{flow-vort}^{(2)}$ (shown here for $c=0.005$, $\kappa=100$ and $c\cdot De=0.9$) a particle's orientation trajectories that start close to the flow-gradient plane (solid lines) approach the flow direction either on the same or the opposite side of the gradient-vorticity plane. Trajectories further away from the flow-gradient plane (dashed lines) approach the stable limit cycle near the vorticity axis. Due to a saddle and stable node close to each other on the flow-gradient plane and a stable limit cycle near the vorticity axis, another saddle-node emerges on the faster eigen-direction of the stable node. The gray surface is the unit sphere, i.e., the orientation space.} \label{fig:TrajectoriesR5}}
\end{figure}

\begin{figure}
	\centering	
	\subfloat{\includegraphics[width=0.6\textwidth]{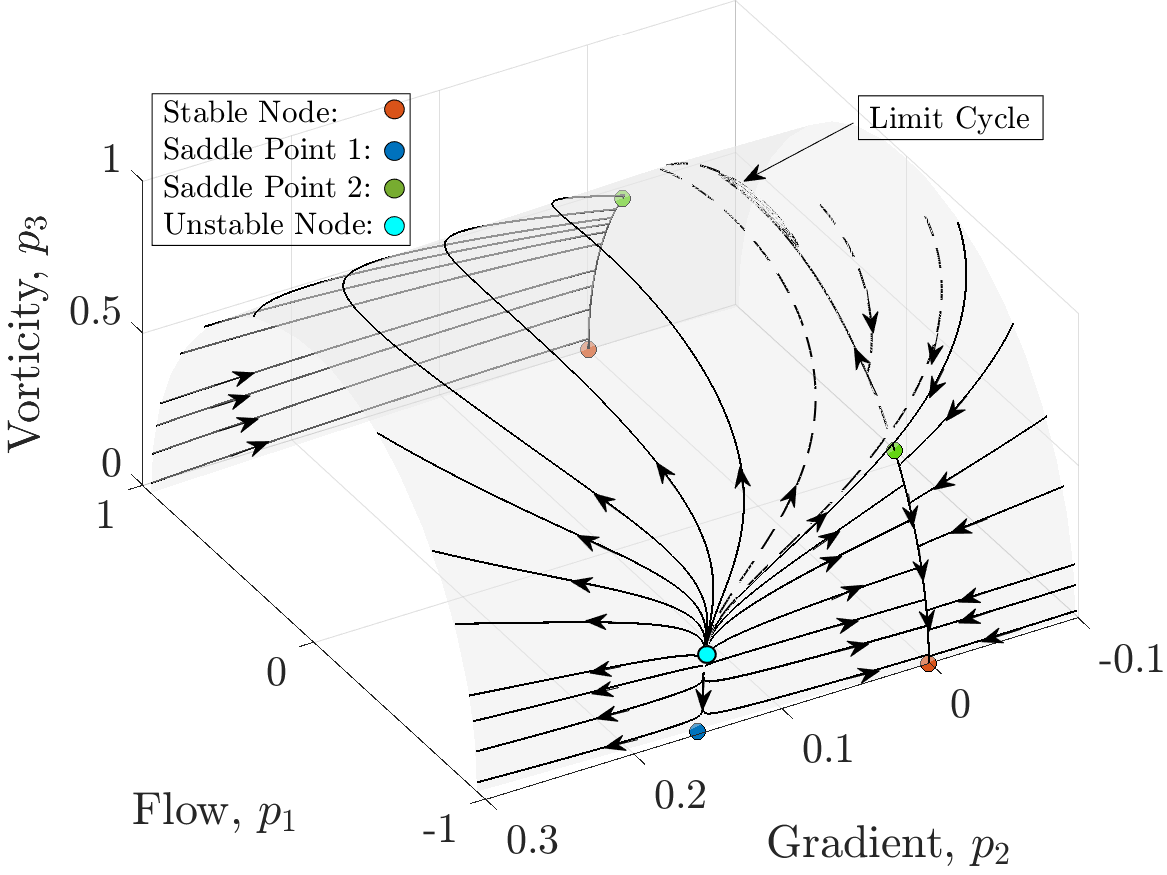}}
	\caption {In $R_{flow-vort}^{(3)}$ (shown here for $c=0.005$, $\kappa=100$ and $c\cdot De=1.2$) the behavior of a particle's orientation trajectories is similar to that in $R_{flow-vort}^{(2)}$ shown in figure \ref{fig:TrajectoriesR5}. The primary difference between $R_{flow-vort}^{(2)}$ and $R_{flow-vort}^{(3)}$ is that, in the former, the fixed point on the flow-gradient plane further from the flow direction is an unstable fixed point (figure \ref{fig:TrajectoriesR5}), while it is a saddle-node in the latter (shown here).  Additionally, in  $R_{flow-vort}^{(3)}$, there is an unstable node at the intersection of the stable manifold of the saddle points in flow-gradient and flow-vorticity plane. Trajectories with solid lines end up at one of the fixed points near the flow direction, and those with dashed lines end in the stable limit cycle near the vorticity axis. \label{fig:TrajectoriesR6}}
\end{figure}

  The locations of fixed points corresponding to $\tilde{p}_{2,\text{flow}}^{0-}$ and $\tilde{p}_{2,\text{flow}}^{0--}$ in the orientation space are $\begin{bmatrix}
 \pm\sqrt{1-(\tilde{p}_{2,\text{flow}}^{0-})^2}&\pm\tilde{p}_{2,\text{flow}}^{0-} &0
\end{bmatrix}$ and  $\begin{bmatrix}
\pm\sqrt{1-(\tilde{p}_{2,\text{flow}}^{0--})^2}&\pm\tilde{p}_{2,\text{flow}}^{0--} &0
\end{bmatrix}$. These locations match the corresponding fixed points in the numerically integrated trajectories in $R_{vort}^{(2)}$, $R_{flow-vort}^{(2)}$, and, $R_{flow-vort}^{(3)}+R_{flow}$ shown in plots of figures \ref{fig:TrajectoriesR4}, \ref{fig:TrajectoriesR5}, and, \ref{fig:TrajectoriesR6} respectively. We mark the analytical locations of these fixed points with two different colored markers on the flow-gradient plane ($p_3=0$) in figures \ref{fig:TrajectoriesR4}, \ref{fig:TrajectoriesR5}, and, \ref{fig:TrajectoriesR6} and show the nearby trajectories to approach/ leave these locations in the fashion described by the linear stability theory discussed above.

Numerical integration of the governing equation shown in figure \ref{fig:TrajectoriesR4} confirms the existence of the stable fixed point and unstable node predicted by linear stability analysis. In agreement with the $p_2-p_3$ phase plane analysis above, the particle starting near the flow gradient plane in $R_{vort}^{(2)}$ approaches the flow direction. It then monotonically departs the flow-gradient plane (FGP) along the flow-vorticity plane. The primary difference between $R_{vort}^{(1)}$ and $R_{vort}^{(2)}$ is that the particle leaves the FGP spirally in the former (figure \ref{fig:TrajectoriesR2}) and monotonically in the latter (figure \ref{fig:TrajectoriesR4}). As mentioned earlier, these trajectories are reminiscent of that in the experimental observations of \citet{bartram1975particle}. We can only make qualitative comparisons with \citet{bartram1975particle} due to unknown non-Newtonian properties of their viscoelastic fluid. Furthermore, the velocity field around the $\kappa=9.1$ particles used in their experiments is likely to have a quantitative difference from the SBT approximations used in our theory.

 In $R_{flow-vort}^{(2)}$, the numerically integrated trajectories shown in figure \ref{fig:TrajectoriesR5} reveal the presence of stable and unstable nodes in the flow-gradient plane at the locations  predicted by the linear stability analysis. Similarly, the existence of a stable node close to the flow direction and a saddle point further away on the flow-gradient plane is confirmed from the numerical integration shown in figure \ref{fig:TrajectoriesR6} for the parameters chosen in $R_{flow-vort}^{(3)}$. In $R_{flow-vort}^{(2)}$ and $R_{flow-vort}^{(3)}$ the trajectories starting close to the flow-gradient plane approach a stable orientation near the flow direction (stable fixed point at $\tilde{p}_{2,\text{flow}}^{0-}$).


\subsection{Global particle orientation dynamics}\label{sec:GlobalFlow}
In section \ref{sec:LogRollingandTumbling} we analyzed the equations for the rotational motion of a particle at the vorticity axis and within the flow-gradient plane (FGP). We found that viscoelasticity does not lead to a nonzero $\dot{\mathbf{p}}$ at the vorticity axis and it does not change the log-rolling rotational velocity. In the flow-gradient plane, we observed that a small amount of viscoelasticity (small $c\cdot De$) leads to a larger orbit time period. Further increasing $c\cdot De$ leads to an infinite period bifurcation and birth of two fixed points, thereby leading to particle migration towards the flow direction when in the FGP. In section \ref{sec:NearVorticity}, by linearizing the governing equation at the vorticity axis, we showed that the vorticity axis is an unstable spiral for $c\cdot De<\kappa$. For $c\cdot De>\kappa$, it is an unstable node. Therefore, a small perturbation of the particle orientation from the vorticity axis leads to departure from the axis. This was corroborated by the orientation trajectories obtained by numerical integration of the system of governing equations. By analyzing the system and its fixed points (when they exist) near the flow direction in the FGP and performing numerical integration of the trajectories starting near but not on the FGP in section \ref{sec:NearFlow}, we determined the orientational dynamics in the region nearby the flow gradient plane. The particle either spirals away from or towards the FGP when there are no fixed points in FGP. When the fixed points in FGP arise, the one closer to the flow direction is either a saddle point (with its stable manifold in the FGP) or a stable node. Therefore, the particle first migrates towards the flow direction nearly parallel to the FGP. Then it may either settle near the flow direction or monotonically drift away from the FGP along the flow-vorticity plane. The boundaries in the $c\cdot De-\kappa$ space separating these different qualitative behaviors were shown in figure \ref{fig:PhasecDekapa} for $c=0.005$.  The qualitative nature of these boundaries is not sensitive to the exact value of $c$. As we observed in figures \ref{fig:TrajectoriesR2}, \ref{fig:TrajectoriesR3}, \ref{fig:TrajectoriesR4}, \ref{fig:TrajectoriesR5} and \ref{fig:TrajectoriesR6}, there are other invariant features (stable limit cycle, unstable limit cycle, saddle point, and unstable node) that arise in the orientation space off the FGP and the vorticity axis. These features may be viewed as a result of the orientation space being restricted to a unit sphere.

{In $R_{vort}^{(1)}$ (figure \ref{fig:PhasecDekapa}), which extends to arbitrarily small (but finite) viscoelasticity (small $c\cdot De$), the vorticity axis is a spiral source, and the flow-gradient plane is an unstable limit cycle. Therefore, at least one stable attractor must exist between these two regions on the unit sphere. In the case of a Newtonian fluid, the particle follows an initial condition dependent Jeffery orbit, i.e., one of a concatenation of neutrally stable periodic orbits centered at the vorticity axis. Therefore, at small (but finite) $c\cdot De$ the simplest change in phase-plane dynamics leads to a stable limit cycle around and near one of these limit cycles, as shown in figure \ref{fig:TrajectoriesR2}. As shown by linear stability analysis at the vorticity direction in section \ref{sec:NearVorticity} the rate of deviation of trajectories away from the vorticity axis is driven by the $\mathcal{O}(1/\kappa^{2})$ flow generated by the force doublet and $\mathcal{O}(1/\kappa^{2})$ Stokeslet. Thus, a slender particle starting at any orientation between the vorticity axis and the flow-gradient plane leads to a final orientation behavior where the particle undergoes periodic motion very close to and around the vorticity axis.}

In $R_{flow-vort}^{(1)}$ (figure \ref{fig:PhasecDekapa}), the FGP changes from being an unstable to a stable limit cycle. Therefore, an unstable invariant object or a repeller in the form of an unstable limit cycle exists between the stable limit cycle near the vorticity direction and the FGP, as shown in figure \ref{fig:TrajectoriesR3}. Thus in $R_{flow-vort}^{(1)}$, a particle with an initial orientation between the unstable limit cycle and FGP spirals to the FGP, where it undergoes tumbling motion. A particle with an initial orientation between the unstable limit cycle above the FGP and the stable limit cycle near the vorticity axis spirals towards the stable limit cycle near the vorticity axis. A particle released at an arbitrarily small (but finite) angle from the vorticity axis spirals outwards. In both these cases, eventually, the particle undergoes perpetual periodic motion on the stable limit cycle close to the vorticity axis.

In $R_{vort}^{(2)}$ (figure \ref{fig:PhasecDekapa}), the FGP contains two fixed points as shown in figure \ref{fig:TrajectoriesR4}. One is an unstable node, and the other is a saddle point with its unstable manifold perpendicular to the FGP. Thus, no other invariant feature (attractor or repeller) is needed between the FGP and the stable limit cycle near the vorticity axis, which is the only stable invariant object on the orientation space. Hence a particle starting at an arbitrarily small angle from the vorticity axis or the FGP ends up in a periodic motion very close to the vorticity axis.

Similar to $R_{vort}^{(1)}$, $R_{vort}^{(2)}$ and $R_{flow-vort}^{(1)}$ discussed above, there is a stable limit cycle around the vorticity axis in $R_{flow-vort}^{(2)}$ and $R_{flow-vort}^{(3)}$. Therefore, a particle with an initial orientation arbitrarily close to the vorticity direction (but not on the vorticity axis) eventually undergoes a periodic motion around the vorticity axis. This stable limit cycle exists for all regions except for $R_{flow}^{(1)}$ and $R_{flow}^{(2)}$ and it will be further analyzed in section \ref{sec:StableLC}.

In $R_{flow-vort}^{(2)}$ (figure \ref{fig:PhasecDekapa}) the fixed point in the FGP closest to the flow direction is a stable node and the other fixed point in the FGP is an unstable node as shown in figure \ref{fig:TrajectoriesR5}. Hence, a saddle node exists near the flow-vorticity plane to repel the particle orientation trajectories towards the stable limit cycle near the vorticity axis on one side and the stable fixed point near the flow direction on the FGP on the other. This saddle point receives trajectories emanating from the unstable point on the FGP. Thus, depending upon the initial orientation of the particle, it may eventually either obtain a stable orientation near the flow direction or undergo a periodic motion close to the vorticity axis.

The dynamics in $R_{flow-vort}^{(3)}$ (figure \ref{fig:PhasecDekapa}) are similar to those in $R_{flow-vort}^{(2)}$, but the unstable fixed point in the FGP in $R_{flow-vort}^{(2)}$ changes to a saddle-node in $R_{flow-vort}^{(3)}$ with no qualitative change to the other invariant objects present in $R_{flow-vort}^{(2)}$. Therefore, to allow smooth and continuous phase flow, an unstable node occurs at the intersection of the saddle points in the FGP and near the flow-vorticity plane, as shown in figure \ref{fig:TrajectoriesR6}. At a given $\kappa$, the value of $c\cdot De$ in $R_{flow-vort}^{(3)}$ is larger than that in $R_{flow-vort}^{(2)}$ (figure \ref{fig:PhasecDekapa}) and the separation between two fixed points increases with increasing $c\cdot De$ as shown by figure \ref{fig:accuratefixpointsonfg}. Due to this increased separation and the unstable node above the flow gradient plane, more of the phase flow is directed toward the FGP in $R_{flow-vort}^{(3)}$  than in $R_{flow-vort}^{(2)}$. Thus, the basin of attraction for the stable fixed point near the flow direction is larger in $R_{flow-vort}^{(3)}$ as compared to $R_{flow-vort}^{(2)}$. In other words more of the initial particle orientations lead to a stable final orientation near the flow direction in the  $R_{flow-vort}^{(3)}$  than in $R_{flow-vort}^{(2)}$ as observed by comparing the proportion of solid lines in figures \ref{fig:TrajectoriesR5} and \ref{fig:TrajectoriesR6}. When the stable limit cycle near the vorticity axis exists, the saddle point on the flow-vorticity plane is important in determining the proportions of initial orientations leading to the flow alignment and towards the periodic motion around the vorticity axis. Its location will be further analyzed in section \ref{sec:SaddleonFV}.

The $c\cdot De|_\text{cut-off}^\text{Limit}$ boundary in figure \ref{fig:PhasecDekapa}, beyond which the stable limit cycle near the vorticity direction does not exist, may be viewed as arising due to the interaction between the saddle-node near the flow-vorticity plane and the stable limit cycle near the vorticity direction as these move in the orientation space upon increasing $c\cdot De$. We will discuss this in section \ref{sec:FlowAlignment}.

\subsubsection{Stable Limit cycle around the vorticity axis}\label{sec:StableLC}
The previous study of \cite{leal1975slow} concerning a slender particle in simple shear flow of a second-order fluid predicts that viscoelasticity (although incorrect in attributing it to the second instead of the first normal stress difference)  drives the particle orientation towards the vorticity axis. He only consider the Stokeslet fluid flow (of order $\mathcal{O}(1/\log(\kappa))$ and $\mathcal{O}(1/\log(\kappa)^2)$) from the slender body theory. This velocity field is proportional to $p_2$ and hence has no effect when the particle is in the flow-vorticity plane. However, our study shows that a stable limit cycle exists around the vorticity axis due to the competition between polymer-driven torque arising from the doublet flow causing the particle to spiral away from the vorticity axis and that from the Stokeslet flow at {$\mathcal{O}(1/\log(\kappa))$} causing it move towards the region around vorticity.

\begin{figure}
	\centering
	\includegraphics[width=0.55\linewidth]{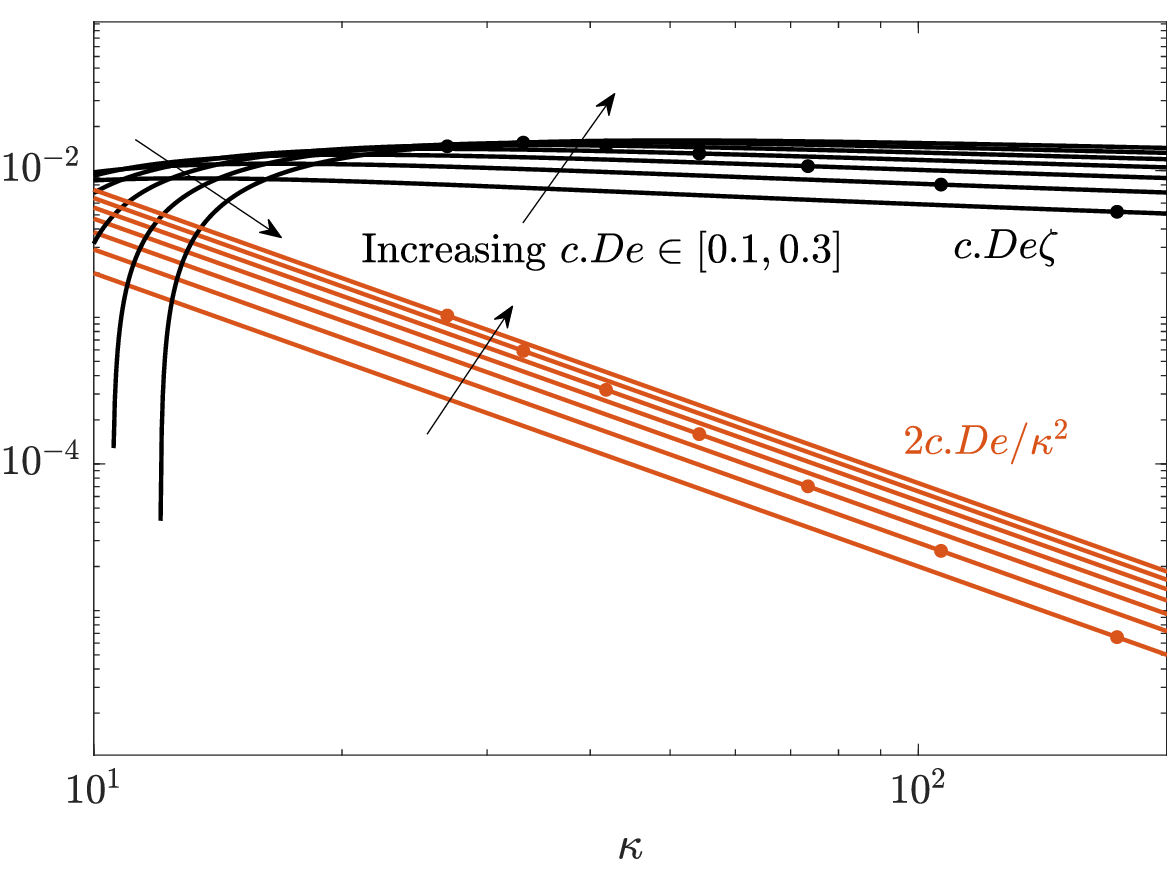}
	\caption{{The variation with $\kappa$ of the rate of deviation of orientation trajectories away from the vorticity axis, $2cDe/\kappa^2$ (orange), and the flow gradient plane, $cDe\zeta$ (black), in $R_{vort}^{(1)}$ ($b^2>0$) for a few values of $c\cdot De$ at $c=0.005$. On each curve, $b^2>0$ is to the left of the solid markers. An increasing gap between orange and black curves, with almost horizontal black curves, with $\kappa$ indicates an larger increase in the departure rate of the orientation trajectories from the flow gradient plane than that from the vorticity axis, implying a shift of the stable limit cycle closer to the vorticity axis.}}
	\label{fig:zetaand2cdebykapasq}
\end{figure}

The real part of the eigenvalues of the fixed point at the vorticity axis in $R_{vort}^{(1)}$ is $2cDe/\kappa^2$ (equation \eqref{eq:EigVort}), and the rate at which trajectories leave the flow-gradient plane is $cDe\zeta$ (equation \eqref{eq:ClosedFormNearFlow}). Figure \ref{fig:zetaand2cdebykapasq} shows the variation of $2cDe/\kappa^2$ and $cDe\zeta$ with $\kappa$ for a few values of $c\cdot De$ at $c=0.005$. From this figure, we conclude that for $\kappa\gtrapprox15$, trajectories leave the flow gradient plane much faster than the outward spiraling rate from the vorticity axis. Hence, the stable limit cycle shifts closer to the vorticity axis as $\kappa$ increases. Furthermore, as $\kappa$ increases beyond $\kappa\approx20$, $cDe\zeta$ remains almost constant while the rate of spiralling from the vorticity axis, $\kappa^{-2}$, decreases. Therefore, the stable limit cycle approaches the vorticity axes at larger $\kappa$.

\begin{figure}
	\centering
	\subfloat{\includegraphics[width=0.49\textwidth]{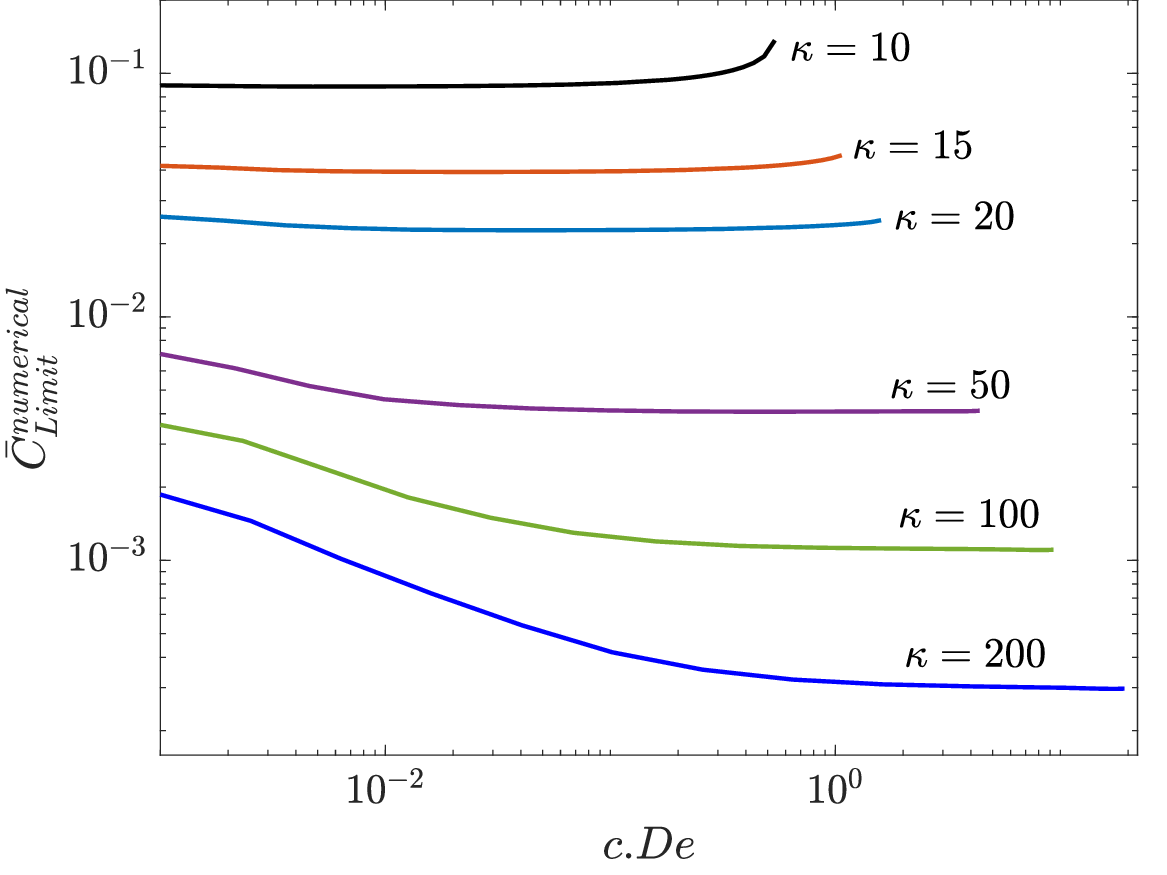}}\hfill
	\subfloat{\includegraphics[width=0.49\textwidth]{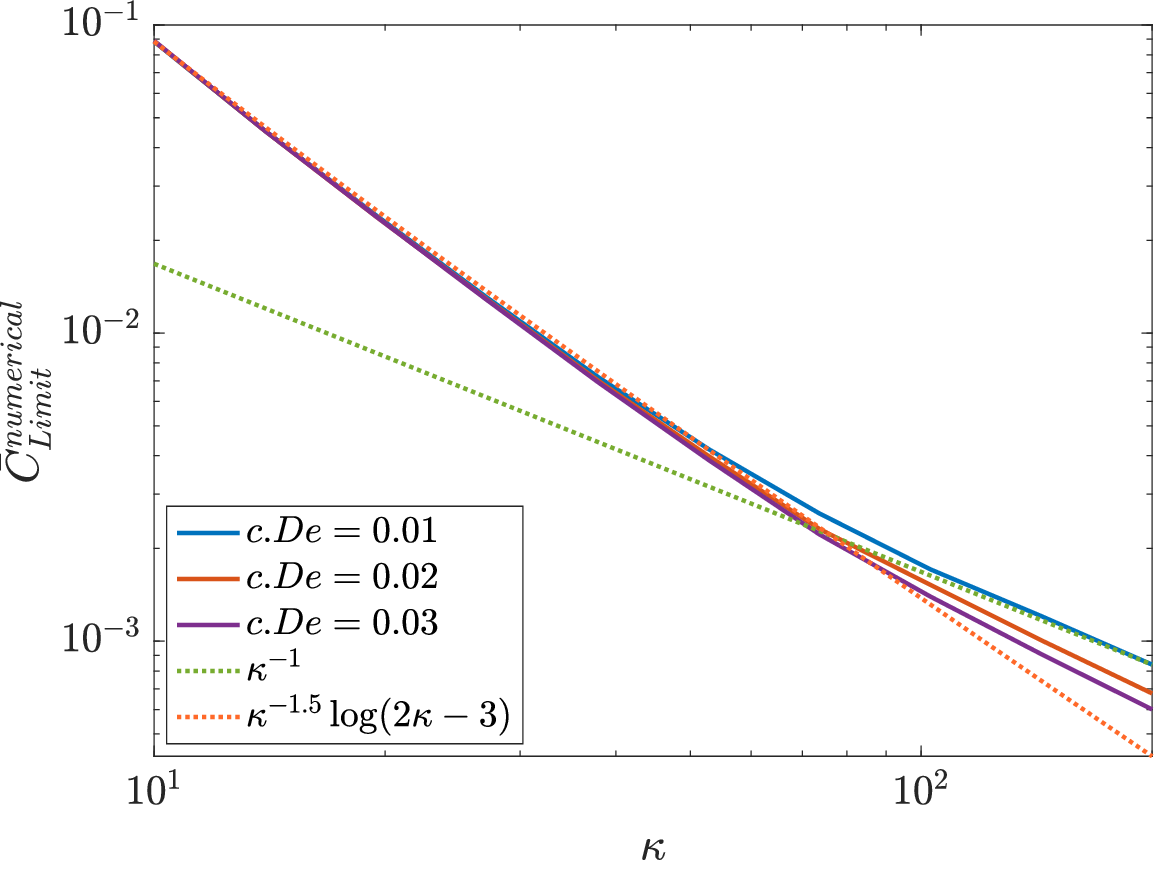}}
	\caption{{Left: The location of the stable limit cycle at different $c\cdot De$ and $\kappa$ at $c=0.005$ (nearly identical curves are obtained for other polymer concentrations in the range $0.01\le c\le0.2$) indicated by the average $C=\sqrt{p_2^2+p_1^2/\kappa^2}/{p_3}$, $\bar{C}_\text{Limit}^\text{numerical}$ on the limit cycle. $C=0$ is the vorticity axis and $C=\infty$ is the flow-gradient plane. Right: $\bar{C}_\text{Limit}^\text{numerical}$ on the limit cycle vs. $\kappa$ at low $c\cdot De$.}}
	\label{fig:caveragediffkapa}
\end{figure}

We quantify the location of the limit cycle observed in the numerical integration of equation \eqref{eq:FullOrbitEquationNewtExact}. For this purpose we transform this equation into the $C-\tau$ coordinate system \citep{leal1971effect}, using,
\begin{equation}
C=\frac{\sqrt{p_2^2+p_1^2/\kappa^2}}{p_3}, \tau=\frac{p_1}{p_2},
\end{equation}
to obtain,
\begin{eqnarray}
\frac{d C}{dt}=cDe\tau^2\Big(\frac{4(1+C^2)}{C\kappa^4}-\frac{C}{(4\log(2\kappa)-3)(\kappa^2+\tau^2)}\Big)p_2^2+cDe\mathcal{O}(p_2^4),\\
\frac{d \tau}{dt}=1+cDe\tau p_2^2\Big(\frac{\tau^2}{4\log(2\kappa)-3}+4\frac{\kappa^2+\tau^2}{C^2\kappa^4}\Big)+\mathcal{O}(c\cdot De)
\end{eqnarray}
In the Newtonian limit, ${d C}/{dt}=0$ and the particle trajectory is determined by its initial condition $C(t)=C(0)=C_N$. Each $C_N=[0,\infty]$ represents a different Jeffery orbit. $C=0$ represents the log-rolling motion, i.e., the particle rotating about its major axis  which is aligned with the vorticity axis, and $C=\infty$ represents the major axis rotating within the flow-gradient plane. $\tau=t$ represents the phase on the Jeffery orbit. A periodic orbit can thus be ascertained using a single parameter $C$ that repeats periodically. In the range, $10\le\kappa\le200$, for $c\cdot De\ge0.01$ we evolve the system of equations for an initial condition starting near the vorticity direction, at $C=10^{-7}$. We evolve the equations for each choice of $c\cdot De$ and $\kappa$ until the change in the period-averaged $C$ across 100 successive Jeffery periods is less than $10^{-5}$. Finally, we define $\bar{C}_\text{Limit}^\text{numerical}$ as the average $C$ over the last 100 Jeffery periods. $\bar{C}_\text{Limit}^\text{numerical}$ vs. $c\cdot De$ for a few values of $\kappa\in[10,200]$ is shown in the left panel of figure \ref{fig:caveragediffkapa}.

$\bar{C}_\text{Limit}^\text{numerical}$ is smaller than $0.1$ for $\kappa>10$, decreases with $\kappa$ and varies slightly with $c\cdot De$ at each $\kappa$. For $\kappa>50$ $\bar{C}_\text{Limit}^\text{numerical}$ is less than $0.01$ for a range of $c\cdot De$. Therefore, the limit cycle is close to the vorticity axis. Beyond a certain value of $c\cdot De=c\cdot De|_\text{cut-off}^\text{Limit}$, the limit cycle does not exist.  This marks the boundary between $R_{flow-vort}^{(2)}$ and $R_{flow}^{(1)}$ for a given $\kappa$, as shown in figure \ref{fig:PhasecDekapa}. $c\cdot De|_\text{cut-off}^\text{Limit}$ increases with $\kappa$ and for $10\le\kappa\le200$ we find,
\begin{equation}
c\cdot De|_\text{cut-off}^\text{Limit}\approx 0.1056\kappa-0.5183.
\end{equation}
At larger $\kappa$, the stable limit cycle is closer to the vorticity axes and exists for a larger range of $c\cdot De$. Hence, marked as $R_{flow-vort}^{(1)}$ and $R_{flow-vort}^{(2)}$ in figure \ref{fig:PhasecDekapa}, there is a significant range of $c\cdot De$ at large $\kappa$ where two possible types of orientation dynamics are possible: a steady state orientation close to the flow direction and a periodic orbit near the vorticity direction. For $\kappa\lessapprox20$, increasing $c\cdot De$ from very small values does not affect the position of the limit cycle much. However, as $c\cdot De$ approaches $c\cdot De|_\text{cut-off}^\text{Limit}$, the limit cycle moves away from the vorticity axis. For $\kappa\gtrapprox 20$, the limit cycle moves towards the vorticity axis upon increasing $c\cdot De$ from very small values. However, beyond a given $c\cdot De$, up to $c\cdot De|_\text{cut-off}^\text{Limit}$ the limit cycle's angular position is not influenced by $c\cdot De$. These features are reflected by the trends in $\bar{C}_\text{Limit}^\text{numerical}$ in the left panel of figure \ref{fig:caveragediffkapa}. At low $c\cdot De$, the limit cycle is very close to one of the degenerate Jeffery orbits of the Newtonian case, i.e., $C$ does not change much on the limit cycle. The variation of $\bar{C}_\text{Limit}^\text{numerical}$ with $\kappa$ at low $c\cdot De=0.01, 0.02 $ and 0.03 is shown in the right panel of figure \ref{fig:caveragediffkapa}, where we can observe $\bar{C}_\text{Limit}^\text{numerical}$ to scale as $\kappa^{-1.5}\log(2\kappa-3)$ for $10\le\kappa\le100$ and approximately $\kappa^{-1}$ for larger $\kappa$.

We find the particle orientation trajectories to approach this stable limit cycle close to the vorticity axis instead of spiraling towards the vorticity axis as concluded from the experiments of \cite{gauthier1971particle} and \cite{iso1996orientation1}. \cite{gauthier1971particle} claim the approach toward the vorticity axis by extrapolating the observed data. In some of the experimental observations of \cite{iso1996orientation1} at small elasticity after undergoing initial spiraling, the particle oscillates around the vorticity axis without settling at a steady orientation. This may indicate the limit cycle discussed above. The theoretical rate of spiraling away from the vorticity axis  ($2cDe/\kappa^2$) is small (figure \ref{fig:zetaand2cdebykapasq}), and small imperfections from the Couette cell might have led to secondary flows in the experiments. Thus, the existence and size of the stable limit cycle around the vorticity axis must be further tested in future experiments and numerical simulations.

\subsubsection{Saddle node near the flow-vorticity plane in $R_{flow-vort}^{(2)}$, $R_{flow-vort}^{(3)}$, $R_{flow}^{(1)}$ and $R_{flow}^{(2)}$}\label{sec:SaddleonFV}
\begin{figure}
	\centering	
	\subfloat{\includegraphics[width=0.49\textwidth]{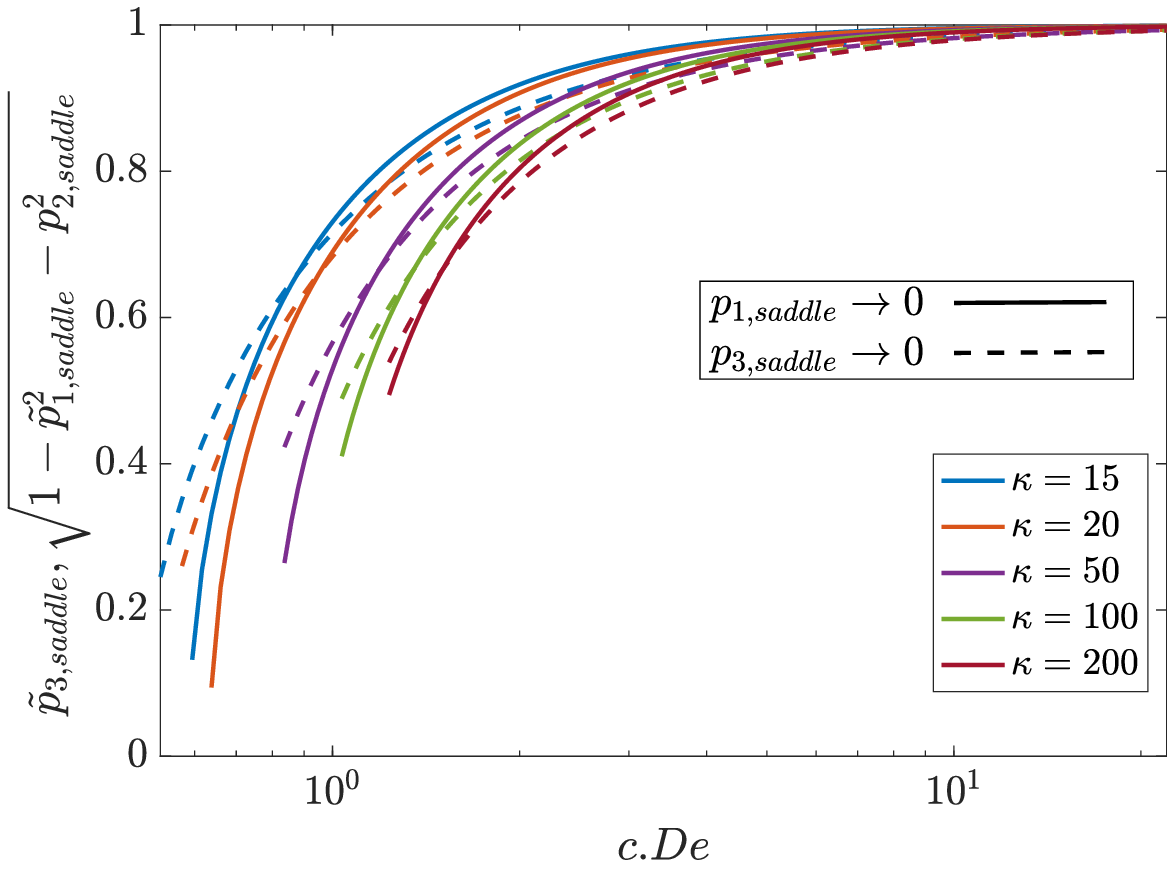}}\hfill
	\subfloat{\includegraphics[width=0.49\textwidth]{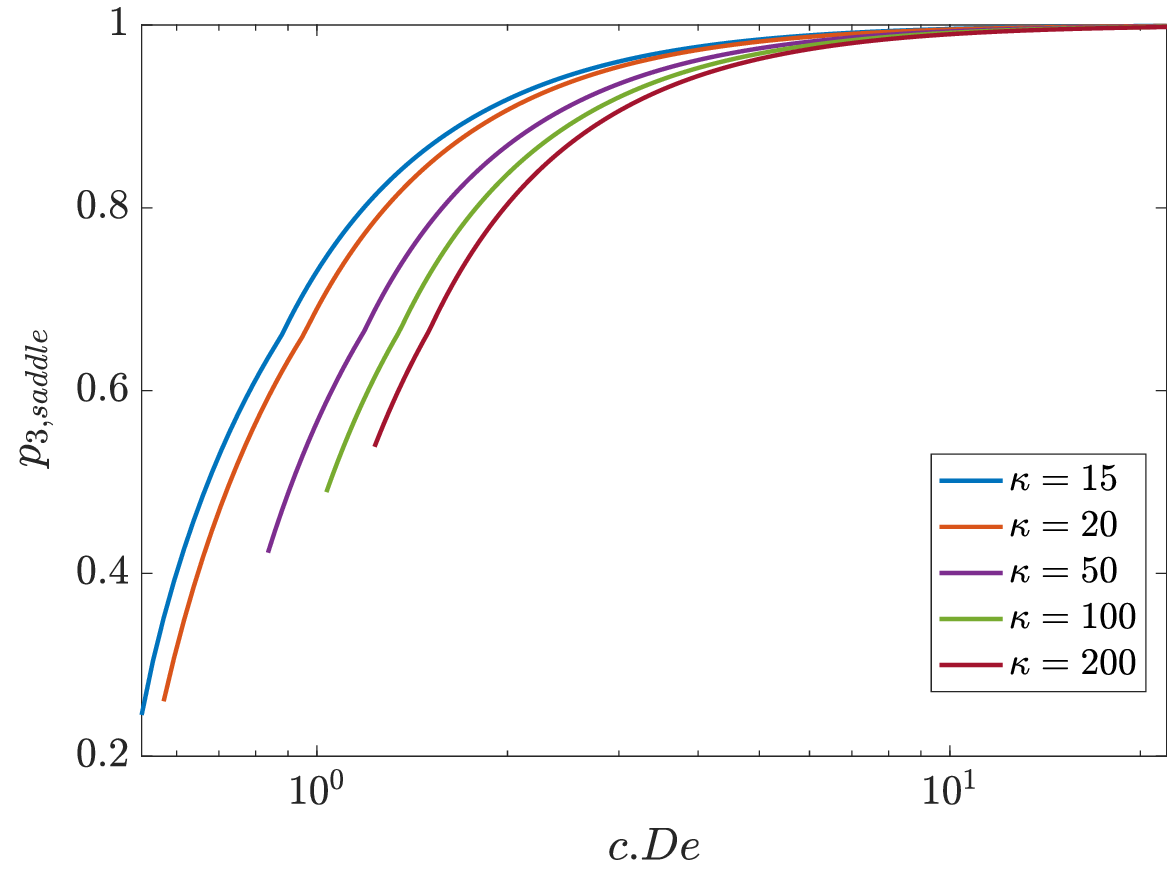}}
	\caption{{$p_3$ coordinate of the saddle point in the flow-vorticity plane in $R_{flow-vort}^{(2)}$ and $R_{flow-vort}^{(3)}$. Left: Solid lines (the graph of $\sqrt{1-\tilde{p}_\text{1,saddle}^2-p_\text{2,saddle}^2}$ from equation \eqref{eq:P1Saddle} and \eqref{eq:P2Saddle}) provide the more accurate $p_3$ location of the saddle point when it is closer to the vorticity axis ($p_\text{3,saddle}\approx1$) and dashed lines (the graph of equation \eqref{eq:P3Saddle}) represent the more accurate $p_3$ location of the saddle point when it is closer to the flow axis ($p_\text{3,saddle}\approx0$)\label{fig:p3saddle}. Right: most accurate location of the saddle point constructed from the left figure using the cross over point between the solid and dashed lines for each $\kappa$ i.e. using equation \eqref{eq:SaddleConditional}.}}
\end{figure}

From the trajectories of particle orientation in $R_{flow-vort}^{(2)}$ and $R_{flow-vort}^{(3)}$ shown in figures \ref{fig:TrajectoriesR5} and \ref{fig:TrajectoriesR6} we find the stable limit cycle around the vorticity axis and the stable node on the flow gradient plane near the flow axis to be separated by a saddle point. The saddle point arises on the boundary between $R_{vort}^{(2)}$ and $R_{flow-vort}^{(2)}$ when the saddle point near the flow direction on the flow-gradient plane changes into a stable node. It also arises on the boundary between $R_{flow-vort}^{(3)}$ and $R_{flow-vort}^{(1)}$ when the unstable limit cycle in the flow-gradient plane vanishes. The saddle point persists in the region $R_{flow-vort}^{(3)}$ when approaching it by increasing $c\cdot De$ from $R_{flow-vort}^{(2)}$. In the regions $R_{flow-vort}^{(2)}$ and $R_{flow-vort}^{(3)}$, the unstable direction of the saddle point leads to the limit cycle on one side, and the stable node on the other. Based on numerical evidence from figures \ref{fig:TrajectoriesR5} and \ref{fig:TrajectoriesR6} we assume the saddle point, $\mathbf{p}_\text{saddle}=\begin{bmatrix}
{p}_\text{1,saddle}&{p}_\text{2,saddle}&{p}_\text{3,saddle}
\end{bmatrix}$, to approximately lie at the same $p_2$ coordinate as the stable node in the flow gradient plane,
\begin{equation}
{p}_\text{2,saddle}={p}_{2,\text{flow}}^{0-}+{p}_\text{2,saddle}',\hspace{0.2in} {p}_\text{2,saddle}' \ll {p}_{2,\text{flow}}^{0-}
\end{equation}
where ${p}_{2,\text{flow}}^{0-}$ is given by equation \eqref{eq:FixedPointsNearFlow}. From $\dot{p}_\text{2,saddle}=0$, we obtain,
\begin{equation}
{p}_\text{2,saddle}=\frac{\sqrt{-b^2}c\cdot De+(4\log(2\kappa)-6)b^2}{c\cdot De(1-p_{1,saddle})-\sqrt{-b^2}(4\log(2\kappa)-6)}.\label{eq:P2Saddle}
\end{equation}
From equation \eqref{eq:FullOrbitEquationNewtExact}, $\dot{p}_\text{1,saddle}=\dot{p}_\text{3,saddle}=0$ is equivalent to,
\begin{equation}
p_\text{1,saddle} \alpha|_{p_1=p_\text{1,saddle},p_3=p_\text{3,saddle}}= p_\text{2,saddle}/(c\cdot De).
\end{equation}
{A general solution to the above equation is intractable, but we can solve it in the limit of small $p_\text{1,saddle}$ or small $p_\text{3,saddle}$. We denote the value of $p_\text{1,saddle}$ in the limit of small $p_\text{1,saddle}$ as $\tilde{p}_\text{1,saddle}$ and the value of $p_\text{3,saddle}$ in the limit of small $p_\text{3,saddle}$ as $\tilde{p}_\text{3,saddle}$.} We obtain,
\begin{equation}
\tilde{p}_\text{1,saddle}=\frac{\begin{split}
	&-8 (c\cdot De)^2 f^2+16 c\cdot De f^3 \sqrt{-b^2}\\&-\sqrt{
		\begin{split}
		&4c\cdot De f \left(-(c\cdot De)^2 \kappa^2+2 (c\cdot De) f \kappa^2 \sqrt{-b^2}+4 f^2\right) \\&4\left(-3 \pi  (c\cdot De)^2 \kappa^2+2 c\cdot De f \left(8 f+3 \pi  \kappa^2 \sqrt{-b^2}\right)+12 \pi  f^2\right)\\&+128 c\cdot De f^3 ((c\cdot De)-2 f \sqrt{-b^2})^2\end{split}}
	\end{split}
}{(c\cdot De) \left(3 \pi  (c\cdot De)^2 \kappa^2-2 (c\cdot De) f \left(8 f+3 \pi  \kappa^2 \sqrt{-b^2}\right)-12 \pi  f^2\right)},\label{eq:P1Saddle}
\end{equation}
and,
\begin{equation}
\tilde{p}_\text{3,saddle}=-\frac{\begin{split}&3 \pi  (c\cdot De) \left((c\cdot De)^2 \kappa^2-2 c\cdot De f \kappa^2 \sqrt{-b^2}-4 f^2\right)\\&+\sqrt{\begin{split}&9 \pi ^2 (c\cdot De)^2 \left(-(c\cdot De)^2 \kappa^2+2 c\cdot De f \kappa^2 \sqrt{-b^2}+4 f^2\right)^2\\&+\frac{1}{2}( c\cdot De)^3 \kappa^2 \left(3 \pi ^2 c\cdot De-2  f \left(3 \pi ^2 \sqrt{-b^2}+2\right)\right)\\&+\frac{1}{2} c\cdot Def^2\left(f (16-64 c\cdot De \sqrt{-b^2})+c\cdot De \left(8 \kappa^2 \sqrt{-b^2}-12 \pi ^2\right)\right) \\&\left(3 \pi ^2 (c\cdot De)^2 \kappa^2-2 c\cdot De f \left(16 f+3 \pi ^2 \kappa^2 \sqrt{-b^2}\right)-4 f^2 \left(16 f \sqrt{-b^2}+3 \pi ^2\right)\right)\end{split}}\end{split}}{2 \left(\frac{3}{4} \pi ^2 c\cdot De \left((c\cdot De)^2 \kappa^2-2 c\cdot De f \kappa^2 \sqrt{-b^2}-4 f^2\right)-8 c\cdot De f^2 ((c\cdot De)+2 f \sqrt{-b^2})\right)},\label{eq:P3Saddle}
\end{equation}
where $f=2\log(2\kappa)-3$ and $\sqrt{-b^2}$ is in equation \eqref{eq:bsqrandzeta}. The variation of $\tilde{p}_\text{3,saddle}$ and $\sqrt{1-\tilde{p}_\text{1,saddle}^2-{p}_\text{2,saddle}^2}$ with $c\cdot De$  for a few $\kappa$ is shown in the left panel of figure \ref{fig:p3saddle}.
From either of these approximations, we find that the saddle point moves closer to the vorticity axis as $c\cdot De$ is increased or $\kappa$ is reduced. Both approximations yield a similar location for the saddle point. The best approximation for the location of the saddle point for any $c\cdot De$ and $\kappa$, is $\mathbf{p}_\text{saddle}=\begin{bmatrix}
{p}_\text{1,saddle}&{p}_\text{2,saddle}&{p}_\text{3,saddle}
\end{bmatrix},$ where,
\begin{align}\begin{split}
{p}_\text{1,saddle}=&\tilde{p}_\text{1,saddle},\hspace{0.1in} {p}_\text{3,saddle}=\sqrt{1-\tilde{p}_\text{1,saddle}^2-{p}_\text{2,saddle}^2},\hspace{0.2in} \text{if }|\tilde{p}_\text{1,saddle}|\le|\tilde{p}_\text{3,saddle}|\\
{p}_\text{1,saddle}=&\sqrt{1-{p}_\text{2,saddle}^2-\tilde{p}_\text{3,saddle}^2},\hspace{0.1in}{p}_\text{3,saddle}=\tilde{p}_\text{3,saddle},\hspace{0.2in} \text{if } |\tilde{p}_\text{1,saddle}|>|\tilde{p}_\text{3,saddle}|
\end{split},\label{eq:SaddleConditional}\end{align}
and $p_2$ is from equation \ref{eq:P2Saddle}.
We show this saddle point's more accurate location of $p_3$ in the right panel of figure \ref{fig:p3saddle}. The location of this analytically predicted saddle point, shown with a green marker in figures \ref{fig:TrajectoriesR5}, \ref{fig:TrajectoriesR6} and \ref{fig:TrajectoriesVanishingLimitCycle}, agrees well with that inferred from the of numerically integrated trajectories.

In $R_{flow-vort}^{(2)}$ and $R_{flow-vort}^{(3)}$, the unstable manifold of this saddle point leads to two stable attractors, i.e., either a limit cycle near the vorticity axis or a stable node near the flow direction. Therefore, the location of the saddle point partially dictates the relative sizes of  the basins of attraction of these stable attractors. As $c\cdot De$ increases, due to the changing location of the saddle point, $\mathbf{p}_\text{saddle}$, the particle is more likely to attain a final orientation towards the flow axis as compared to approaching the limit cycle around the vorticity axis (see figures \ref{fig:p3saddle} and compare the proportion of the solid lines in \ref{fig:TrajectoriesR5} and \ref{fig:TrajectoriesR6}). The saddle point also persists in the large $c\cdot De$ regions $R_{flow}^{(1)}$ and $R_{flow}^{(2)}$ as shown in the different plots in figure \ref{fig:TrajectoriesVanishingLimitCycle}.

\subsubsection{{$R_{flow}^{(1)}$ and $R_{flow}^{(2)}$}: Very large $c\cdot De$ guarantees flow alignment}\label{sec:FlowAlignment}
\begin{figure}
	\centering	
	\subfloat{\includegraphics[width=0.49\textwidth]{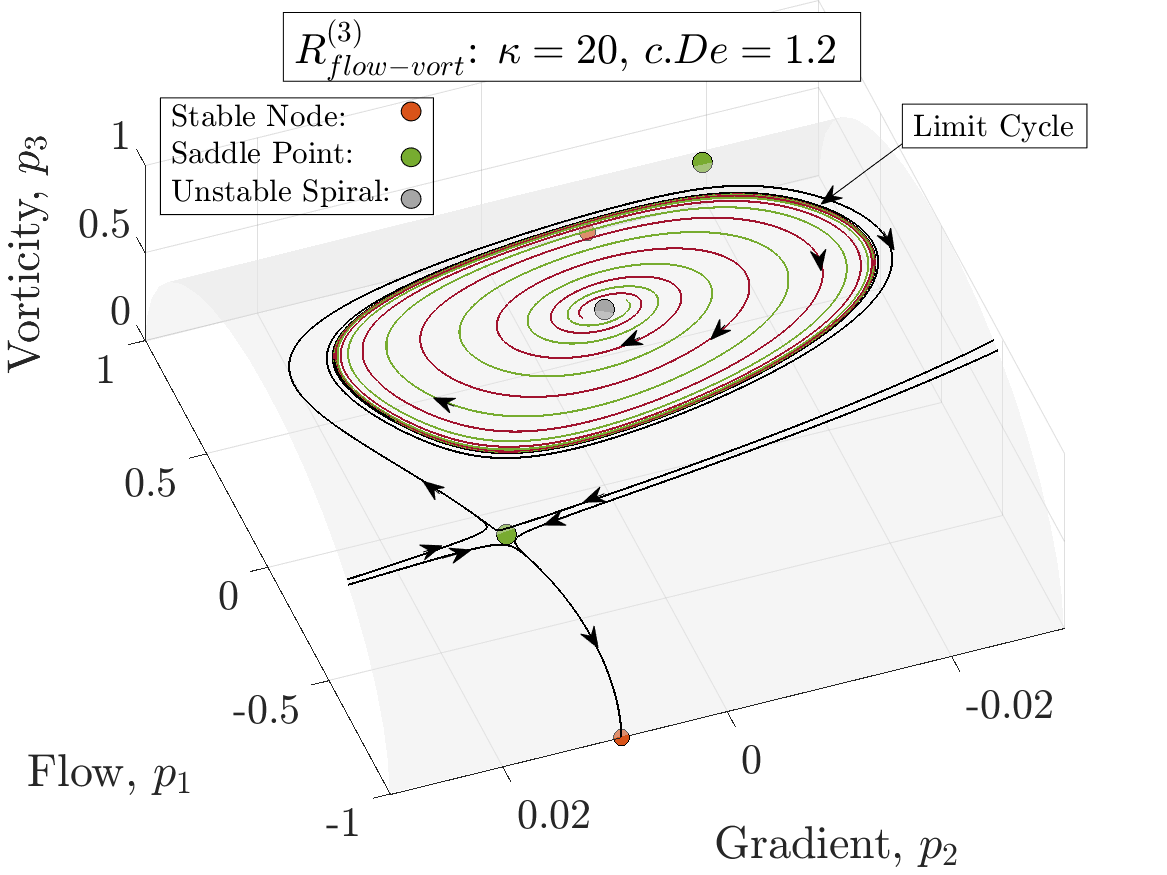}}\hfill
	\subfloat{\includegraphics[width=0.49\textwidth]{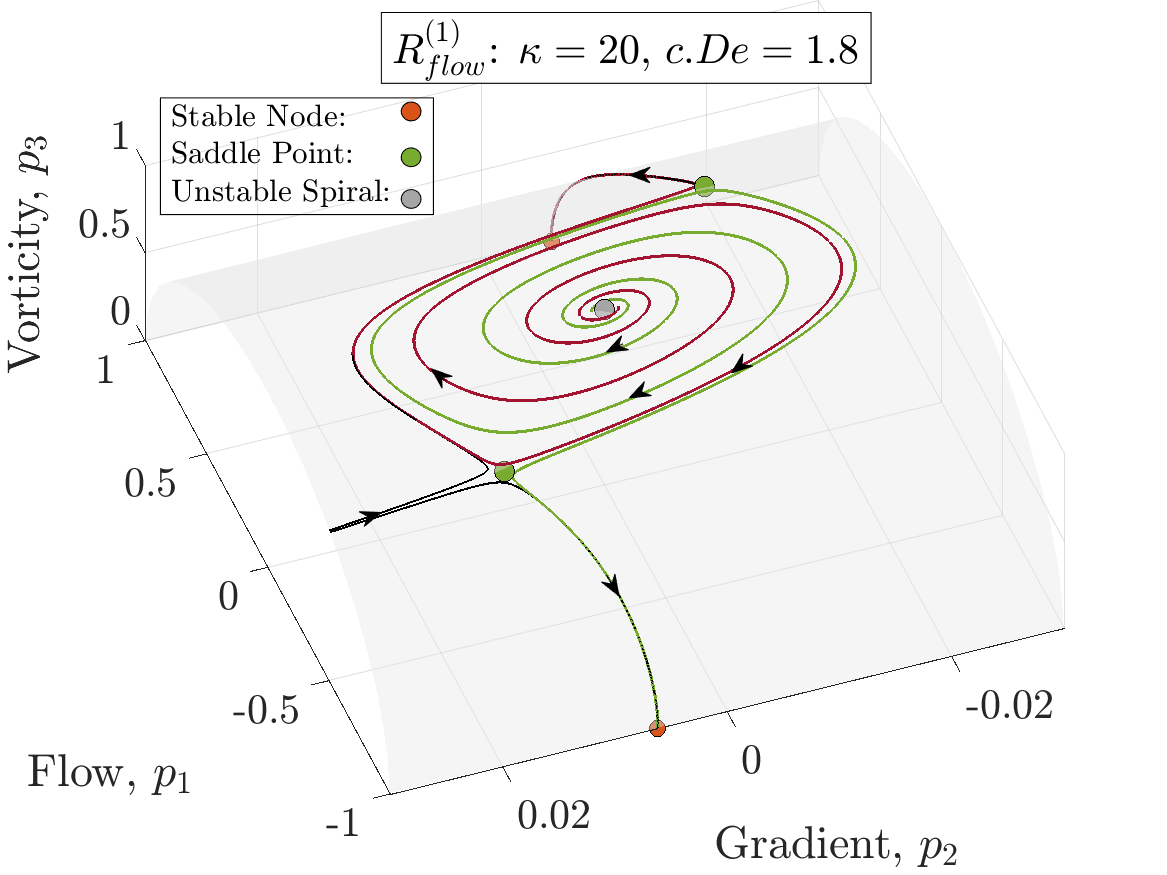}}\hfill	\subfloat{\includegraphics[width=0.49\textwidth]{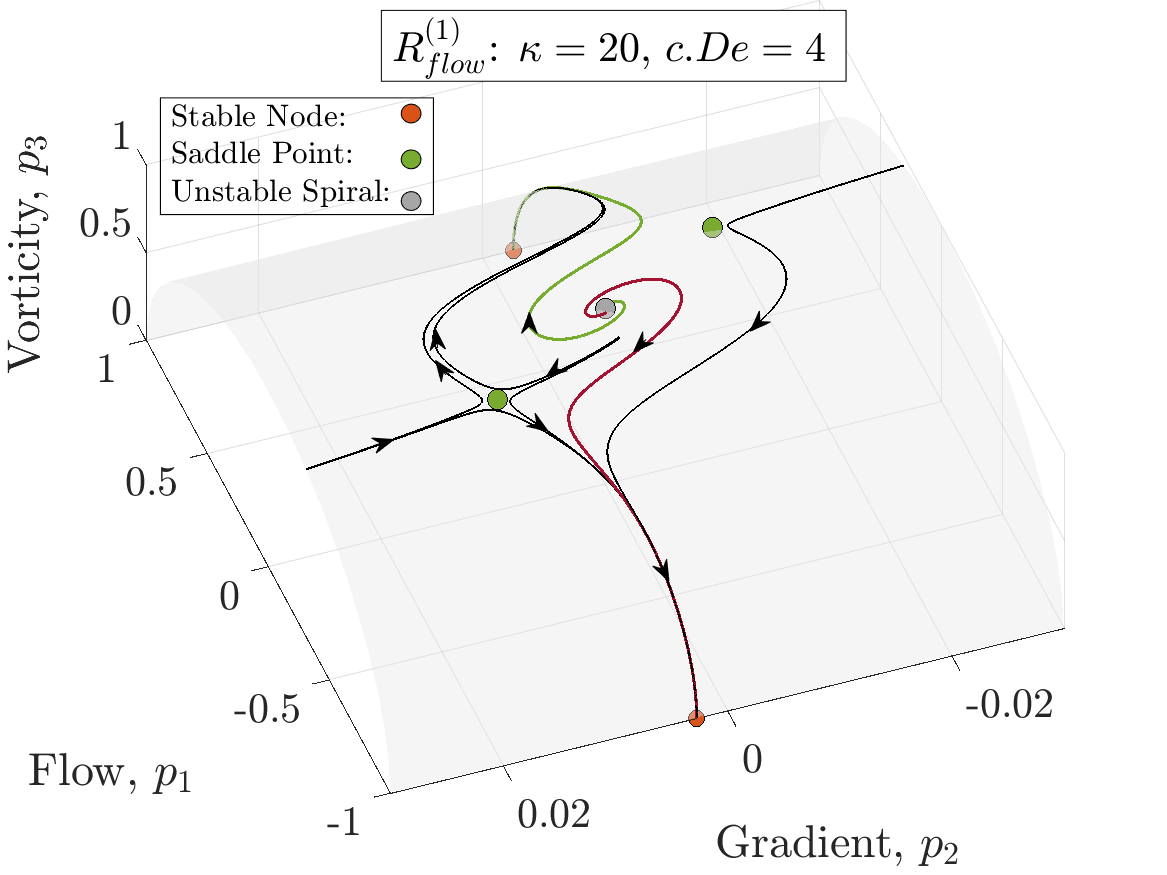}}\hfill	
	\subfloat{\includegraphics[width=0.49\textwidth]{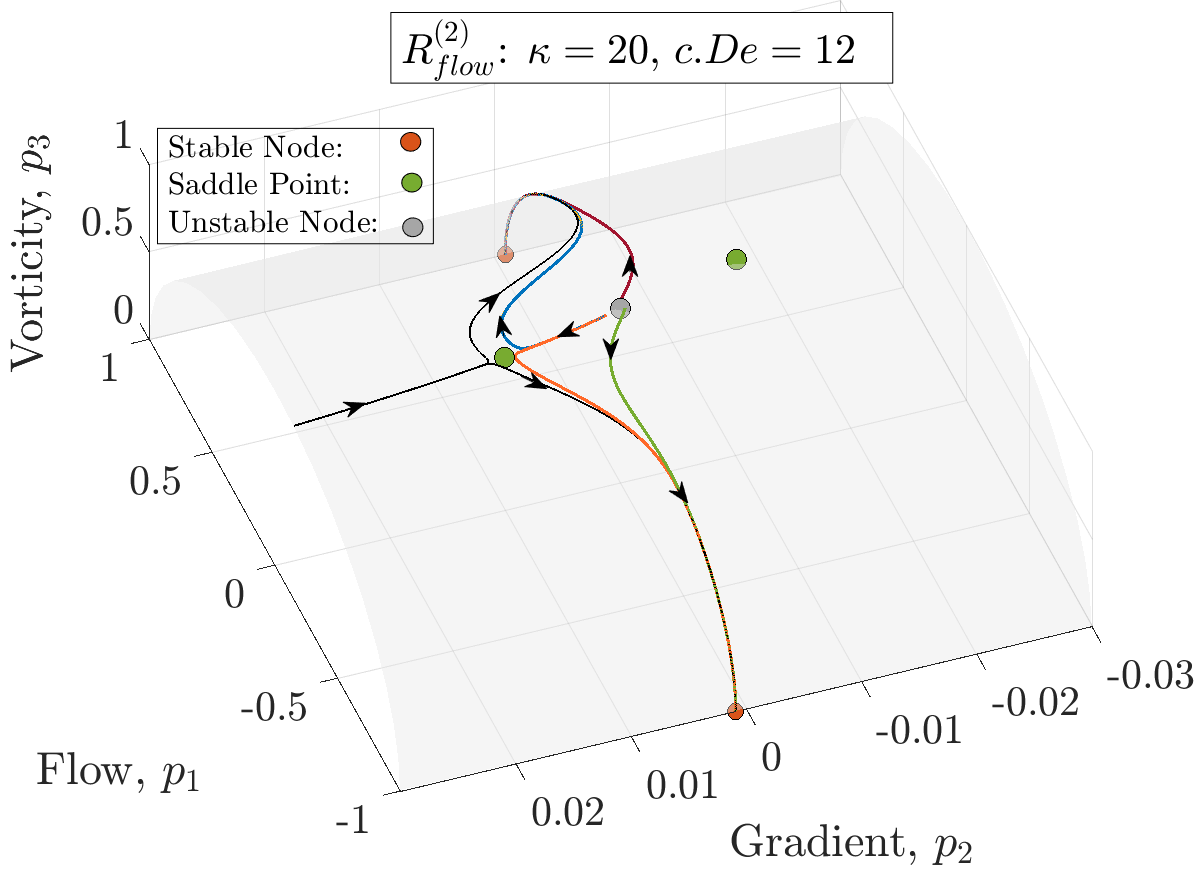}}
	\caption {Trajectories at large $c\cdot De$ for $\kappa=20$ in $R_{flow-vort}^{(3)}$, $R_{flow}^{(1)}$ and $R_{flow}^{(2)}$ zoomed near the flow-vorticity plane. At very large $c\cdot De$ (corresponding to regions $R_{flow}^{(1)}$ and $R_{flow}^{(2)}$), the stable limit cycle does not exist around the vorticity axis and a particle starting anywhere apart from vorticity axis ends up being nearly flow aligned.\label{fig:TrajectoriesVanishingLimitCycle}}
\end{figure}

In section \ref{sec:StableLC}, we found that for $\kappa\gtrapprox20$, the stable limit cycle moves closer to the vorticity axis upon increasing $c\cdot De$. The saddle point near the flow-vorticity plane that exists in regions shown in figure \ref{fig:PhasecDekapa} with $c\cdot De$ equal and larger than that in $R_{flow-vort}^{(2)}$ also moves towards the vorticity axis upon increasing $c\cdot De$ as discussed in section \ref{sec:SaddleonFV}. The saddle point moves towards the vorticity axis faster than the limit cycle. Beyond a $\kappa$-dependent value of $c\cdot De$, when the saddle point is too close to the limit cycle, the latter ceases to exist, making the stable fixed point near the flow direction the only stable attractor. This boundary is marked as $c\cdot De|_\text{cut-off}^\text{Limit}$ in figure \ref{fig:PhasecDekapa}. This implies that a particle with any initial orientation other than the vorticity axis ends up being aligned near the flow direction. Plots corresponding to $c\cdot De=1.2$ and 1.8 for $\kappa=20$   in figure \ref{fig:TrajectoriesVanishingLimitCycle} show the vanishing of the stable limit cycle near the vorticity axis upon going from $R_{flow-vort}^{(3)}$ to $R_{flow}^{(1)}$. For $c\cdot De=1.2$ ($R_{flow-vort}^{(3)}$) the saddle point is just outside the limit cycle, whereas for $c\cdot De=1.8$ ($R_{flow}^{(1)}$), corresponding to a location close to the $R_{flow-vort}^{(3)}-R_{flow}^{(1)}$ boundary, the limit cycle does not exist and instead the stable manifolds of the saddle point originate from the unstable spiral at  the vorticity axis. The unstable manifold of the saddle point leads to the stable node near the flow direction. The saddle point persists for larger $c\cdot De$ in $R_{flow}^{(1)}$ and also $R_{flow}^{(2)}$ as show by the $c\cdot De=4$ and 12 plots at $\kappa=20$ respectively in figure \ref{fig:TrajectoriesVanishingLimitCycle}. As also discussed in section \ref{sec:NearVorticity} (left panel of figure \ref{fig:TrajectoriesR1R6R7}) in $R_{flow}^{(2)}$, a particle leaving the vorticity axis changes from spiralling to monotonic trajectories as the vorticity axis changes from an unstable spiral to an unstable node when going from $R_{flow}^{(1)}$ to $R_{flow}^{(2)}$. From the $c\cdot De=4$ and 12 panels of figure  \ref{fig:TrajectoriesVanishingLimitCycle} at $\kappa=20$, the proximity of the saddle point to the vorticity axis can also explain the breakdown of spiral motion near the vorticity axis in $R_{flow}^{(2)}$.

\section{Conclusion}\label{sec:Conclusions}
Using a regular perturbation expansion in polymer concentration, $c$, the balance of torques on the particle surface at each order in $c$, and the velocity field generated by a slender prolate spheroid (obtained from the slender body theories of \cite{batchelor1970slender} and \cite{cox1971motion}), we develop a theory to characterize the orientation dynamics of a freely rotating (torque-free) particle in  simple shear flow of a viscoelastic fluid with small polymer concentration, $c$, and an arbitrary polymer relaxation time in the absence of inertia. This theory predicts a wide variety of orientation behaviors that are qualitatively similar to previous experimental observations of \citet{gauthier1971particle,bartram1975particle,iso1996orientation,iso1996orientation1}.

In the absence of inertia, for a viscoelastic fluid, where the fluid stress is a sum of the solvent and the polymer or non-Newtonian stress, the momentum equation can be decomposed into a Newtonian and a non-Newtonian part. The Newtonian part comprises the particle's motion in a Newtonian fluid undergoing the imposed flow and leads to the Newtonian stress field. The non-Newtonian part has zero velocity at the boundaries and comprises the balance of the divergence of the polymer induced solvent stress and the divergence of the polymer stress. Thus, three physically distinct stress mechanisms impacting the particle surface are the elastic or polymer stress, the polymer-induced solvent stress, and the Newtonian stress. The decomposition of torques in this way can be useful in obtaining further insights into the particle motion in a viscoelastic or a non-Newtonian fluid where the fluid stress can be decomposed into a Newtonian and a non-Newtonian part. In addition to the non-Newtonian momentum equation, different stress components are coupled with the torque-free boundary condition and the polymer constitutive equation. The balance of torques generated by the three stresses leads to the appropriate angular velocity to allow torque-free particle motion. The boundary conditions for the motion of the particle and the imposed flow are contained in the Newtonian part of the momentum equation. Hence, the torque generated from the corresponding stress is termed the motion induced solvent torque (MIST). The polymer constitutive equation is driven by the sum of the velocity field from the Newtonian and non-Newtonian momentum components. The elastic torque is a function of the polymer stress on the particle surface. By definition, the polymer-induced solvent torque (PIST) is the antisymmetric first moment of the polymer-induced solvent stress over the particle surface . However, using a generalized reciprocal theorem and the non-Newtonian momentum equation, we can express the PIST directly as a volume integral of the polymer stress.

Using a regular perturbation in the polymer concentration, $c$, we obtain the equations for particle's orientation dynamics in a small $c$ viscoelastic fluid subjected to a simple shear flow. At $\mathcal{O}(1)$, the elastic and the polymer induced solvent stress are zero. Thus, the particle orientation dynamics are the same as that in the simple shear flow of a Newtonian fluid, i.e., the particle undergoes \cite{jeffery1922motion} rotations. At $\mathcal{O}(c)$, all three mechanisms mentioned above lead to a finite torque. The $\mathcal{O}(c)$ Newtonian stress is that on a particle rotating (at $\mathcal{O}(c)$ velocity) in a quiescent Newtonian fluid. Therefore, the $\mathcal{O}(c)$ rotation rate is the one that generates enough $\mathcal{O}(c)$ MIST to balance the PIST and the elastic torque. The total $\mathcal{O}(c)$ rotation rate may be decomposed as the sum of $\dot{\mathbf{p}}^{(1)}_\text{Elastic}$ and $\dot{\mathbf{p}}^{(1)}_\text{PIST}$ i.e. the $\mathcal{O}(c)$ elastic and PIST generated rotation rates. The leading order velocity field is taken from the slender body theory of Batchelor \citep{batchelor1970slender} and Cox \citep{cox1971motion}, and this drives the $\mathcal{O}(c)$ polymer stress. As both the elastic torque and the PIST can be expressed as a function of the polymer stress, the $\mathcal{O}(c)$ rotation rate of the particle is obtained from the analytical velocity field from the slender body theory.

The polymer relaxation time ($\lambda$) is non-dimensionalized with the shear rate ($\dot{\gamma}$) to yield Deborah number, $De(=\lambda\dot{\gamma})$ ($c = 0$ or $De = 0$ implies a Newtonian fluid). We find $\dot{\mathbf{p}}^{(1)}_\text{Elastic}$, to be independent of $De$ and initially consider $\dot{\mathbf{p}}^{(1)}_\text{PIST}$, separately in the low and high $De$ limits. In the low $De$ limit, $\dot{\mathbf{p}}^{(1)}_\text{Elastic}$ and  $\mathcal{O}(De^0)$ $\dot{\mathbf{p}}^{(1)}_\text{PIST}$ are the rotation rates due to the additional torque from the rate of strain and fluid pressure respectively of the Newtonian fluid with an additional viscosity $c$. Therefore, in the low $De$ limit, the sum of $\dot{\mathbf{p}}^{(1)}_\text{Elastic}$ and the $\mathcal{O}(De^0)$ contribution to $\dot{\mathbf{p}}^{(1)}_\text{PIST}$ is identical to the rotation due to the Newtonian torque. Since the Newtonian torque is already balanced to be zero at the leading order in $c$, the first viscoelastic effect on the rotation rate for $De \ll 1$ arises at $\mathcal{O}(c. De)$ and is entirely from the polymer-induced solvent stress. In the high $De$ limit, $\dot{\mathbf{p}}^{(1)}_\text{PIST}\sim\mathcal{O}(De)$ and is hence $\mathcal{O}(De)$ larger than $\dot{\mathbf{p}}^{(1)}_\text{Elastic}$. Therefore, the particle-induced polymer stress is the primary agent changing the particle dynamics in both the low and high $De$ limits. By interpolating the rotation rate in $De$ between the large and small (up to $\mathcal{O}(De)$) $De$ limits, we obtain a uniformly valid equation for the particle orientation. We analyze the effect of particle aspect ratio, $\kappa$ and viscoelastic fluid parameters, polymer concentration ($c$), and relaxation rate ($De$) on this equation valid for any $De$.

Depending upon $c\cdot De$ and particle aspect ratio ($\kappa$), several qualitatively different particle orientation behaviors are possible. At constant $c\cdot De$, $c$ affects the behavior only quantitatively. In a Newtonian fluid ($c\cdot De=0$), a particle undergoes periodic motion on an initial condition dependent \cite{jeffery1922motion} orbit around the vorticity axis. As expected from symmetry, the vorticity axis and the flow-gradient plane (FGP) remain invariant within the orientational space (unit sphere). Viscoelasticity (for all $c\cdot De$ and $\kappa$) does not affect the rotation rate of the log-rolling state on the vorticity axis. In the FGP, even very small $c\cdot De$ creates a bottleneck near the flow direction where the particle slows down in proportion to $c\cdot De$.  Thus viscoelasticity increases the period of tumbling motion within the FGP. At $c\cdot De=cDe_{crit}=(4\log(2\kappa)-6)/\kappa$, an infinite period bifurcation occurs, and two fixed points arise in the FGP near the flow direction. Therefore, for $c\cdot De>cDe_{crit}$, the tumbling motion within the FGP does not exist, and a particle initially oriented within FGP ultimately aligns near the flow direction.

In a Newtonian fluid, the phase flow of the dynamical system of the  particle's orientational drift is of equal and opposite magnitude in the vorticity direction about the flow-vorticity plane. Thus the \cite{jeffery1922motion} orbits, in that case, are symmetric about the flow-vorticity plane. For $0<c\cdot De<cDe_{crit}$ and $\kappa>\kappa_{FV1}$, the phase flow remains qualitatively similar, but the magnitude of the phase velocity in the vorticity direction is larger on the downstream side of the phase-flow. Thus, a particle starting close to (but not in) the flow-gradient plane spirals towards the vorticity direction. $\kappa_{FV1}\approxeq17$ for $c=0.005$ and slightly reduces with $c$. For $\kappa<\kappa_{FV1}$, spiraling away from the FGP occurs for a $c\cdot De$ up to a value slightly less than $cDe_{crit}$ indicated by the $\zeta=0$ curve in figure \ref{fig:PhasecDekapa}.  $R_{vort}^{(1)}$ in figure \ref{fig:PhasecDekapa} is the region where the particle spirals away from the FGP. It continues to spiral towards the vorticity axis as it moves away from the FGP until it comes close to the axis. A particle with an initial orientation arbitrarily close to, but not on, the vorticity axis spirals away from the vorticity axis. Thus, in $R_{vort}^{(1)}$, a particle ultimately undergoes a periodic motion in a limit cycle near the vorticity axis. The spiraling towards the vorticity axis due to viscoelasticity is observed in the experiments of \cite{gauthier1971particle} and \cite{iso1996orientation,iso1996orientation1}.

For $\kappa<\kappa_{FV1}$ and between $c\cdot De$ corresponding to $\zeta=0$ and $c\cdot De=cDe_{crit}$, marked as $R_{flow-vort}^{(1)}$ in figure \ref{fig:PhasecDekapa}, the phase velocity in the vorticity direction downstream of the flow-vorticity plane and close to the FGP changes direction as compared to the $R_{vort}^{(1)}$ region discussed above or the case for a Newtonian fluid. Thus, a particle with an initial orientation close to the FGP spirals into the FGP, where it ultimately undergoes tumbling motion (albeit with a larger period than in Newtonian fluid). A particle starting further away from the FGP or near the vorticity axis spirals towards the periodic orbit (similar to $R_{vort}^{(1)}$) near the vorticity axis.

The phase space for the dynamical system of a particle in a Newtonian fluid consists of a concatenation of neutral centers. The periodic orbit at small angles away from the vorticity axis for the case of a viscoelastic fluid ($c\cdot De>0$) is a stable limit cycle. It is the only stable attractor in the phase space in $R_{vort}^{(1)}$. As $c\cdot De$ increases at large $\kappa\gtrapprox50$ in the small $c\cdot De$ regime, the limit cycle shrinks as it goes towards the vorticity axis. The size of the limit cycle is not affected by $c\cdot De$ in the small $c\cdot De$ regime for  $\kappa\lessapprox20$. The previous studies by \cite{leal1975slow} using the slender body theory at low $De$ have considered only the $\mathcal{O}(1/(2\log(2\kappa)-3))$ velocity disturbance and found viscoelasticity to cause a slender particle to spiral towards the vorticity axis. However, the polymeric torque due to the force doublet and $\mathcal{O}(1/\kappa^2)$ Stokeslet, $\mathcal{O}(1/\kappa^2)$ velocity disturbance from slender body theory of \cite{cox1971motion} that accounts for the fluid velocity disturbance the particle makes when in the flow-vorticity plane ($p_2=0$), makes the vorticity axis an unstable spiral. Thus, a stable limit cycle occurs from the competition of polymeric torques from the disturbance at $\mathcal{O}(1/\kappa^2)$ and  the Stokeslet flow at order $\mathcal{O}(p_2/(2\log(2\kappa)-3))$. Unlike \cite{leal1975slow} who attributes it to second normal stress difference, our theory predicts the effect of viscoelasticity in a second order fluid to arise from the first normal stress difference of the fluid. In $R_{vort}^{(1)}$, the FGP is an unstable limit cycle and, in $R_{flow-vort}^{(1)}$, it is a stable limit cycle. In both cases, the stable limit cycle exists near the vorticity axis. Thus, in $R_{flow-vort}^{(1)}$, there is an unstable limit cycle between the two stable ones, and there are two possible final orientation behaviors, tumbling within the FGP and periodic motion close to the vorticity axis.

 For $c\cdot De>cDe_{crit}=(4\log(2\kappa)-6)/\kappa$, the fixed points in the FGP leads to monotonic particle drift toward or away from the FGP instead of spiraling motion. For $\kappa>\kappa_{FV1}$ when the fixed points first appear at $cDe_{crit}$ the one closer to the flow direction is a saddle node, and the other is an unstable node. Thus a particle starting close to the FGP migrates along the stable manifold of the saddle point up towards the flow direction and then monotonically drifts along the flow-vorticity plane towards the periodic orbit close to the vorticity axis. The region where this qualitative behavior occurs is marked as $R_{vort}^{(2)}$ in figure \ref{fig:PhasecDekapa} and the orientation trajectories are similar to the experimental observations of \cite{bartram1975particle}. Also, for $\kappa>\kappa_{FV1}$, upon increasing $c\cdot De$ beyond a certain value indicated by the $\lambda_{2,\text{flow}}^{0-}=0$ boundary in figure \ref{fig:PhasecDekapa}, the saddle point changes into a stable node and hence a particle with an initial orientation close to FGP ultimately becomes stably flow aligned. This is marked as $R_{flow-vort}^{(2)}$ in figure \ref{fig:PhasecDekapa}. In this region, depending upon the initial orientation, a particle finally obtains either a stable orientation close to the flow direction or undergoes periodic motion close to the vorticity axis. For $c\cdot De$ beyond the $\lambda_{2,\text{flow}}^{0--}=0$ boundary of figure \ref{fig:PhasecDekapa} for $\kappa>\kappa_{FV1}$, or for $c\cdot De>cDe_\text{crit}$ when $\kappa<\kappa_{FV1}$, the fixed point in the FGP away from the flow direction is a saddle node with its stable manifold perpendicular to the FGP. In the part of this region marked as $R_{flow-vort}^{(3)}$ in figure \ref{fig:PhasecDekapa}, the final particle behavior is similar to that in $R_{flow-vort}^{(2)}$ discussed above. But, due to a saddle node instead of an unstable fixed point further from the flow direction in the FGP, a greater proportion of the initial orientations in $R_{flow-vort}^{(3)}$ than in $R_{flow-vort}^{(2)}$ lead the particle towards a flow aligned state. Upon increasing $c\cdot De$ beyond $c\cdot De|_{cut-off}^{Limit}$, in the regions marked as $R_{flow}^{(1)}$ and $R_{flow}^{(2)}$, the particle at all initial orientations apart from the log-rolling state at the vorticity axis lead to a flow-aligned state. This is because the limit cycle close to the vorticity axis does not exist and the vorticity axis is an unstable spiral ($R_{flow}^{(1)}$) or node ($R_{flow}^{(2)}$). The flow-alignment of the particle was observed in the experiments of \cite{iso1996orientation} at larger elasticity ($c$).

We are able to obtain qualitative agreement with the low shear rate experiments of \citet{gauthier1971particle} and \citet{iso1996orientation1} which correspond to low $c\cdot De$ regime ($R_{vort}^{(1)}$ in figure \ref{fig:PhasecDekapa}). In these experiments, the particle spirals towards the vorticity axis, and our theory predicts spiraling towards the stable limit cycle near the vorticity axis. While the authors \citep{gauthier1971particle,iso1996orientation1} state that the particle approaches the vorticity axis, this behavior is not shown. Moreover, in the time series plot of angle with the vorticity axis in the results of \citet{iso1996orientation1} a deviation from zero and a small oscillatory behavior can be observed, indicating the possibility of the existence of a stable limit cycle near the vorticity axis in their experiments. The high shear rate experiments of \citet{bartram1975particle} were performed by releasing a $\kappa=9.1$ rod close to the gradient axis. The particle initially travels nearly parallel to the flow-gradient plane and approaches the flow direction. Upon perturbing, it moves out of the flow-gradient plane along the flow-vorticity plane in a monotonic fashion before it starts spiraling near the vorticity axis. This is qualitatively similar to the orientation dynamics in the $R_{vort}^{(2)}$ region shown in figure \ref{fig:PhasecDekapa} where a saddle point exists near the flow direction (stable manifold in the flow-gradient plane) and a stable limit cycle is near vorticity axis. Our theory is however unable to fully explain the high shear rate experiments of \citet{iso1996orientation1} performed for $\kappa=34.4$, $De=3$ and $c=0.39$. \citet{iso1996orientation1} observe the particle to remain at an angle between 5 to 60$^\circ$ from the vorticity axis near the flow-vorticity plane after spiraling or moving monotonically away from the FGP. The regions $R_{flow-vort}^{(2)}$, $R_{flow-vort}^{(3)}$, $R_{flow}^{(1)}$, and $R_{flow}^{(2)}$ shown in figure \ref{fig:PhasecDekapa}) do have an attractor near the flow-vorticity plane, but it is a saddle point. From equation \eqref{eq:SaddleConditional} ${p}_\text{3,saddle}=0.7$ for $c\cdot De=1.17$ and $\kappa=34.4$ i.e. the saddle point is at an angle of 25.2$^\circ$ from vorticity axis. In our theory, the absence of a stable fixed point in the flow-vorticity plane between the vorticity and flow directions may be due to neglecting shear thinning, finite $c$ effects, finite polymer length, and polymer entanglement. We use an Oldroyd-B equation to model the polymer stress. Finite polymer length may be captured by the FENE-P model and polymer entanglement likely at larger $c$ by the Giesekus model \cite{bird1987dynamics}. Future numerical investigations may be used to test our theory and further clarify the previous experimental findings.

Our findings have a major implication in using dilute (low $c$) polymeric liquids to achieve desired properties such as strength and anisotropy of products manufactured from dilute fiber-filled suspensions mentioned in the introduction. At very low $c\cdot De$, all the fibers will eventually have orientations close to the vorticity axis. Therefore, adding a small polymer concentration to the fluid in roll-to-roll manufacturing can lead to low resistance films with higher anisotropy and hence better quality. Since the limit cycle in the low $c\cdot De$ regime becomes closer to the vorticity axis as $c\cdot De$ increases, the anisotropy can be tuned by changing the shear rate or polymer relaxation time (De is the product of shear rate and polymer relaxation time). Even higher anisotropy can be obtained if very large $De$ can be achieved since at a large $c\cdot De$, the fibers with all initial orientations ultimately align near the flow direction. For moderate values of $c\cdot De$, the flow field that precedes a period of simple shear may pre-orient the fibers somewhat closer to the flow-gradient plane or the vorticity axis. This will determine whether the fibers lie in the basin of attraction of attractor near the flow direction or the attractor near the vorticity axis. The polymer stresses in the simple shear flow can then drive the particles to a single final orientation leading to highly aligned fibers.

\section*{Acknowledgement}
This work was supported by NSF grants 1803156 and 2206851 and NASA grant 80NSSC23K0348.\\

Declaration of Interests: The authors report no conflict of interest.
\bibliographystyle{jfm}
\bibliography{MainDocument}

\end{document}